\documentclass[11pt]{article}
\usepackage{comment}
\usepackage{hyperref}
\usepackage{enumitem}
\usepackage{amsmath,amssymb,epsf,cite,graphicx,subfigure}
\numberwithin{equation}{section}
\newcommand{\beq}{\begin{equation}}
\newcommand{\eeq}{\end{equation}}
\newcommand{\bea}{\begin{eqnarray}}
\newcommand{\eea}{\end{eqnarray}}
\newcommand{\bdm}{\begin{displaymath}}
\newcommand{\edm}{\end{displaymath}}
\newcommand{\nn}{\nonumber}

\newcommand{\Rcal}{\mathcal{R}}

\def\<{\langle}
\def\>{\rangle}

\def\a{\alpha}

\def\L{\Lambda}
\def\O{\Omega}

\def\S{\Sigma}

\def\cm{{\cal M}}

\def\co{{\cal O}}

\def\car{{\cal R}}

\newcommand{\gz}{g^{(0)}}
\newcommand{\go}{g^{(1)}}

\newcommand{\tr}{\textrm{tr}}

\newcommand{\arctanh}{\mathop{\rm arctanh}\nolimits}

\def \nn {{\mathbb N}}

\def\N{{\mathcal{N}}}
\def\A{{\mathcal{A}}}
\def\Am{\mathcal{A}_{\textrm{min}}}

\def\CP{\mathbb{CP}}
\def\RP{\mathbb{RP}}
\def\Sp{\mathbb{S}}

\def\nn{\nonumber}

\def\see{S_{\textrm{EE}}}
\def\seuc{S_{\textrm{Euc}}}
\def\sbr{S_{\textrm{brane}}}
\def\seez{S_{\textrm{EE}}^{(0)}}
\def\seeo{S_{\textrm{EE}}^{(1)}}
\def\seh{S_{\textrm{EH}}}

\def\sbrren{S_{\textrm{brane}}^{\textrm{ren}}}
\def\sct{S_{\textrm{CT}}}
\def\seuc{S^{\textrm{euc}}_{\textrm{brane}}}
\def\sqz{S_q^{(0)}}
\def\sqo{S_q^{(1)}}

\def\fz{F^{(0)}}
\def\fo{F^{(1)}}
\def\sz{S^{(0)}}
\def\so{S^{(1)}}

\def\ceo{c^{(1)}}
\def\cdz{c_d^{(0)}}
\def\cdo{c_d^{(1)}}
\def\cto{C_T^{(1)}}
\def\reps{r_{\varepsilon}}

\def\xeps{x_{\varepsilon}}
\def\Ocal{{\mathcal{O}}}
\def\vol{\textrm{vol}}
\def\Hy{\mathbb{H}}
\def\sech{\textrm{sech}\,}

\def\tanh{\textrm{tanh}\,}
\def\rp{r_{\parallel}}
\def\rn{r_{\perp}}

\def\Zk{\mathbb{Z}_k}
\def\DW{\textrm{diff}\rtimes\textrm{Weyl}}

\begin{document}
\baselineskip=15.5pt
\pagestyle{plain}
\setcounter{page}{1}

\begin{flushright}
{YITP-SB-13-27
\\
DAMTP-2013-54}
\end{flushright}

\vskip 0.8cm

\begin{center}
{\LARGE \bf Holography, Entanglement Entropy, and Conformal Field Theories with Boundaries or Defects}
\vskip 0.5cm

Kristan Jensen$^{1,2,a}$ and Andy O'Bannon$^{3,b}$

\vspace{0.1cm}

{\it ${}^1$ Department of Physics and Astronomy, University of Victoria,\\ Victoria, BC V8W 3P6, Canada\\}

\vspace{0.1cm}

{\it ${}^2$ C. N. Yang Institute for Theoretical Physics, SUNY, \\ Stony Brook, NY 11794-3840, United States\\}

\vspace{.1cm}

{\it ${}^3$ Department of Applied Mathematics and Theoretical Physics, University of Cambridge,\\
Cambridge CB3 0WA, United Kingdom  \\}

\vspace{0.1cm}

{\tt  ${}^a$kristanj@max2.physics.sunysb.edu, ${}^b$A.OBannon@damtp.cam.ac.uk \\}

\medskip

\end{center}

\vskip0.3cm

\begin{center}
{\bf Abstract}
\end{center}
We study entanglement entropy (EE) in conformal field theories (CFTs) in Minkowski space with a planar boundary or with a planar defect of any codimension. In any such boundary CFT (BCFT) or defect CFT (DCFT), we consider the reduced density matrix and associated EE obtained by tracing over the degrees of freedom outside of a (hemi-)sphere centered on the boundary or defect. Following Casini, Huerta, and Myers, we map the reduced density matrix to a thermal density matrix of the same theory on hyperbolic space. The EE maps to the thermal entropy of the theory on hyperbolic space. For BCFTs and DCFTs dual holographically to Einstein gravity theories, the thermal entropy is equivalent to the Bekenstein-Hawking entropy of a hyperbolic black brane. We show that the horizon of the hyperbolic black brane coincides with the minimal area surface used in Ryu and Takayanagi's conjecture for the holographic calculation of EE. We thus prove their conjecture in these cases. We use our results to compute the R\'enyi entropies and EE in DCFTs in which the defect corresponds to a probe brane in a holographic dual.
\newpage

\tableofcontents

\section{Introduction, Summary, and Outlook} \label{S:intro}

\subsection{Introduction and Summary}

Conformal field theories (CFTs) are of central importance in many branches of physics, two prominent examples being critical phenomena and string theory. These examples also illustrate the importance of studying CFTs on spaces with boundaries: any critical system in a laboratory will necessarily have finite size, and in string theory an open string worldsheet has a boundary. Furthermore, these examples highlight the importance of defects in CFTs. For example, the Kondo effect admits a description as a renormalization group (RG) flow from one (1+1)-dimensional CFT to another, triggered by a point-like impurity~\cite{Affleck:1995ge}. Such a description applies not only to Fermi liquids doped with magnetic impurities~\cite{Affleck:1995ge} but also to some cases of D-brane decay~\cite{Schomerus:2002dc}.

By definition, the correlation functions of a CFT in $d$-dimensional Minkowski spacetime are invariant under the action of the conformal group $SO(2,d)$, which includes the Poincar\'e group as a subgroup, as well as dilatations and special conformal transformations. A boundary CFT (BCFT) is a CFT on a space with a boundary, with boundary conditions that preserve dilatations. A defect CFT (DCFT) is a CFT deformed by a defect, of spatial co-dimension $n$, that preserves dilatations. In what follows we exclusively consider CFTs in Minkowski space\footnote{Everything we will do, both in field theory and in holography, could be straightforwardly generalized to BCFTs and DCFTs in $\mathbb{R} \times \Sp^{d-1}$, where $\mathbb{R}$ denotes time.} with boundaries or defects that preserve planar symmetry. A planar defect preserves the $SO(2,d-n) \times SO(n) \subset SO(2,d)$ whose generators leave the defect's position unchanged~\cite{Cardy:1984bb,McAvity:1993ue,McAvity:1995zd}. A planar boundary preserves the same symmetry as a planar $n=1$ defect.

In a quantum system, such as a CFT, BCFT, or DCFT, one measure of quantum entanglement between subsystems is entanglement entropy (EE). To define an EE we artificially divide the Hilbert space into two subspaces and then trace over the states in one subspace to obtain a reduced density matrix $\rho$ in the other subspace. The EE, $\see$, is the von Neumann entropy of $\rho$,
\beq
\label{eedef}
\see \equiv - \tr \left( \rho \ln \rho \right).
\eeq
We will divide the Hilbert space into subspaces describing states in different spatial regions. In a time-independent state we can choose a fixed time slice without loss of generality, and then divide space into a region $\cm$ and its complement $\overline{\cm}$, with these two separated by an ``entangling surface'' $\S$. For such a division, eq.~\eqref{eedef} gives us the EE of the degrees of freedom in $\cm$ due to their quantum entanglement with the degrees of freedom in $\overline{\cm}$.

The EE is a special case of R\'enyi entropy~\cite{renyi1,renyi2}, $S_q$, defined in terms of $\rho$ as
\beq
\label{E:sqdef}
S_q \equiv \frac{1}{1-q} \, \ln \left[ \tr \left( \rho^q \right) \right]\,,
\eeq
where $q$ is a non-negative integer. We obtain $\see$ by analytically continuing $q$ to non-integer values and then taking the $q\to 1$ limit of $S_q$. Other limits of $S_q$ provide additional information about the eigenvalue spectrum of $\rho$. We can extract the total number of non-vanishing eigenvalues of $\rho$ from the $q \to 0$ limit, and the largest eigenvalue of $\rho$ from the $q \to \infty$ limit. Extracting such information from limits of $S_q$ requires caution because $S_q$ is not necessarily analytic in $q$.

EE has many possible uses. For example, in some cases EE can measure the number of degrees of freedom in a quantum field theory. For a CFT in even $d$, for a spherical $\Sigma$ one contribution to the EE is proportional to the central charge~\cite{Holzhey:1994we,Casini:2011kv}, defined from the coefficient of the Euler density in the trace anomaly. For $d=2$ and $d=4$ proofs exist~\cite{Zamolodchikov:1986gt,Komargodski:2011vj} that these central charges are strictly non-increasing along an RG flow from an ultra-violet (UV) fixed point to an infra-red (IR) fixed point, a necessary condition for anything that counts degrees of freedom. In other $d$, the proposal that EE may count degrees of freedom remains conjectural~\cite{Myers:2010xs,Myers:2010tj,Liu:2012eea,Casini:2012ei,Klebanov:2012va}.

For a BCFT or DCFT in $d=2$, the EE of an interval containing the boundary or symmetric about the defect includes a contribution independent of the size of the interval, called the boundary entropy or impurity entropy~\cite{Affleck:1991tk,Calabrese:2004eu,Azeyanagi:2007qj,Affleck2009}, which is strictly non-increasing along an RG flow from one BCFT or DCFT to another~\cite{Affleck:1991tk,Friedan:2003yc}. Intuitively, the boundary or impurity entropy counts degrees of freedom localized at the boundary or impurity. Whether a boundary or defect entropy with such properties can be defined when $d>2$, or can be extracted from an EE, remain open questions.

In general, however, $\see$ and $S_q$ are difficult to calculate even in free quantum field theories. Fortunately, for some CFTs the Anti-de Sitter/CFT (AdS/CFT) correspondence~\cite{Maldacena:1997re} provides a relatively simple way to compute $\see$, and in some cases $S_q$. AdS/CFT is a holographic duality: it equates certain weakly-coupled theories of quantum gravity on $(d+1)$-dimensional AdS space, $AdS_{d+1}$, with certain strongly-coupled $d$-dimensional CFTs ``living'' at the $AdS_{d+1}$ boundary. The isometry group of $AdS_{d+1}$ is $SO(2,d)$, which is dual to the conformal group. In the best-understood examples, the CFTs are non-abelian gauge theories in the 't Hooft large-$N$ limit and with large 't Hooft coupling. In these cases, the dual gravity theory is usually a string theory in the semiclassical limit, where it is well-approximated by classical supergravity with small Newton's constant $G$, for example when $d=4$ typically $G \propto 1/N^2$. The classical action of supergravity consists of an Einstein-Hilbert term plus terms for matter fields. In what follows we consider only such Einstein theories of gravity, unless stated otherwise.

For Einstein theories of gravity, Ryu and Takayanagi (RT) conjectured that to compute the EE in a time-independent state using holography, in the bulk gravity theory we must determine the time-independent surface of spatial codimension one with minimal area $\Am$ that approaches $\Sigma$ at the $AdS_{d+1}$ boundary~\cite{Ryu:2006bv,Ryu:2006ef}. Specifically, RT conjecture that $\see$ is
\beq
\label{rt}
\see = \frac{\Am}{4 G}.
\eeq
Eq.~\eqref{rt} represents only the leading contribution to EE in the large-$N$ and large-coupling limits. For example, if $G \propto 1/N^2$, then the $\see$ in eq.~\eqref{rt} will generically be order $N^2$. RT's proposal has been applied to study various questions and conjectures about EE, including whether EE can count of degrees of freedom in arbitrary $d$~\cite{Myers:2010xs,Myers:2010tj}. 

Compelling evidence in support of RT's conjecture continues to accumulate~\cite{Lewkowycz:2013nqa}, including rigorous proofs for $d=2$~\cite{Hartman:2013mia,Faulkner:2013yia}. Casini, Huerta, and Myers (CHM)~\cite{Casini:2011kv} also provided a proof in some special cases, as we now briefly review. CHM began by proving a remarkable statement about the $\rho$ for spherical $\S$ in CFTs. For any CFT in any $d$, when $\Sigma$ is a sphere of radius $R$, CHM showed that $\rho$ is identical to the thermal density matrix of the same CFT on a hyperbolic space, with radius of curvature $R$, at the temperature $1/(2\pi R)$. As a result, $\see$ maps to a thermal entropy and $S_q$ is proportional to the difference of free energies of the CFT on the hyperbolic space at temperatures $1/(2\pi R)$ and $1/(2 \pi R q)$~\cite{Hung:2011nu}. To prove these facts, CHM rely only on conformal symmetry and the high degree of symmetry of a spherical $\Sigma$.

For CFTs holographically dual to an Einstein gravity theory, CHM also translated their mapping to a change of foliation of $AdS_{d+1}$. In particular, CHM showed that the ``hyperbolic slicing'' of $AdS_{d+1}$ reveals a black brane with a hyperbolic horizon of radius $R$ and Hawking temperature $1/(2\pi R)$. Of course, the spacetime is merely $AdS_{d+1}$, and the horizon is an artifact of the coordinate choice, being the horizon seen by an observer with a particular acceleration~\cite{Emparan:1999gf}. Via holography, the Bekenstein-Hawking entropy of the black brane is the thermal entropy of the dual CFT on hyperbolic space, and hence is the $\see$ of a spherical region of radius $R$ of the CFT in Minkowski space. Crucially, CHM observe that the black brane's horizon coincides precisely with the minimal surface whose $\Am$ is in eq.~\eqref{rt}. CHM thus rigorously prove RT's conjecture in these special cases of spherical $\S$ in CFTs. Moreover, the difference in the free energy of the black brane at temperatures $1/(2\pi R)$ and $1/(2\pi R q)$ gives the R\'enyi entropies $S_q$~\cite{Hung:2011nu}.

In this paper we extend CHM's results to BCFTs and DCFTs. In section~\ref{S:CHMcft} we show that for a BCFT in any $d$ with a planar boundary or a DCFT in any $d$ with a planar defect of codimension $n$, when $\Sigma$ is a (hemi-)sphere centered on the boundary or defect the reduced density matrix $\rho$ is idential to the thermal density matrix of the BCFT or DCFT in a hyperbolic space. As a result, $\see$ maps to the thermal entropy and $S_q$ is proportional to a difference of free energies of the same BCFT or DCFT on hyperbolic space, as in CHM's mapping. Our proof relies on symmetry alone and in particular does not depend on the existence of a holographic dual. In section~\ref{S:CHMads} we consider BCFTs and DCFTs holographically dual to Einstein gravity theories. We write the most general metric that is asymptotically locally $AdS_{d+1}$ but has only $SO(2,d-n) \times SO(n)$ isometry, dual to the conformal group of the BCFT or DCFT. We determine the slicing that reveals a hyperbolic horizon, and demonstrate that the horizon coincides with the surface of minimal area $\Am$.

In section~\ref{S:probes} we consider a class of DCFTs dual to Einstein gravity theories in which the defect is described by a brane of codimension $n$. To be precise, we write a classical action including only an Einstein-Hilbert term, negative cosmological constant, and a brane action consisting of a tension, which we denote $\mu_n/(16 \pi G)$, times the brane's worldvolume. Such an action often arises in string and M-theory as an effective action in the near-horizon limit of various string and brane intersections, when we can neglect the other fields sourced by the strings or branes, such as the dilaton or Ramond-Ramond (RR) form fields. In many of the string and M-theory examples, a number $N_f$ of branes is typically dual to $N_f$ fields in the fundamental representation of the gauge group, \textit{i.e.}\ flavor fields, restricted to propagate along a defect of codimension $n$~\cite{Karch:2000gx,DeWolfe:2001pq,Karch:2002sh,Constable:2002xt}. For such flavor branes typically $\mu_n/G \propto N_f N$, and so if for example $G \propto 1/N^2$, then $\mu_n \propto N_f/N$.

When $\mu_n =0$ the bulk solution is simply $AdS_{d+1}$. When $\mu_n$ is non-zero, the brane back-reacts, deforming the spacetime and producing a metric with $SO(2,d-n)\times SO(n)$ isometry. In that case the bulk theory is dual to a DCFT. If $\mu_n$ is non-zero but small, in units of the curvature scale, then we can take the probe limit: expand all quantities in $\mu_n$ and retain terms up to linear order. In that case the bulk spacetime is an undeformed $AdS_{d+1}$ probed by the brane. In many string and M-theory examples, our effective action is only valid in the probe limit: often, at higher order in $\mu_n$ the brane sources fields that break the $SO(2,d-n) \times SO(n)$ isometry.

The probe limit is a useful simplification: for many field theory obervables we can compute the order $\mu_n$ contribution holographically just from the probe brane action in an undeformed ($\mu_n=0$) background. What about EE, however? Na\"ively, to compute the order $\mu_n$ contribution to EE we must calculate the order $\mu_n$ change in $\Am$, which requires calculating the order $\mu_n$ change in the metric. Precisely such calculations were performed in ref.~\cite{Chang:2013mca}, for branes with $n=1$ and $n=0$ (a space-filling brane) with any $d$, including two examples from string theory with $d=4$.

In section~\ref{S:bottomup} we show that for any brane described by our effective action, and for a spherical $\S$ centered on the defect, we can compute the order $\mu_n$ contribution not only to $\see$ but also to $S_q$ without ever computing back-reaction or a minimal area surface, using our results from section~\ref{S:dcfts}. More specifically, our method uses two key ingredients. The first ingredient is our mapping of the reduced density matrix to the thermal density matrix in hyperbolic space, so that our goal is then to compute the order $\mu_n$ contribution to the free energy and thermal entropy in hyperbolic space. In the holographic dual we thus consider the probe brane in the hyperbolic black brane background. The second ingredient is the fact that at order $\mu_n$ the free energy can receive only two contributions, one from the brane's action, evaluated in the undeformed background, and one from a boundary term proportional to the order $\mu_n$ correction to the metric. We can show on general grounds that the latter vanishes. As a result, we can obtain the order $\mu_n$ contribution to both $\see$ and $S_q$ simply by evaluating the probe brane action in the undeformed hyperbolic black brane background, rather than by computing back-reaction and the minimal area surface. For $n=1$ and $n=0$ our results for $\see$ agree perfectly with those of ref.~\cite{Chang:2013mca} (as mentioned already in ref.~\cite{Chang:2013mca}), providing two non-trivial checks of our method. Our results for $\see$ for other $n$, and our results for $S_q$, are novel.

In section~\ref{S:topdown} we apply our results to probe branes in $AdS_{d+1}$ backgrounds of string and M-theory. We begin by enumerating the criteria that a brane in string or M-theory must satisfy for our effective action to be reliable in the probe limit, and hence for our method to work. For example, one criterion is that the only contribution to the (Einstein-frame) stress-energy tensor at order $\mu_n$ is from the brane~\cite{Chang:2013mca}. We then apply our method to obtain the order $\mu_n$ contribution to $\see$ in four examples that meet the criteria. The first two are the string theory examples of ref.~\cite{Chang:2013mca}, with which we find perfect agreement. We then consider two examples in M-theory of DCFTs in $d=3$ with $n=1$ and $n=2$. 

What if a brane's order $\mu_n$ back-reaction preserves $SO(2,d-n) \times SO(n)$ isometry, but our effective action is not reliable in the probe limit? That is, what if we cannot neglect the other fields sourced by the brane? In such cases our mapping of $\see$ to a thermal entropy in hyperbolic space remains valid. At order $\mu_n$ that thermal entropy will receive contributions from the probe brane action and from boundary terms, each of which is proportional to an order $\mu_n$ correction to a bulk field sourced by the brane. As in cases above, the boundary term involving the order $\mu_n$ correction to the metric vanishes. The boundary terms involving other fields may be non-zero, however. If they are non-zero, then we cannot compute the order $\mu_n$ contributions to $\see$ and $S_q$ from the probe brane action alone. Indeed, we conclude section~\ref{S:topdown} with an explicit example where such a boundary term is non-zero. We consider an $n=0$ brane in an $AdS_4$ background of type IIA supergravity whose back-reaction preserves $AdS_4$. We argue that the RR two-form sourced by the brane cannot be neglected in Einstein-frame stress-energy tensor at order $\mu_n$. We then show, using the fully back-reacted solution, that in the thermal entropy the boundary term involving the order $\mu_n$ correction to the RR two-form is indeed non-zero, and contributes to $\see$.

In short, our method has two requirements. First, the back-reacted solution must have $SO(2,d-n) \times SO(n)$ isometry to order $\mu_n$, so that our mapping of $\see$ to a thermal entropy on hyperbolic space is valid. Second, any boundary terms involving the order $\mu_n$ corrections to bulk fields sourced by the brane must vanish, at least to order $\mu_n$. Proving these does not necessarily require computing back-reaction. In particular, showing that a simple action, including just the metric, cosmological constant, and brane, is a reliable effective description is sufficient to satisfy both requirements.

\subsection{Outlook}

Looking to the future, we hope that our results may be helpful for studying many questions about EE in BCFTs and DCFTs, both with and without holographic duals.

For instance, our results may be useful for generalizing boundary or impurity entropy to $d>2$, especially in conjunction with existing proposals motivated by holography~\cite{Yamaguchi:2002pa,Takayanagi:2011zk,Fujita:2011fp,Nozaki:2012qd}. Indeed, in our two examples from M-theory of DCFTs with $d=3$ and $n=1$ or $n=2$, the order $\mu_n$ contribution to $\see$ takes a form very suggestive of a defect or impurity entropy, respectively. In particular, in our $d=3$ and $n=1$ example the defect contribution to $\see$ resembles that of a CFT in $d=2$, allowing us to identify a central charge associated with the defect degrees of freedom.

CHM's mapping of EE to thermal entropy on a hyperbolic space is valid for \textit{any}\ CFT, including CFTs with holographic duals even beyond Einstein gravity. In particular, for a classical gravity theory in $AdS_{d+1}$ whose action includes the Einstein-Hilbert term plus higher-derivative terms, such as Gauss-Bonnet gravity, we can use Wald's entropy formula to compute the thermal entropy associated with the hyperbolic horizon, and so compute $\see$. CHM's mapping thus provides access to higher-derivative corrections to RT's prescription, which typically correspond  in the CFT to finite-coupling corrections to $\see$~\cite{Hung:2011nu,Galante:2013wta}. Similarly, quantum corrections to the hyperbolic black brane entropy will correspond to $1/N$ corrections to $\see$~\cite{Barrella:2013wja}. A precise proposal for the form of the leading $1/N$ correction appears in ref.~\cite{Faulkner:2013ana}. Our results may help to compute such higher-curvature and/or quantum corrections in the duals of BCFTs and DCFTs.

Our results may also be useful for studying the relationship between EE and black hole horizons. In particular, we have in mind the ``ER=EPR'' conjecture~\cite{Maldacena:2013xja}, the proposal that maximally-entangled Einstein-Podolsky-Rosen (EPR) pairs are connected by Einstein-Rosen (ER) bridges. In fact, one of the current authors has already used our results to demonstrate that the holographic dual of an EPR pair, namely a certain kind of probe string in $AdS_{d+1}$, indeed exhibits an ER bridge~\cite{Jensen:2013ora}, providing a concrete example of the ER=EPR conjecture.

EE can also help to characterize states of matter. For example, EE may characterize topological order when local order parameters are insufficient~\cite{Kitaev:2005dm,Levin:2006zz}, including in topologically-ordered gapped states, \textit{i.e.}\ topological insulators (TIs). Any TI described by a free electron Hamiltonian is characterized completely by ``edge modes,'' gapless modes localized at the interface between the TI and the vacuum, or more generally any other TI state. Some important open questions are: what TI states are possible in interacting systems, \textit{i.e.}\ systems that admit no description in terms of free electrons? Are such interacting TIs still characterized completely by their edge modes? To address such questions, holographic duals of interacting TIs have been constructed using branes in string theory~\cite{HoyosBadajoz:2010ac,Estes:2012nx}. The properties of the edge modes in these holographic TIs remain mysterious. For any TI, gapping the edge modes and integrating them out leads at low energy to a DCFT. At that level, our results could help determine what role, if any, EE might play in characterizing interacting TI states, including holographic TIs and their mysterious edge modes.

EE may also help to characterize non-Fermi liquids. Specifically, EE may be able to reveal ``hidden'' Fermi surfaces that are otherwise difficult to detect~\cite{Ogawa:2011bz,Huijse:2011ef}, such as quark Fermi surfaces, which have no obvious gauge-invariant definition. Holography provides access to many strongly-interacting non-Fermi liquids. In particular, many probe brane systems describe non-Fermi liquids with various bizarre properties: no broken symmetries, no obvious Fermi surface yet a zero sound excitation~\cite{Karch:2008fa}, a non-zero extensive entropy at zero temperature~\cite{Karch:2008fa,Karch:2009eb}, an unusual moduli space~\cite{Chang:2012ek,Ammon:2012mu}, and so on. Calculations of EE using our results could help to characterize such states, perhaps by revealing ``hidden'' Fermi surfaces.

\section{CHM for BCFTs and DCFTs} \label{S:dcfts}

We begin with some reminders about spurionic invariances and global symmetries of CFTs, BCFTs, and DCFTs, as these will play a key role in what follows. Consider some quantum degrees of freedom coupled to a set of fixed external sources. For example, in a quantum field theory with a Lagrangian description, we can imagine a functional integral representation of the generating functional, and treat the metric, masses, couplings, etc., as sources for various operators. We define a theory by choosing the values of these sources at some energy scale. To compute correlators, we vary the generating functional with respect to the sources and then set the sources to our chosen values.

Spurionic invariances are transformations under which the sources might change, but the value of the generating functional remains invariant, up to anomalies. A useful analogy comes from the integral calculus: given an integral (generating functional) that depends on some parameters (sources), if we change the integration variables (transform the fields), then the form of the integrand may change, but the value of the integral will not. Global symmetries are the subgroup of the spurionic invariances that leave the sources unchanged, \textit{i.e.}\ they map the theory to itself. In the calculus analogy, a global symmetry would be a change of integration variables that leaves the form of the integrand unchanged.

Consider a CFT in $d$-dimensional Minkowski space. Such a theory has a number of spurionic invariances, including in particular the semi-direct product of diffeomorphisms with Weyl transformations, $\DW$. The global symmetries of a CFT are the subgroup of $\DW$ that preserves the Minkowski metric, $\eta_{\mu\nu}$, \textit{i.e.}\ the subgroup generated by the conformal Killing vectors of the Minkowski metric. That subgroup is called the conformal group $SO(2,d)$, whose generators are those of the Poincar\'e group plus dilatations and special conformal transformations.

BCFTs and DCFTs are defined by their global symmetries, as follows. A DCFT in $d$-dimensional Minkowski space with a codimension-$n$ planar defect is defined as a theory invariant under the defect conformal group, $SO(2,d-n)\times SO(n)$, which includes: (i.) translations along the defect, (ii.) boosts along the defect, (iii.) rotations in the $\mathbb{R}^{d-n}$ occupied by the defect, (iv.) rotations in the $\mathbb{R}^{n}$ transverse to the defect, (v.) dilatations (acting on all of the $d$ coordinates), and (vi.) inversions through points in the defect~\cite{Cardy:1984bb,McAvity:1993ue,McAvity:1995zd}. A combination of (i.) and (vi.) then produce special conformal transformations along the defect. A BCFT in $d$-dimensional Minkowski space with a planar boundary is defined as a theory invariant under the boundary conformal group $SO(2,d-1)$, which is identical to the global symmetry of a DCFT with a codimension-one defect.

Crucially, notice that to define a BCFT or DCFT we must first specify a CFT and then introduce a boundary or defect. The reason is simple: the boundary or defect conformal symmetry includes scale invariance and special conformal transformations, and if the bulk theory is not invariant under these, then the theory with boundary or defect cannot be invariant under them. In short, to define a BCFT or DCFT, a necessary, though not sufficient, ingredient is a bulk CFT.

What are the spurionic invariances of a BCFT or DCFT? Na\"ively we might expect that a BCFT or DCFT cannot have the full $\DW$ spurionic invariance of a CFT, but that is not the case. A BCFT or DCFT has precisely the same $\DW$ spurionic invariance as the underlying CFT. Indeed, given a CFT, we can obtain a BCFT or DCFT simply by imposing an extra constraint: we demand that the global symmetries are the subgroup of $\DW$ that not only preserve the Minkowski metric but also preserve the location of the boundary or defect~\cite{Cardy:1984bb,McAvity:1993ue,McAvity:1995zd}. The global symmetry is then reduced from $SO(2,d)$ to the boundary or defect conformal group, where the ``broken'' generators of $SO(2,d)$ change the location of the boundary or defect (for example, imagine performing a translation in a direction normal to a defect). In other words, the broken generators go from being global symmetries of the CFT to spurionic invariances of the BCFT or DCFT. The full $\DW$ group remains as a spurionic invariance, however.

We are interested in the EE between two regions of a space in CFTs, BCFTs, and DCFTs. Generically, such an EE diverges due to correlations among short-distance, or equivalently UV, modes near the arbitrarily sharp division $\S$ between the two regions of space. We must therefore introduce a cutoff, smearing $\S$ over a region of extent $\varepsilon$ much smaller than any other scale in the system. The EE will then exhibit divergences as we remove the cutoff, $\varepsilon \to 0$, whose precise form depends on $d$ and $\S$. In particular, for a CFT in $d$ spacetime dimensions, the EE for spherical $\S$ of radius $R$ takes the form~\cite{Ryu:2006ef,Ryu:2006bv}
\beq
\label{E:SeeSphere}
\see = s_{d-2} \, \frac{R^{d-2}}{\varepsilon^{d-2}} + \hdots +  \left\{ \begin{array}{ll}  s_1 \frac{R}{\varepsilon} + s_0 + \Ocal\left(\varepsilon\right), & \quad d \, \text{  odd}, \\ & \\
s_2 \frac{R^2}{\varepsilon^2} + s_L \, \ln \left( \frac{2R}{\varepsilon}\right) + \tilde{s}_0 + \Ocal\left(\varepsilon^2\right), & \quad d \, \text{  even},
\end{array}\right.
\eeq
where $s_{d-2}$, $s_1$, $s_2$, $s_L$, etc. are independent of $R$- and $\varepsilon$ but depend on the details of the CFT. As with any other UV-divergent quantity in field theory, we must be careful if we want to extract physical information from EE. In particular, we could renormalize EE by introducing covariant counterterms to cancel any $\varepsilon \to 0$ divergences. The information that would remain would be physical and ``universal,'' meaning independent of the details of the renormalization. In odd $d$ the universal information is $s_0$, which is proportional to the partition function of the Euclidean theory on $\mathbb{S}^d$~\cite{Myers:2010xs,Myers:2010tj}. In $d=3$, the current conjecture is that the partition function at the UV fixed point of an RG flow is $\geq$ that at the IR fixed point~\cite{Jafferis:2011zi}. In even $d$ the universal information is $s_L$, which is propoprtional to a central charge, defined from the coefficient of the Euler density in the trace anomaly~\cite{Holzhey:1994we,Ryu:2006ef,Casini:2011kv}. In $d=2$ and $d=4$, that central charge is strictly non-increasing along an RG flow from the UV to the IR~\cite{Zamolodchikov:1986gt,Komargodski:2011vj}. Notice that in even $d$, the constant term $\tilde{s}_0$ in eq.~\eqref{E:SeeSphere} is cutoff-dependent and hence not universal: because of the $\ln (2R/\varepsilon)$ term we can change $\tilde{s}_0$ simply by rescaling $\varepsilon$, for example.

In a BCFT or DCFT, if we want to extract from EE some physical information that might characterize a boundary or defect, for example by counting degrees of freedom localized near the boundary or defect, then we must be especially careful. To see why, consider an example: a CFT in $d=3$ coupled as a defect to a CFT in $d=4$, with $\S$ a sphere centered on the defect. Indeed, let us consider the case where the two CFTs are arbitrarily weakly coupled to one another, so that to good approximation the EE is simply the linear combination of the $d=3$ and $d=4$ EEs in eq.~\eqref{E:SeeSphere},
\beq
\label{E:seedefectexample}
\see \approx s_2 \, \frac{R^2}{\varepsilon^2} + s_1 \, \frac{R}{\varepsilon}  + s_L \ln \left( \frac{2R}{\varepsilon}\right) + \tilde{s}_0 + s_0 + \Ocal\left(\varepsilon\right).
\eeq
For a CFT in $d=3$, $s_0$ is physical, but because of the $\ln (2R/\varepsilon)$ term in eq.~\eqref{E:seedefectexample} we can change $\tilde{s}_0 + s_0$ simply by rescaling $\varepsilon$, so in this DCFT $s_0$ is not physical. On the other hand, $s_L$ is physical, but ``knows'' nothing about the defect. Indeed, in our example $s_L$ will be the same as in the CFT in $d=4$, because the coupling to the defect is arbitrarily weak. How can we extract physical information about a boundary or defect from EE?

Luckily, an answer has already been proposed for BCFTs and DCFTs in $d=2$~\cite{Affleck:1991tk,Calabrese:2004eu,Azeyanagi:2007qj,Affleck2009}, where the boundary or defect is point-like. First, compute the EE for the BCFT or DCFT, with $\cm$ an interval of length $2R$ either containing the boundary or centered on the defect. In a DCFT in $d=2$ the result, $\see^{(\textrm{DCFT})}$, takes the form
\beq
\label{E:dcfteed2}
\see^{(\textrm{DCFT})} = s_L \, \ln \left( \frac{2R}{\varepsilon}\right) + \tilde{s}_0^{(\textrm{DCFT})} + \Ocal\left(\varepsilon^2\right),
\eeq
while in a BCFT the result, $\see^{(\textrm{BCFT})}$, has the same form, but with $s_L \to \frac{1}{2} s_L$ and $ \tilde{s}_0^{(\textrm{DCFT})} \to  \tilde{s}_0^{(\textrm{BCFT})}$. Second, compute the EE with $\cm$ an interval of length $2R$ infinitely far from the boundary or defect, or equivalently compute the EE of an interval in the underlying CFT. In either case the result, $\see^{(\textrm{CFT})}$, will be simply that of the CFT without boundary or defect,
\beq
\see^{(\textrm{CFT})} = s_L \, \ln \left( \frac{2R}{\varepsilon}\right) + \tilde{s}_0^{(\textrm{CFT})} + \Ocal\left(\varepsilon^2\right),
\eeq
where $s_L$ is identical to that in eq.~\eqref{E:dcfteed2} but $\tilde{s}_0^{(\textrm{CFT})}$ is not necessarily identical to $\tilde{s}_0^{(\textrm{DCFT})}$ or $\tilde{s}_0^{(\textrm{BCFT})}$. Finally, take the difference $(\see^{(\textrm{DCFT})} - \see^{(\textrm{CFT})})$ or $(\see^{(\textrm{BCFT})} - \frac{1}{2}\see^{(\textrm{CFT})})$ and send $\varepsilon \to 0$. The result is $(\tilde{s}_0^{(\textrm{DCFT})} - \tilde{s}_0^{(\textrm{CFT})})$ or $(\tilde{s}_0^{(\textrm{BCFT})} - \frac{1}{2}\tilde{s}_0^{(\textrm{CFT})})$, which is physical, and indeed is proportional to the impurity or boundary entropy~\cite{Affleck:1991tk,Calabrese:2004eu,Azeyanagi:2007qj,Affleck2009}. In short, we can extract physical information about the boundary or defect by computing the \textit{change} in EE due to the boundary or defect. Crucially, in this procedure we must use the same cutoff $\varepsilon$ in both the BCFT or DCFT and in the underlying CFT, otherwise the non-universal terms will not cancel in the difference of EEs.

Such a procedure is straightforward to generalize to $d>2$ and spherical $\S$.\footnote{We thank A.~Karch for useful discussions on this topic.} For a DCFT in $d$ spacetime dimensions with a planar defect of any codimension, compute the EE for a spherical $\S$ of radius $R$ centered on the defect, using cutoff $\varepsilon$, $\see^{(\textrm{DCFT})}$, and then compute the EE for spherical $\S$ of radius $R$ in the underlying CFT, $\see^{(\textrm{CFT})}$, using the same cutoff. We can then define a defect entropy $S_{\textrm{defect}}$ as
\beq
S_{\textrm{defect}} \equiv \see^{(\textrm{DCFT})} - \see^{(\textrm{CFT})}.
\eeq
For a BCFT in $d$ spacetime dimensions with planar boundary, compute the EE with $\S$ a hemisphere centered on the boundary, using cutoff $\varepsilon$, $\see^{(\textrm{BCFT})}$. We can then define a boundary entropy $S_{\partial}$ as
\beq
S_{\partial} \equiv \see^{(\textrm{BCFT})} - \frac{1}{2}\see^{(\textrm{CFT})}.
\eeq
The defect or boundary entropies so defined will contain physical information either in constants such as $(\tilde{s}_0^{(\textrm{DCFT})} - \tilde{s}_0^{(\textrm{CFT})})$ or $(\tilde{s}_0^{(\textrm{BCFT})} - \tilde{s}_0^{(\textrm{CFT})})$, or in the coefficient $(\tilde{s}_L^{(\textrm{DCFT})} - \tilde{s}_L^{(\textrm{CFT})})$ or $(\tilde{s}_L^{(\textrm{BCFT})} - \tilde{s}_L^{(\textrm{CFT})})$ of a $\ln \left( 2R/\varepsilon\right)$ factor. Whether that physical information is strictly non-increasing along an RG flow from a UV fixed point to an IR fixed point, similar to the boundary or impurity entropy in $d=2$~\cite{Affleck:1991tk,Friedan:2003yc}, is an important question that we will leave for future research.

For simplicity, in what follows we will focus exclusively on $\see^{(\textrm{DCFT})}$ and $\see^{(\textrm{BCFT})}$, rather than $S_{\textrm{defect}}$ and $S_{\partial}$, although we must keep in mind that, strictly speaking, only the latter contain well-defined physical information.

\subsection{The Field Theory Story}
\label{S:CHMcft}

We will now prove that for a BCFT or DCFT with $\S$ a (hemi-)sphere centered on the boundary or defect, the reduced density matrix $\rho$ is identical to the thermal density matrix of the BCFT or DCFT on a hyperbolic space. We will closely follow CHM's proof for CFTs~\cite{Casini:2011kv}. Indeed, CHM's proof relies crucially on the $\DW$ spurionic invariance of a CFT, and since BCFTs and DCFTs possess the same spurionic invariance, we can simply repeat CHM's arguments, keeping track of what happens to the boundary or defect along the way.

We consider BCFTs and DCFTs in $d$-dimensional Minkowski space, with metric
\beq
\eta_{\mu \nu} \, dX^{\mu} dX^{\nu} = - dt^2 + dr^2 + r^2 g_{\Sp^{d-2}},
\eeq
where $X^0 = t$ is the time coordinate, $r^2 = (X^1)^2 + (X^2)^2 + \ldots + (X^{d-1})^2$ is the spatial radial coordinate, and $g_{\Sp^{d-2}}$ is the metric of a unit-radius $(d-2)$-sphere, $\Sp^{d-2}$. We will consider only a planar boundary at $X^{d-1}=0$ or planar defect of codimension $n$ extended along $X^0, X^1, \ldots, X^{d-n-1}$ and sitting at the origin of the transverse directions, $X^{d-n} = X^{d-n+1} = \ldots = X^{d-1}=0$. For now we will consider a boundary or defect with at least one spatial direction: for a BCFT we take $d>2$ and for a DCFT we take $n < d-1$. In BCFTs with $d=2$ or DCFTs with $n = d-1$, the boundary or defect is point-like, and so breaks translational symmetry in all spatial directions. These are special cases that we will defer to the end of this subsection. We place the origin $r=0$ within the boundary or defect, so in a BCFT the space includes only half of the $\Sp^{d-2}$.

In a DCFT we choose $\cm$ to be a solid ball, or in a BCFT a solid half-ball, of radius $R$ centered on the boundary or defect at time $t=0$, so that $\S$ is the surface $r=R$ at time $t=0$. Tracing over the degrees of freedom outside of $\cm$, we obtain the reduced density matrix $\rho$, with associated $\see$ and $S_q$ given by eqs.~\eqref{eedef} and~\eqref{E:sqdef}.

The causal diamond of $\cm$ is the set of points with $\{r +t \leq R \}\cap \{ r-t \leq R\}$. Following CHM, we will map the causal diamond of $\cm$ to a Rindler wedge~\cite{Casini:2011kv}. Explicitly, we perform a diffeomorphism $X^{\mu} \to x^{\mu}(X)$, where
\beq
\label{E:diffeo1}
x^{\mu}(X) = \frac{X^{\mu}-(X\cdot X) C^{\mu}}{1-2X\cdot C+(X\cdot X)(C\cdot C)}+2R^2 C^{\mu}, \qquad C^{\mu}\partial_{\mu} \equiv \frac{1}{2R}\partial_1,
\eeq
where $X^1$ is a direction along the boundary or defect, such that the metric becomes
\beq
\eta_{\mu\nu} \, dX^{\mu}dX^{\nu} = \Omega(x)^2 \, \eta_{\mu\nu} \, dx^{\mu}dx^{\nu}, \qquad \Omega = \left(1+2x\cdot C + (x\cdot x)( C\cdot C)\right)^{-1},
\eeq
followed by a Weyl transformation to remove the $\Omega(x)^2$ factor from the metric. In other words, we perform a translation along the boundary or defect, to place the origin at the edge of the sphere $\Sigma$, then an inversion through the origin, then another translation along the boundary or defect, and finally a Weyl transformation. These transformations together form an element of $\DW$ that is a global symmetry of the BCFT or DCFT.

Under the diffeomorphism in eq.~\eqref{E:diffeo1}, in a DCFT $\cm$'s causal diamond maps to the Rindler wedge $\Rcal$ given by $x^{\pm} \equiv x^1 \pm x^0 \geq 0$. In particular the boundary of $\cm$'s causal diamond maps to the horizon of $\Rcal$, $x^{\pm}=0$, $\S$ maps to the plane $x^+=x^-=0$, and the defect maps to the submanifold $x^{d-n}=..=x^{d-1}=0$, and so wraps a Rinder wedge. In a BCFT, $\cm$'s causal diamond maps to a Rindler wedge which ends on the boundary $x^{d-1}=0$, which is itself a Rindler wedge.

The vacuum state in $\cm$'s causal diamond maps to the vacuum state in $\car$, and so the reduced density matrix $\rho$ is equal to the density matrix of the theory in $\car$. We can describe $\mathcal{R}$ via a family of uniformly accelerating observers by defining new coordinates $z$ and $\tau$ through $x^{\pm} \equiv z e^{\pm \tau/R}$, so that the metric becomes
\beq
\label{E:gRindler}
dx^+dx^- + \sum_{i=2}^{d-1} dx^i dx^i = -\frac{z^2}{R^2}d\tau^2 + dz^2 + \sum_{i=2}^{d-1} dx^i dx^i\,.
\eeq
For field theories without defects or boundaries, Unruh taught us~\cite{Unruh:1976db} that these accelerated observers experience the Rindler vacuum as a thermal state with temperature $T_0=1/(2\pi R)$. That is, $\rho$ is a thermal density matrix on $\mathcal{R}$ where time evolution is generated by $\partial_{\tau}$. To show that this remains true for theories with defects or boundaries, let us present a modern version of Unruh's argument (see ref.~\cite{Belin:2013dva} for similar arguments). To begin, we regard the density matrix $\rho$ on $\mathcal{R}$ as a reduced density matrix obtained by tracing out degrees of freedom in $x^1<0$. We can then give $\rho$ a functional integral representation, as follows. First we Wick-rotate time, $x^0 \to - i x^d$ so that the space becomes a Euclidean $\mathbb{R}^d$. We then impose the boundary conditions that the state approaches some prescribed values $\chi^{\pm}$ as $x^d \to 0^{\pm}$ in the $x^1>0$ region. The corresponding matrix element of $\rho$, which we denote $\rho(\chi^+,\chi^-)$, is then proportional to a functional integral, with the given boundary conditions, of the weight factor $\exp(-S_E)$, with $S_E$ the Euclidean action. (We fix the overall constant by demanding $\text{tr}\,\rho=1$.) Crucially, our defects and boundaries are rotationally invariant in the $x^1$-$x^d$ plane. The entire system is thus rotationally invariant in this plane, and so we can interpret $\rho(\chi^+,\chi^-)$ as an element of a thermal density matrix, treating the angle in the $x^1$-$x^d$ plane as Euclidean time. Finally, we observe that this is the same thermal density matrix we obtain if we Wick-rotate and compactify the Rindler time $\tau$ in eq.~\eqref{E:gRindler} with coordinate periodicity $2\pi R$. This demonstrates that $\rho$ is a thermal density matrix with time evolution generated by $\partial_{\tau}$. Note that if the defects or boundaries were not invariant under $\tau$-translations, then we would have no such thermal field theory interpretation. 

The metric in eq.~\eqref{E:gRindler} is conformal to $\mathbb{R} \times \Hy^{d-1}$,
\beq
\label{E:rindtohyp}
-\frac{z^2}{R^2}d\tau^2 + dz^2 + \sum_{i=2}^{d-1} dx^i dx^i = \frac{z^2}{R^2}\left[ -d\tau^2 + \frac{R^2}{z^2}\left(dz^2 + \sum_{i=2}^{d-1} dx^i dx^i\right)\right] = \frac{z^2}{R^2}\left( -d\tau^2 + R^2 g_{\mathbb{H}^{d-1}}\right),
\eeq
where $g_{\mathbb{H}^{d-1}}$ is the metric of a unit-radius $\Hy^{d-1}$. In a DCFT, the defect is extended along an equatorial hyperboloid inside of $\mathbb{H}^{d-1}$. In a BCFT, the $\mathbb{H}^{d-1}$ ends on the equatorial hyperboloid $x^{d-1}=0$. Our final step is to perform a Weyl transformation to eliminate the overall factor $z^2/R^2$ in the final equality in eq.~\eqref{E:rindtohyp}. The value of the BCFT or DCFT's generating functional is invariant under such a Weyl transformation, \textit{i.e.}\ Weyl transformations are spurionic invariances of BCFTs and DCFTS, as we discussed above. Moreover, after the Weyl transformation $\partial_{\tau}$ remains the generator of time evolution. We thus conclude that the thermal partition function of the theory on $\Rcal$ at temperature $T_0=1/(2\pi R)$ is identical to the thermal partition function of the theory on $\mathbb{R} \times \Hy^{d-1}$ at the same temperature. As a result, the reduced density matrix for spherical $\S$ of radius $R$ of the theory in Minkowski space is equivalent to the thermal density matrix of the theory on $\mathbb{R} \times \Hy^{d-1}$, with temperature $T_0=1/(2 \pi R)$ and $\Hy^{d-1}$ radius $R$. This completes our proof.

As discussed in ref.~\cite{Casini:2011kv}, we can also go directly from $\cm$'s causal diamond to $\mathbb{R} \times \Hy^{d-1}$, without detouring through the Rindler wedge $\Rcal$, via the change of coordinates
\beq
\label{direct}
r = \frac{R \, \sinh u}{\cosh u + \cosh \left(\tau/R\right)}\,, \qquad t = \frac{R \, \sinh \left(\tau/R\right)}{\cosh u +  \cosh \left(\tau/R\right)}\,,
\eeq
where $u\in (0,\infty)$ and $\tau \in (-\infty,\infty)$. In particular,
\beq
\label{directlimits}
\tau \rightarrow \pm \infty: \quad (t,r) \rightarrow (\pm R,0)\,, \qquad \textrm{and} \qquad u \rightarrow \infty: \quad (t,r) \rightarrow (0,R)\,,
\eeq
that is, the extreme limits of $\tau$ and $u$ are the corners of $\cm$'s causal diamond. In the new coordinates, the Minkowski metric takes the form
\beq
\label{directmetric}
\eta_{\mu\nu} \, dX^{\mu}dX^{\nu} = \Omega^2 \left(- d\tau^2 + R^2 g_{\Hy^{d-1}}\right), \qquad \Omega = \left(\cosh u + \cosh \left(\tau/R\right)\right)^{-1},
\eeq
where here we write the metric of a unit-radius $\Hy^{d-1}$ as
\beq
\label{E:hypmetric}
g_{\Hy^{d-1}} = du^2 + \sinh^2 u \, g_{\Sp^{d-2}}.
\eeq
We can then again simply perform a Weyl transformation and invoke the spurionic invariance of the BCFT or DCFT to reach the same conclusion as above. The change of coordinates in eq.~\eqref{direct} will be especially useful to us in section~\ref{S:probes}.

A number of consequences immediately follow from the equivalence of the reduced density matrix $\rho$ of the theory in Minkowski space with the thermal density matrix of the theory in $\mathbb{R} \times \Hy^{d-1}$. Clearly the EE with spherical $\S$ will be equivalent to the thermal entropy of the theory in $\mathbb{R} \times \Hy^{d-1}$. Moreover, as discussed in ref.~\cite{Hung:2011nu}, the R\'enyi entropies $S_q$ are related to the free energy $F(T)$ of the theory on $\mathbb{R} \times \Hy^{d-1}$ at temperature $T$ as
\beq
\label{E:SrenyiF}
S_q = \frac{q}{1-q}\frac{1}{T_0}\left( F(T_0)-F\left( \frac{T_0}{q}\right)\right).
\eeq
This form of $S_q$ will be especially useful to us in subsection~\ref{S:renyi}.

Let us now return to the special cases of point-like defects and boundaries. In a DCFT with $n=d-1$, the defect is a point-like impurity, which we take to sit at the origin of the spatial coordinates, $X^1 = X^2 = \ldots = X^{d-1}=0$. The impurity worldline is thus a straight line in the $t$ direction. After the diffeomorphism in eq.~\eqref{E:diffeo1}, the impurity is extended in the new time coordinate $x^0$ and sits at the origin of all spatial coordinates except $x^1$, where it sits at $x^1 = \sqrt{R^2 + (x^0)^2}$. In terms of the coordinates $z$ and $\tau$ the impurity sits at $z = R$ and evolves in $\tau$, that is, the impurity uniformly accelerates in the Rindler wedge. Switching to time coordinate $\tau$, we find a time-independent equilibrium state: the impurity sits still at $z=R$. We can thus define a thermal partition function on $\Rcal$. The remaining arguments are unchanged: we identify the reduced density matrix $\rho$ with the thermal density matrix on $\Rcal$ at temperature $T_0$, Weyl-transform to $\mathbb{R} \times \Hy^{d-1}$, where the impurity sits at $z=R$ inside $\Hy^{d-1}$, and then invoke the spurionic invariance of the DCFT to conclude that $\rho$ is equivalent to the thermal density matrix on $\mathbb{R} \times \Hy^{d-1}$. Notice that $\Hy^{d-1}$ is maximally symmetric, so nothing is special about $z=R$, \textit{i.e.}\ we can move the impurity to any other point with no change to our conclusion. Indeed, already in the Rindler wedge nothing was special about $z=R$. The crucial property was the impurity's uniform acceleration. For a BCFT in $d=2$, where the boundary is a single point, all of the arguments above apply, with trivial modifications, except the boundary is obviously a special point that cannot be moved.

\subsection{The Gravity Story} \label{S:CHMads}

\subsubsection{Review: CHM's Proof in AdS}
\label{S:CHMpureAdS}

We will use various ``slicings'' of $AdS_{d+1}$, meaning various foliations of $AdS_{d+1}$ that make manifest different subgroups of its $SO(2,d)$ isometry. For example, the metric of unit-radius $AdS_{d+1}$ in Poincar\'e slicing is
\beq
\label{adspoin}
g_{AdS_{d+1}} = \frac{1}{z^2} \left( dz^2 - dt^2 + dr^2 + r^2 g_{\Sp^{d-2}} \right),
\eeq
where $z$ is the $AdS_{d+1}$ radial coordinate, with the $AdS_{d+1}$ boundary at $z \to 0$ and the Poincar\'e horizon at $z \to \infty$. A Poincar\'e slice is a surface of fixed $z$. An $AdS_{d+1}$ spacetime of radius $L$ has metric  $L^2 g_{AdS_{d+1}}$.

Recall that to define a finite metric on the $AdS_{d+1}$ boundary, and hence for the dual CFT, we must specify a ``defining function''~\cite{Witten:1998qj}, as follows. Any asymptotically $AdS_{d+1}$ metric will have a second-order pole at the $AdS_{d+1}$ boundary. To extract a finite metric at the boundary we must multiply the $AdS_{d+1}$ metric by a defining function, a function that has a second-order zero at the $AdS_{d+1}$ boundary but is otherwise arbitrary, and then restrict to the boundary. A particular slicing of $AdS_{d+1}$ may naturally suggest a particular defining function. For example, in Poincar\'e slicing with radius $L$ the natural defining function is $z^2/L^2$. With that choice of defining function, the dual CFT is defined on Minkowski spacetime.

In what follows we will need to perform $\DW$ transformations in the CFT, BCFT, or DCFT. In holography, $\DW$ transformations are realized in the bulk gravity theory by large diffeomorphisms, \textit{i.e.} diffeomorphisms with support at the $AdS_{d+1}$ boundary. For example, consider Poincar\'e-sliced $AdS_{d+1}$, with $X^{\mu}$ denoting all coordinates except $z$, so $\mu = 0, 1, \ldots, d$. To implement a diffeomorphism in the dual CFT we must perform a large diffeomorphism: $X^{\mu} \to X'^{\mu}(X)$ with $z$ unchanged, at least near $z=0$. To implement a Weyl transformation, we must perform a large diffeomorphism whose non-trivial effect near $z =0$ is to map $z \to z' = \Omega(X)^{-1} \, z$ with $X^{\mu}$ unchanged. If we leave the defining function $z^2/L^2$ unchanged, then the metric at the boundary acquires an overall factor of $\Omega(X)^2$, as expected. Alternatively, we can perform no diffeomorphisms at all, but change the defining function, $z^2/L^2 \to \O(X)^2 z^2/L^2$, which clearly has the same effect. The $SO(2,d)$ conformal group, which is a subgroup of $\DW$, is dual to the $SO(2,d)$ isometry of the $AdS_{d+1}$ metric, which is a subgroup of the large diffeomorphisms.

For a static spacetime such as $AdS_{d+1}$, RT's prescription to compute $\see$ is to choose a fixed value of time $t$, which we choose without loss of generality to be $t=0$, and then to determine the surface of minimal area $\Am$ that approaches $\Sigma$ as $z\to 0$.\footnote{In string theory, we must use the Einstein-frame metric to compute $\Am$ in RT's prescription.} In our case, $\Sigma$ is a sphere $\Sp^{d-2}$ of radius $R$. The minimal area surface will wrap the entire $\Sp^{d-2}$ and trace a curve in the quadrant spanned by $z$ and $r$ (recall $z \in (0,\infty)$ and $r \in [0,\infty)$). Upon parameterizing that curve as $z(r)$, we can write the area functional $\A$ as
\beq
\label{afunc}
\A = L^{d-1}\vol (\Sp^{d-2})\int dr \, \frac{r^{d-2}}{z^{d-1}} \, \sqrt{1 + \left(\partial_r z(r)\right)^2},
\eeq
where we performed the integration over the $\Sp^{d-2}$ to obtain the volume factor $\vol (\Sp^{d-2})$. When $d=2$, $r$ is not a radial coordinate: $r \in (-\infty,\infty)$. To include this case in eq.~\eqref{afunc}, we define $\vol(\mathbb{S}^0)=2$ and restrict to $r \in [0,\infty)$. Upon variation of $\A$, we obtain an Euler-Lagrange equation for $z(r)$, whose solution obeys~\cite{Ryu:2006bv,Ryu:2006ef}
\beq
\label{zsol}
z(r)^2 + r^2 = R^2.
\eeq
In other words, the $z(r)$ that extremizes $\A$ looks like a hemisphere of radius $R$ in $AdS_{d+1}$, centered on the boundary. Indeed, as $z \to 0$, the solution for $z(r)$ clearly approaches the surface $r=R$, which is precisely $\Sigma = \Sp^{d-2}$, as required. To obtain the value of $\Am$, we insert the solution for $z(r)$ into eq.~\eqref{afunc} and integrate in $r$. That integral diverges due to the infinite volume of $AdS_{d+1}$, so we introduce a cutoff at $z = \varepsilon$, which via eq.~\eqref{zsol} becomes a cutoff $\reps$ on the $r$ integration,
\beq
\label{E:reps}
\reps \equiv \sqrt{R^2-\varepsilon^2} = R - \frac{\varepsilon^2}{2 R} + \Ocal\left(\varepsilon^4/R^3\right).
\eeq
Indeed, in principle we should have implemented the cutoff before varying $\A$, since minimizing a divergent quantity is nonsensical. With the cutoff in place we find
\beq
\label{adsamin}
\Am = L^{d-1}\vol(\Sp^{d-2})  \int_0^{\reps} dr \, \frac{Rr^{d-2}}{\left(R^2-r^2\right)^{d/2}}.
\eeq
To obtain $\see$ via eq.~\eqref{rt}, we multiply by $1/(4G)$,
\beq
\label{E:adsEE}
\see =L^{d-1} \frac{\vol(\Sp^{d-2})}{4G}\int_0^{\reps}dr\,\frac{Rr^{d-2}}{\left(R^2-r^2\right)^{d/2}},
\eeq 
and must then perform the $r$ integration. The result takes the form expected for an EE in a $d$-dimensional CFT~\cite{Ryu:2006bv,Ryu:2006ef}, eq.~\eqref{E:SeeSphere}.

If we plug the solution for $z(r)$ in eq.~\eqref{zsol} into $\A$, then by definition we obtain a local extremum, but is it the \textit{global} minimum $\Am$? Could some other solution produce a smaller value of $\A$? We will encounter similar questions in subsection~\ref{S:CHMdcftbcft}, when we compute $\see$ in BCFTs and DCFTs using holography, so we would like a proof for the global minimization of $\A$ that we can easily adapt to the BCFT and DCFT cases. We have actually found several such proofs, all of which are easy to adapt to the cases in subsection~\ref{S:CHMdcftbcft}, for example we can use a special conformal transformation~\cite{Ryu:2006bv,Ryu:2006ef}. In appendix~\ref{A:globmin} we present the quickest proof we have found.

Let us now review CHM's proof of RT's proposal. We consider an $AdS_{d+1}$ spacetime of radius $L$. To go from the $AdS_{d+1}$ metric in Poincar\'e slicing to the $AdS_{d+1}$ metric in hyperbolic slicing, we change coordinates~\cite{Casini:2011kv}:
\begin{subequations}
\label{pointohyp}
\begin{align}
\label{ztohyp}
z &= \frac{R L}{v \cosh u + \cosh \left(\tau/R\right) \sqrt{v^2 - L^2}}\,,
\\
\label{rtohyp}
r &= \frac{R \, v \, \sinh u}{v \cosh u + \cosh \left(\tau/R\right) \sqrt{v^2 - L^2}}\,,
\\
\label{ttohyp}
t &= \frac{R \, \sinh \left(\tau/R\right) \sqrt{v^2 - L^2}}{v \cosh u +  \cosh \left(\tau/R\right) \sqrt{v^2 - L^2}}\,,
\end{align}
\end{subequations}
where $v \in [L,\infty)$, $u\in [0,\infty)$, and $\tau \in (-\infty,\infty)$, and all $\Sp^{d-2}$ coordinates remain unchanged. The $AdS_{d+1}$ metric then takes the form
\beq
\label{adshyp}
L^2 \, g_{AdS_{d+1}} = \frac{dv^2}{f(v)} - f(v) \, \frac{L^2}{R^2} \, d\tau^2 + v^2 \, g_{\mathbb{H}^{d-1}}, \qquad f(v) = \frac{v^2}{L^2} - 1,
\eeq
where $g_{\Hy^{d-1}}$ is of the form in eq.~\eqref{E:hypmetric}.

In hyperbolic slicing, we can approach the $AdS_{d+1}$ boundary, $z \to 0$, in three different ways. First, we can fix $\tau$ and $u$ and take $v \to \infty$, in which case the $r$ and $t$ in eq.~\eqref{pointohyp} approach those of eq.~\eqref{direct}. At the boundary of $AdS_{d+1}$, the hyperbolic slicing thus only covers the causal diamond of $\cm$. In other words, eq.~\eqref{pointohyp} is not only a coordinate change but also a coordinate restriction, \textit{i.e.}\ the metric in eq.~\eqref{adshyp} covers only a patch of $AdS_{d+1}$ . Second, we can fix $v$ and $u$ and take $\tau \to \pm \infty$. Third, we can fix $v$ and $\tau$ and take $u \to \infty$. Each of these last two limits sends $z \to 0$ and sends $r$ and $t$ to a corner of the causal diamond of $\cm$, as specified in eq.~\eqref{directlimits}.

Eq.~\eqref{pointohyp} is a large diffeomorphism, and thus changes the coordinates in the dual CFT. If we continue to use the natural defining function for Poincar\'e slicing, $z^2/L^2$, which from eq.~\eqref{ztohyp} behaves near the $AdS_{d+1}$ boundary as $z^2/L^2 \approx (R \Omega)^2/v^2$ with $\Omega = \left(\cosh u + \cosh\left(\tau/R\right)\right)^{-1}$, then the metric at the $AdS_{d+1}$ boundary remains the Minkowski metric, as written in eq.~\eqref{directmetric}. We implement a Weyl transformation in the CFT by changing to the natural defining function in hyperbolic slicing, $R^2/v^2$, in which case the metric at the $AdS_{d+1}$ boundary becomes $-d\tau^2 + R^2 g_{\Hy^{d-1}}$, that is, the boundary becomes $\mathbb{R} \times \Hy^{d-1}$.

The $AdS_{d+1}$ metric in eq.~\eqref{adshyp} has a horizon at $v_H \equiv L$, the outermost value of $v$ where $f(v)=0$. The spacetime is merely a patch of $AdS_{d+1}$, which is non-singular and has no horizon, so the horizon is an artifact of the hyperbolic slicing. In that sense the horizon at $v_H$ is analogous to a Rindler horizon~\cite{Emparan:1999gf}. The Hawking temperature associated with the horizon at $v_H$ is $T_0=1/(2\pi R)$. The AdS/CFT dictionary then implies that the dual CFT is in a thermal equilibrium state at the temperature $T_0$. The area of the horizon, $\mathcal{A}_H$, is simply the volume of $\Hy^{d-1}$ of radius $L$: denoting the volume of $\Hy^{d-1}$ of unit radius as $\vol(\Hy^{d-1})$, we have
\beq
\label{ahor}
\mathcal{A}_H = L^{d-1} \vol(\Hy^{d-1}) = L^{d-1} \, \vol(\Sp^{d-2}) \int_0^{u_{\varepsilon}} du \, \sinh^{d-2} u,
\eeq
where $u_{\varepsilon}$ is a cutoff at some large but finite value of $u$. The thermal entropy of the hyperbolic black brane, and of the CFT, is then the Bekenstein-Hawking entropy $\mathcal{A}_H/(4 G)$.

Now we come to the crux of CHM's proof: the horizon $v_H$ coincides precisely with the minimal area surface. Indeed, using eqs.~\eqref{ztohyp} and~\eqref{rtohyp}, we find that at the horizon $v_H$,
\beq
\left . \left [ z^2 + r^2 \right] \right|_{v_H} = R^2\,,
\eeq
which is precisely eq.~\eqref{zsol}. Inserting $v_H=L$ into eq.~\eqref{rtohyp} we find $r = R\, \tanh u$, so that $u = \arctanh(r/R)$ and hence
\beq
\label{E:hypvol}
\vol(\Hy^{d-1}) = \vol(\Sp^{d-2}) \int_0^{u_{\varepsilon}} du \, \sinh^{d-2} u = \vol(\Sp^{d-2}) \int_0^{\reps} dr \, \frac{Rr^{d-2}}{\left(R^2-r^2\right)^{d/2}}\,.
\eeq
The integral for $\mathcal{A}_H$ in eq.~\eqref{ahor} will be identical to the integral for $\mathcal{A}_{\rm min}$ in eq.~\eqref{adsamin} if we choose the cutoff $r_{\varepsilon}$ in eq.~\eqref{E:hypvol} to be the same as in eq.~\eqref{adsamin}, which requires $u_{\varepsilon} = \text{arctanh}\left( \sqrt{1-\frac{R^2}{\varepsilon^2}}\right)$. We then have $\mathcal{A}_H = \Am$, which proves RT's conjecture: for the CFT in Minkowski space, when $\Sigma$ is a sphere of radius $R$, $\see$ is identical to the thermal entropy of the same CFT on $\mathbb{R} \times \Hy^{d-1}$ at temperature $T_0 = 1/(2\pi R)$, which via holography is $\mathcal{A}_H/(4G) = \Am/(4G)$.

CHM's mapping of EE for spherical $\S$ to thermal entropy on $\mathbb{R} \times \Hy^{d-1}$ also provides access to the R\'enyi entropies $S_q$, via eq.~\eqref{E:SrenyiF}. We can calculate the free energies $F(T_0)$ and $F(T_0/q)$ via holography as follows. First, while the spacetime eq.~\eqref{adshyp} describes the CFT on $\mathbb{R} \times \Hy^{d-1}$ at temperature $T_0$, we need solutions that describe the CFT on $\mathbb{R} \times \Hy^{d-1}$ at the temperatures $T_0/q$. Fortunately, the $AdS_{d+1}$ metric in eq.~\eqref{adshyp} is just one of a family of black brane metrics, identical in form to that of eq.~\eqref{adshyp} but with
\beq
\label{E:adshypbraneotherT}
f(v) = \frac{v^2}{L^2} - 1 - \frac{m}{v^{d-2}}.
\eeq
Such metrics are solutions of Einstein's equation with negative cosmological constant for any $m$. Only the solution with $m=0$ is non-singular, being simply a patch of $AdS_{d+1}$.  When $m$ is non-zero, a singularity appears at $v=0$, shielded by an event horizon where $f(v_H)=0$, and the spacetime is not equivalent to $AdS_{d+1}$. The metric remains asymptotically $AdS_{d+1}$, however, and with defining function $R^2/v^2$ describes the dual CFT on $\mathbb{R} \times \Hy^{d-1}$ at temperature
\beq
\label{E:adshypbraneotherTvalue}
T = \frac{2v_H^d+(d-2)mL^2}{4\pi L R v_H^{d-1}}.
\eeq
Solutions with $m<0$ describe black holes with $T<T_0$, including an extremal black hole, with $T=0$, when $m=-\frac{2}{d-2} \left(\frac{d-2}{d}\right)^{d/2} L^{d-2}$. The AdS/CFT dictionary equates $F(T)/T$ with the on-shell Euclidean action of the gravity theory~\cite{Witten:1998zw}. Using the black hole solutions with nonzero $m$, we can calculate $F(T_0)$ and $F(T_0/q)$, and hence $S_q$, holographically~\cite{Hung:2011nu}.

\subsubsection{Proof for the Duals of BCFTs and DCFTs}
\label{S:CHMdcftbcft}

We now turn to our main goal in this section: extending CHM's proof to BCFTs and DCFTs. We will first present the metrics that appear in the holographic duals of BCFTs and DCFTs, and then we will present our extension of CHM's proof to those spacetimes.

In the holographic dual of a BCFT or DCFT, the metric must have $SO(2,d-n) \times SO(n)$ isometry, meaning the spacetime must include $AdS_{d+1-n}$ and $\Sp^n$ factors. As a gentle introduction to such spacetimes, let us first present the $AdS_{d+1}$ metric in $AdS_{d+1-n}$ slicing, which makes manifest the $SO(2,d-n)\times SO(n)$ subgroup of the full $SO(2,d)$ isometry. We begin with the unit-radius $AdS_{d+1}$ metric in Poincar\'e slicing, eq.~\eqref{adspoin}, which we re-write as follows:
\begin{align}
\begin{split}
\label{adspoin2}
g_{AdS_{d+1}} & =  \frac{1}{z^2} \left( dz^2 - dt^2 + dr^2 + r^2 \, g_{\Sp^{d-2}} \right) \\ & =  \frac{1}{z^2} \left( dz^2 - dt^2 + d\rp^2 + \rp^2 \, g_{\Sp^{d-n-2}} + d\rn^2 + \rn^2 g_{\Sp^{n-1}} \right), 
\end{split}
\end{align}
where on each Poincar\'e slice we have split the coordinates into two subsets, the $d-n$ coordinates along a fictitious ``defect'' and the $n$ coordinates transverse to the defect. For each subset, we have introduced spherical coordinates, with radial coordinates $\rp$ and $\rn$ in the directions parallel and perpendicular to the defect, respectively. When $n=1$, $\rn$ is simply the coordinate of the single perpendicular direction, in which case $\rn \in (-\infty,\infty)$. Clearly in all cases $\rp^2 + \rn^2 = r^2$. To go from Poincar\'e slicing to $AdS_{d+1-n}$ slicing, we change coordinates:
\beq
\label{pointoads}
z = Z \, \sech x, \qquad \rn = Z \, \tanh x\,,
\eeq
where $Z \in (0,\infty)$ and $x \in [0,\infty)$ when $n>1$, while $x \in (-\infty,\infty)$ when $n=1$. In the new coordinates, the unit-radius $AdS_{d+1}$ metric takes the form
\beq
\label{adsslicing}
g_{AdS_{d+1}} =dx^2 + \cosh^2(x) \, g_{AdS_{d+1-n}} + \sinh^2(x) \, g_{\Sp^{n-1}},
\eeq
where the $AdS_{d+1-n}$ and $\Sp^{n-1}$ subspaces are explicit, with
\beq
\label{E:adsslicecoords}
g_{AdS_{d+1-n}} = \frac{1}{Z^2} \left ( dZ^2 - dt^2 + d\rp^2 + \rp^2 \,\, g_{\Sp^{d-n-2}} \right)\,.
\eeq

\begin{figure}[ht!]
\begin{center}
\includegraphics[height=2.0in]{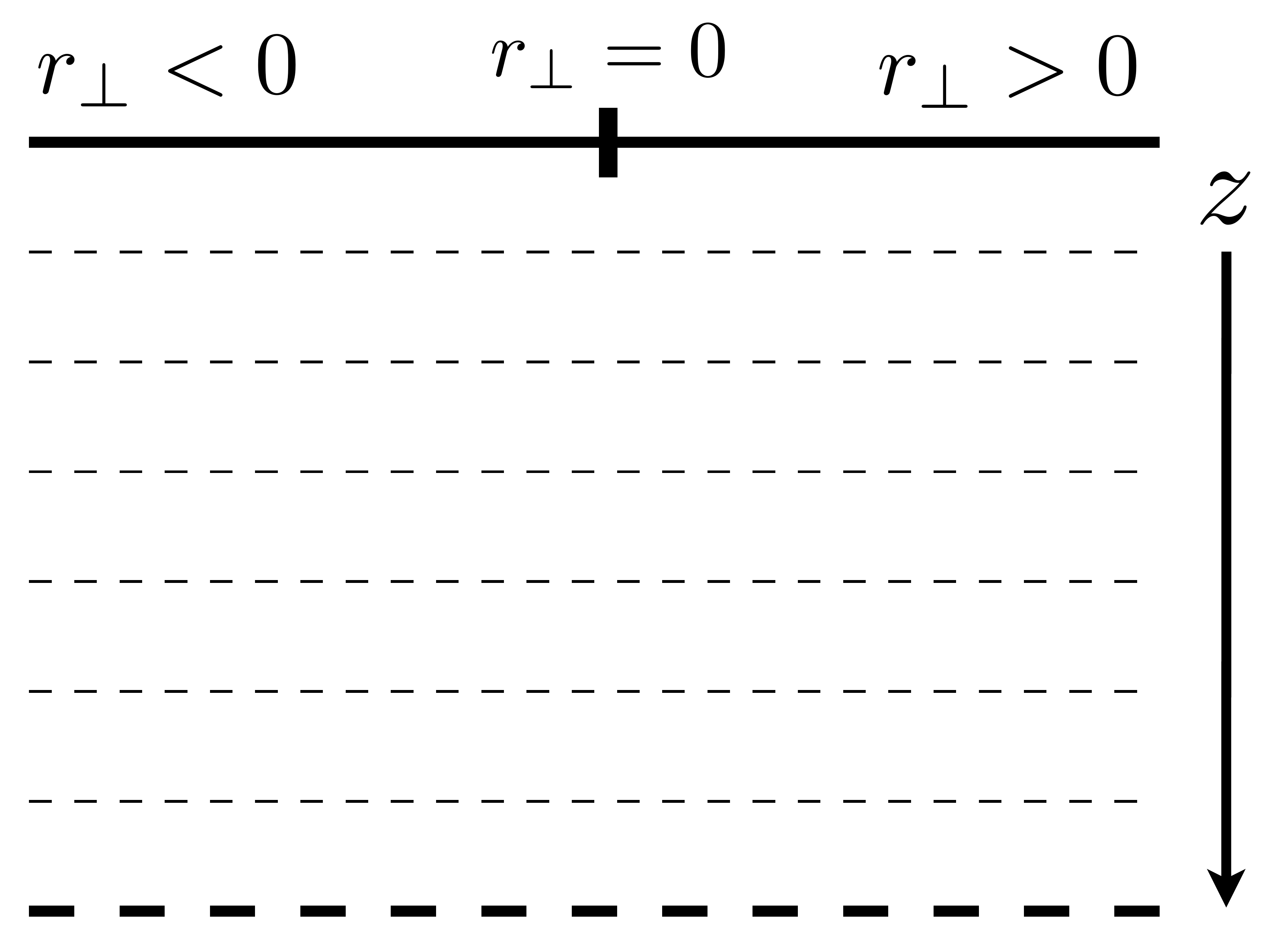} \qquad
\includegraphics[height=2.0in]{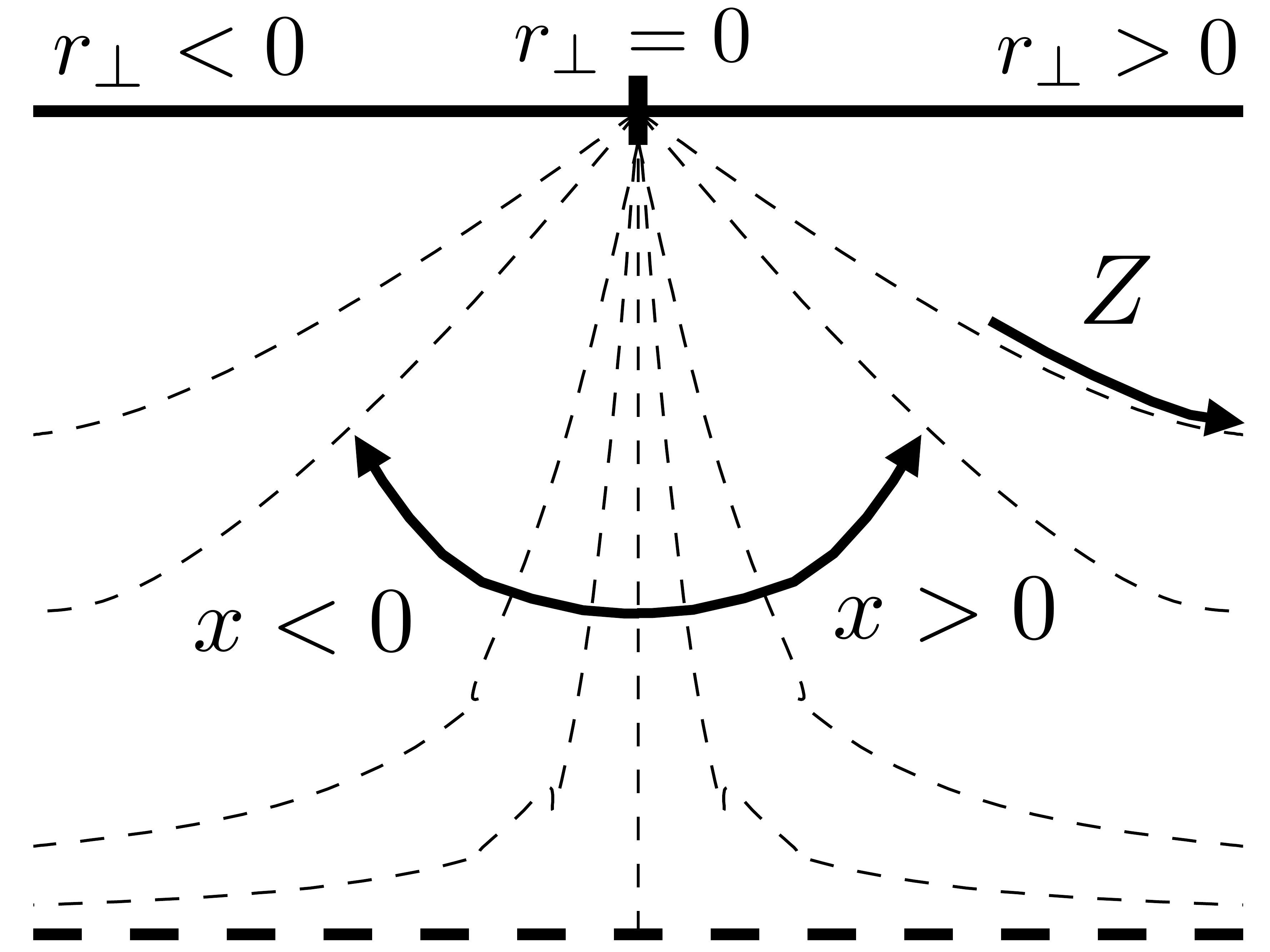} \\
\textbf{(a.)} \hspace{7cm} \textbf{(b.)}
\end{center}
\caption{Cartoons depicting two different ``slicings'' of $AdS_{d+1}$, the Poincar\'e and the $AdS_d$ slicings, which make manifest different subgroups of the $SO(2,d)$ isometry. (Figure adapted from ref.~\cite{Aharony:2003qf}.) \textbf{(a.)} Poincar\'e slicing, with the metric in eq.~\eqref{adspoin}. We suppress all directions except for two: the horizontal axis, $\rn$, and the vertical axis, $z$. The solid horizontal line is the $AdS_{d+1}$ boundary $z \rightarrow 0$, the heavy dashed horizontal line is the Poincar\'e horizon $z \to \infty$, and the thin dashed lines are surfaces of fixed $z$, the Poincar\'e slices, which make manifest the $SO(1,d-1) \in SO(2,d)$. \hspace{1in}
\textbf{(b.)} $AdS_d$ slicing, where the metric is written as in eq.~\eqref{adsslicing} with $n=1$. The $Z$ and $x$ directions are depicted with arrows. The thin dashed lines are surfaces of constant $x$, the $AdS_d$ slices, which make manifest the $SO(2,d-1) \in SO(2,d)$, as approproate for a fictitious $n=1$ defect at $\rn = 0$.}
\label{F:slicings}
\end{figure}

In $AdS_{d+1-n}$ slicing, when $n>1$ we can approach the $AdS_{d+1}$ boundary $z \to 0$ in two ways. First, we can fix $x$ and send $Z \to 0$, meaning we stay within a single $AdS_{d+1-n}$ slice and approach the $AdS_{d+1-n}$ boundary. In the original coordinates, this sends $z \to 0$ and $\rn \to 0$, so we arrive at the boundary at a point on the defect. Second, we can fix $Z$ and send $x \to \infty$, meaning we move through different $AdS_{d+1-n}$ slices, always at fixed $Z$. This sends $z \to 0$ and $\rn \to Z$, so we arrive at the boundary at a point a distance $Z$ away from the defect.  When $n=1$, we can approach the $AdS_{d+1}$ boundary in three ways, as shown in fig.~\ref{F:slicings}. As in the $n>1$ cases, we can fix $x$ and send $Z \to 0$, which takes us to the boundary at a point on the defect. Alternatively, we can fix $Z$ and send $x \to \pm \infty$, which sends $z \to 0$ and $\rn \to \pm Z$, so we arrive at the boundary some distance $Z$ away from the defect, on one side ($\rn \to -Z$) or the other ($\rn \to +Z$). Bear in mind that in all cases the spacetime is just $AdS_{d+1}$, and the ``defect'' is purely fictitious.

For a unit-radius $AdS_{d+1}$ in $AdS_{d+1-n}$ slicing with $n>1$, both $\cosh^2 x$ and $\sinh^2 x$ diverge as $e^{2x}/4$ near the boundary $x \to \infty$, so the natural defining function is $4e^{-2x}$. When $n=1$, the natural defining functions are $4 e^{\mp 2x}$ in the $x \to \pm \infty$ regions, respectively. With these defining functions, when $n>1$ the dual CFT is defined on a spacetime with metric $g_{AdS_{d+1-n}} + g_{\Sp^{n-1}}$ with the defect at the boundary $Z \to 0$ of the $AdS_{d+1-n}$, whereas when $n=1$ the dual CFT is defined on a spacetime consisting of two copies of $AdS_{d+1-n}$ (the two copies we approach as $x \to \pm \infty$ with $Z$ fixed) glued together at their boundaries, where the defect is located. The metric $g_{AdS_{d+1-n}} + g_{\Sp^{n-1}}$ is conformal to the Minkowski metric:
\begin{align}
\begin{split}
\label{adsspconf} 
g_{AdS_{d+1-n}} + g_{\Sp^{n-1}} & =  \frac{1}{Z^2} \left ( dZ^2 - dt^2 + d\rp^2 + \rp^2 g_{\Sp^{d-n-2}} \right) + g_{\Sp^{n-1}}  
\\
& =  \frac{1}{Z^2} \left (- dt^2 + d\rp^2 + \rp^2 g_{\Sp^{d-n-2}} + dZ^2 + Z^2 g_{\Sp^{n-1}} \right)  
\\
& = \frac{1}{\rn^2} \left (- dt^2 + dr^2 + r^2 g_{\Sp^{d-2}} \right),
\end{split}
\end{align}
where in the final equality we used the fact that $\rn^2 \to Z^2$ near the boundary. We can implement the Weyl transformation to the Minkowski metric by switching to the natural defining function in Poincar\'e slicing, $z^2$, which from eq.~\eqref{pointoads} for $n>1$ approaches $z^2 \approx Z^2 (4 e^{-2x}) = \rn^2 (4 e^{-2x})$ as $x \to \infty$ and for $n=1$ approaches $z^2 \approx Z^2 (4 e^{\mp2x}) = \rn^2 (4 e^{\mp 2x})$ as $x \to \pm \infty$. With these defining functions, the boundary is equipped with the flat Minkowski metric.

Now that we understand the $AdS_{d+1-n}$ slicing of $AdS_{d+1}$, let us turn to the gravity dual of a BCFT or DCFT, which will be defined on a spacetime that has only $SO(2,d-n) \times SO(n)$ isometry. In the gravity theory, if the only term on the right-hand-side of Einstein's equation comes from a negative cosmological constant, then the only horizonless spacetime with no conical singularities (from orbifolds, for example) is $AdS_{d+1}$. To obtain the dual to a BCFT or DCFT we must introduce non-trivial matter fields to deform $AdS_{d+1}$ into a spacetime with $SO(2,d-n) \times SO(n)$ isometry. The dual field theory statement, mentioned above, is that in order to define a BCFT or DCFT, a necessary, though not sufficient, ingredient is a CFT. The dual of a BCFT or DCFT will thus have only $SO(2,d-n) \times SO(n)$ isometry, but will nevertheless asymptotically approach $AdS_{d+1}$, at least away from the boundary or defect.

Many examples of such spacetimes exist in the literature, with various $d$ and $n$, obtained as solutions of both ``top-down'' systems, meaning genuine supergravity theories in various dimensions (see for example refs.~\cite{Bak:2003jk,Clark:2005te,D'Hoker:2007xy,D'Hoker:2007xz,D'Hoker:2009gg,Chiodaroli:2009yw,Chiodaroli:2011fn,Berdichevsky:2013ija}), and ``bottom-up'' systems, meaning gravity theories tailored to produce such solutions, but which may nor may not descend from some well-defined string or supergravity theory (see for example refs.~\cite{Takayanagi:2011zk,Fujita:2011fp,Nozaki:2012qd}). In section~\ref{S:probes} we will also present some explicit examples of such spacetimes. In all known examples, the metric can be written in a form similar to the $AdS_{d+1-n}$ slicing of $AdS_{d+1}$:
\beq
\label{genmet}
g = L^2(dx^2+ A(x)^2 \, g_{AdS_{d+1-n}} + B(x)^2 \, g_{\Sp^{n-1}}),
\eeq
that is, $AdS_{d+1-n}$ and ${S}^{n-1}$ are fibered over an interval parameterized by $x$, as required by the $SO(2,d-n)\times SO(n)$ isometry.

In this subsection we will not need explicit expressions for the warp factors $A(x)$ and $B(x)$ in eq.~\eqref{genmet}, but we will need their asymptotics. For the dual of a DCFT, when $n>1$ each of the warp factors $A(x)$ and $B(x)$ diverges as $\exp(x-c_n)/2$ as $x \to \infty$, where $c_n$ is a constant, and when $n=1$ the warp factor $B(x)$ is absent while $A(x)$ diverges as $\exp(\pm (x-c_1^{\pm}))/2$ as $x \to \pm \infty$, where $c_1^{\pm}$ are two constants that need not be identical. The dual of a BCFT will have only a single asymptotically $AdS_{d+1}$ region, say at $x \to \infty$ where $A(x)$ diverges $\exp(x-c_B) /2$, where $c_B$ is a constant, and typically the spacetime will either have a boundary at some finite $x$~\cite{Takayanagi:2011zk,Fujita:2011fp,Nozaki:2012qd} or will cap off smoothly at some finite $x$, as occurs in the dual of the low-energy theory on D3-branes ending on D5- or NS5-branes~\cite{D'Hoker:2007xy,D'Hoker:2007xz}.

With these asymptotics for $A(x)$ and $B(x)$, the metric in eq.~\eqref{genmet} asymptotically approaches $AdS_{d+1}$, written in $AdS_{d+1-n}$ slicing, eq.~\eqref{adsslicing}, at least away from the boundary or defect, as advertised. All of our statements above about defining functions for $AdS_{d+1}$ in $AdS_{d+1-n}$ slicing thus apply to the metric in eq.~\eqref{genmet}. Moreover, the form of the metric in eq.~\eqref{genmet} makes clear that the dual BCFTs and DCFTs are trivially $\DW$ invariant, up to anomalies: in the bulk, large diffeomorphisms act on the BCFT or DCFT as $\DW$ transformations, and changes in the defining functions implement Weyl transformations, just as in a CFT with an $AdS_{d+1}$ dual. For example, with the natural defining functions the dual BCFT or DCFT lives on $AdS_{d+1-n} \times \Sp^{n-1}$, and a change of defining function implements a Weyl transformation to the Minkowski metric.

In the interest of full disclosure, we must mention that in many ten- and eleven-dimensional supergravity duals to BCFTs and DCFTs, the bulk metric is in fact more complicated than that in eq.~\eqref{genmet}~\cite{D'Hoker:2007xy,D'Hoker:2007xz,D'Hoker:2009gg,Chiodaroli:2009yw,Chiodaroli:2011fn,Berdichevsky:2013ija}. Often the bulk spacetime is not simply a direct product of an asymptotically-$AdS_{d+1}$ factor with a compact manifold. Typically the compact manifold is non-trivially fibered over the non-compact part of the spacetime. In such cases the metric only approaches that in eq.~\eqref{genmet} asymptotically. Our analysis will not be immediately applicable to such solutions, however the generalization of our analysis to such solutions is straightforward.

As mentioned above, BCFTs with $d=2$ and DCFTs with $n=d-1$ are special cases because in the field theory the boundary or defect is point-like. The holographic duals are also special cases, for example when $n=d-1$, in the metric of eq.~\eqref{genmet} $AdS_{d+1-n} = AdS_2$ and $\rp$ does not exist. We will defer our discussion of BCFTs with $d=2$ and DCFTs with $n=d-1$ to the end of this subsection. For now, we will consider only BCFTs with $d>2$ and DCFTs with $n<d-1$.

To extend CHM's proof of RT's conjecture to metrics of the form in eq.~\eqref{genmet}, we must perform three steps. First, we must find the minimal area surface that, near the boundary of the bulk spacetime, approaches a (hemi-)sphere of radius $R$ centered on the boundary or defect. Second we must find a hyperbolic slicing of the bulk spacetime exhibiting a horizon, dual to the BCFT or DCFT on $\mathbb{R} \times \Hy^{d-1}$ at temperature $T_0=1/(2\pi R)$. Third, we must show that the hyperbolic horizon coincides with the minimal area surface.

For the first step, finding the minimal-area surface, when the space is $AdS_{d+1}$ we already know the answer, namely the solution $z(r)$ in eq.~\eqref{zsol}. What does the $z(r)$ in eq.~\eqref{zsol} look like in the $AdS_{d+1-n}$ slicing of $AdS_{d+1}$? Using eq.~\eqref{pointoads}, we find
\beq
z^2 + r^2 = z^2 + \rn^2 + \rp^2 = Z^2 \sech^2(x) + Z^2 \tanh^2( x) + \rp^2 = Z^2 + \rp^2 = R^2,
\eeq
that is, all dependence on $x$ disappears, and the form of the solution remains the same, but with $z \to Z$ and $r \to \rp$. The minimal-area surface thus looks like a hemisphere centered at the boundary, but contained entirely within the $AdS_{d+1-n}$ slice. Of course, the solution describes exactly the same surface. In particular, if we approach the $AdS_{d+1}$ boundary by fixing $x$ and taking $Z \to 0$, so that we arrive at the boundary at a point on the defect, then the solution becomes $\rp^2 = R^2$, which describes the part of the sphere that intersects the defect. If we approach the $AdS_{d+1}$ boundary by fixing $Z$ and taking $x \to \infty$ for $n>1$ or $x \to \pm \infty$ for $n=1$, so that we arrive at the boundary a distance $|\rn| = |Z|$ away from the defect, then the solution becomes $\rn^2 + \rp^2 = r^2 = R^2$, which describes the part of the sphere away from the defect.

In fact, we can easily show that $Z^2 + \rp^2 = R^2$ remains a solution with \textit{any} warp factors $A(x)$ and $B(x)$. With the metric in eq.~\eqref{genmet}, the minimal area surface that we want will wrap all of the $\Sp^{n-1}$ and $\Sp^{d-n-2}$ directions and will describe a codimension-one surface in the space of $Z$, $\rp$, and $x$. Upon parameterizing that surface as $Z(\rp,x)$, the area functional $\A$ takes the form
\beq
\label{adssliceA}
\A = \text{vol}\left(\Sp^{n-1}\right) \text{vol}(\Sp^{d-n-2})L^{d-1} \int dx d\rp A(x)^{d-n-1} B(x)^{n-1} \frac{\rp^{d-n-2}}{Z^{d-n-1}} \sqrt{1 + (\partial_{\rp} Z)^2 + \frac{A(x)^2}{Z^2}\left(\partial_x Z \right)^2}.
\eeq
Clearly the Euler-Lagrange equation for $Z(\rp,x)$ will be a complicated non-linear partial differential equation, but if we assume that $\partial_x Z(\rp,x)=0$, then $\A$ becomes
\beq
\label{adssliceA2}
\A = \text{vol}\left(\Sp^{n-1}\right) \text{vol}(\Sp^{d-n-2})L^{d-1} \left[ \int dx \, A(x)^{d-n-1} B(x)^{n-1} \right]\left[ \int d\rp \, \frac{\rp^{d-n-2}}{Z^{d-n-1}} \sqrt{1 + (\partial_{\rp} Z)^2} \right].
\eeq
Crucially, in eq.~\eqref{adssliceA2} the $x$ and $\rp$ dependence factorizes completely. All $x$-dependent factors thus become an overall ``constant'' that drops out of the Euler-Lagrange equation for $Z(\rp)$, which is then just the equation for $Z(\rp)$ in $AdS_{d+1-n}$, of the form in eq.~\eqref{afunc} but with $z \to Z$, $r \to \rp$, and $d \to d-n$. We therefore conclude that, within the class of $x$-independent solutions, $Z(\rp)^2 + \rp^2 = R^2$ is the unique solution producing the minimal value of the $\A$ in eq.~\eqref{adssliceA2}, which is
\beq
\label{adssliceamin}
\Am = \text{vol}\left(\Sp^{n-1}\right) \text{vol}(\Sp^{d-n-2})L^{d-1}\left[ \int d\rp \, \frac{R \, \rp^{d-n-2}}{(R^2-\rp^2)^{(d-n)/2}} \right] \left[ \int dx \, A(x)^{d-n-1} B(x)^{n-1} \right]\,.
\eeq

As in $AdS_{d+1}$, the $\A$ in eq.~\eqref{adssliceA} diverges because the bulk spacetime has infinite volume, so we need cutoffs for $x$ and $\rp$. As discussed at the beginning of this section, in a BCFT or DCFT we should use the same regularization as in the underlying CFT. In the gravity duals of BCFT or DCFTs, in the space spanned by $x$ and $\rp$ we will thus introduce a cutoff surface that corresponds to $z = \varepsilon$ in any asymptotically locally $AdS_{d+1}$ region. We discuss these cutoff surfaces in detail in appendix~\ref{B:cutoffs}. For now we will just assume such a cutoff surface is in place.

So far we have only shown that $Z(\rp)^2 + \rp^2 = R^2$ is a solution for any warp factors $A(x)$ and $B(x)$, and so produces a local extremum of $\A$. Does it produce the \textit{global minimum} $\Am$, however? Could some solution with non-trivial $x$ dependence have smaller area? We can prove rigorously that $Z(\rp)^2 + \rp^2 = R^2$ indeed produces $\Am$ with any $A(x)$ and $B(x)$ in several ways, for example using a special conformal transformation. In appendix~\ref{A:globmin} we present a quick proof that is a straightforward modification of our proof for the global minimization of $\A$ in $AdS_{d+1}$. 

In our extension of CHM's proof to BCFTs and DCFTs, the second step is to find a slicing of the metric in eq.~\eqref{genmet} that describes the dual BCFT or DCFT on $\mathbb{R} \times \Hy^{d-1}$ in thermal equilibrium at temperature $T_0 = 1/(2 \pi R)$. An obvious guess is to perform a hyperbolic slicing of the $AdS_{d+1-n}$ slice itself. Luckily, that guess is correct. With the metric in eq.~\eqref{genmet}, let us perform a change of coordinates and coordinate restriction on the $AdS_{d+1-n}$ slice, identical in form to eq.~\eqref{pointohyp}:
\begin{subequations}
\label{E:defectPoinToHyp}
\begin{align}
Z &= \frac{R L}{V \cosh U + \cosh \left(\tau/R\right) \sqrt{V^2 - L^2}}, \\
\label{E:defectrptohyp}
\rp&= \frac{R \, V \, \sinh U}{V \cosh U + \cosh \left(\tau/R\right) \sqrt{V^2 - L^2}},
\\
t &= \frac{R \, \sinh \left(\tau/R\right) \sqrt{V^2 - L^2}}{V \cosh U +  \cosh \left(\tau/R\right) \sqrt{V^2 - L^2}},
\end{align}
\end{subequations}
where $V \in [L,\infty)$, $U\in[0,\infty)$, and $\tau \in (-\infty,\infty)$, and where we leave unchanged all other coordinates, including $x$, the $\Sp^{n-1}$ coordinates, and all coordinates of the $\Sp^{d-n-2}$ inside $AdS_{d+1-n}$. The metric on the $AdS_{d+1-n}$ slice then takes the form
\beq
\label{adsslicehyp}
L^2 g_{AdS_{d+1-n}} = \frac{dV^2}{f(V)} - f(V) \, \frac{L^2}{R^2} \, d\tau^2 + V^2 \, g_{\mathbb{H}^{d-n-1}}\,, \qquad f(V) = \frac{V^2}{L^2} - 1\,,
\eeq
where $g_{\Hy^{d-n-1}}$ is the metric of a $(d-n-1)$-dimensional unit-radius hyperbolic space,
\beq
\label{E:slicehypmetric}
g_{\mathbb{H}^{d-n-1}} = dU^2 + \sinh^2 U \,\, g_{\Sp^{d-n-2}}\,.
\eeq
All of our statements above about the hyperbolic slicing of $AdS_{d+1}$ apply again here, with the replacement $d+1 \to d + 1-n$. For example, with fixed $x$ we can approach the boundary of $AdS_{d+1-n}$ in three ways: by fixing $\tau$ and $U$ and sending $V \to \infty$, by fixing $\tau$ and $V$ and sending $U \to \infty$, or by fixing $U$ and $V$ and sending $\tau \to \pm \infty$, all of which send $Z \to 0$. These limits reveal that at the $AdS_{d+1-n}$ boundary the hyperbolic slicing covers only intersection of the (hemi-)sphere's causal diamond with the defect or boundary.

With the $AdS_{d+1-n}$ slices in hyperbolic slicing, the metric in eq.~\eqref{genmet} becomes
\beq
\label{E:defectGhyper}
g = A(x)^2 \left[ \frac{dV^2}{f(V)}-f(V)\frac{L^2}{R^2}d\tau^2 + V^2 g_{\mathbb{H}^{d-n-1}}\right] + L^2 \left( dx^2 + B(x)^2 g_{\mathbb{S}^{n-1}}\right)\,,
\eeq
which has a horizon at $V_H \equiv L$, the outermost value of $V$ where $f(V)=0$. The Hawking temperature of the horizon at $V_H$ is $T_0 = 1/(2 \pi R)$. The AdS/CFT dictionary then implies that the dual BCFT or DCFT is in thermal equilibrium at temperature $T_0$. Plugging $V_H=L$ into eq.~\eqref{E:defectrptohyp} we find $\rp = R \, \tanh U$, so using eq.~\eqref{E:hypvol} we can write the area of the horizon as
\beq
\label{E:defecthorarea}
\A_H = \text{vol}\left(\Sp^{n-1}\right) L^{d-1} \vol(\Sp^{d-2}) \left[ \int d\rp \, \frac{R\rp^{d-2}}{(R^2-\rp^2)^{d/2}} \right]\left[ \int dx \, A(x)^{d-n-1} B(x)^{n-1} \right].
\eeq
In eq.~\eqref{E:defecthorarea} we must also introduce the cutoffs on $x$ and $\rp$ discussed in appendix~\ref{B:cutoffs}.

With the $AdS_{d+1-n}$ slice itself in hyperbolic slicing, in what spacetime does the dual BCFT or DCFT live? If we use the natural defining functions for $AdS_{d+1-n}$ slicing, then the metric at the boundary is that of $AdS_{d+1-n} \times \Sp^{n-1}$, as discussed below eq.~\eqref{genmet}, but now the $AdS_{d+1-n}$ is in hyperbolic slicing:
\beq
\label{E:boundarymethyp}
g_{AdS_{d+1-n}} + g_{\Sp^{n-1}} = \frac{1}{L^2} \left[ \frac{dV^2}{f(V)}-f(V)\frac{L^2}{R^2}d\tau^2 + V^2 g_{\mathbb{H}^{d-n-1}}\right] + g_{\mathbb{S}^{n-1}}\,.
\eeq
For our proof, we want instead $\mathbb{R} \times \Hy^{d-1}$, where $\mathbb{R}$ is the time $\tau$ and the $\Hy^{d-1}$ has radius $R$. After a change of coordinates the metric in eq.~\eqref{E:boundarymethyp} is in fact conformal to the metric of $\mathbb{R} \times \Hy^{d-1}$, as we will now show. In eq.~\eqref{E:boundarymethyp}, if we extract a factor of $f(V) \frac{L^2}{R^2}$ from the terms in the square brackets and a factor of $f(V)/R^2$ from the $g_{\Sp^{n-1}}$ term, and we use eq.~\eqref{E:slicehypmetric}, then we find
\beq
\label{E:boundarymethyp2}
g_{AdS_{d+1-n}} + g_{\Sp^{n-1}} = \frac{f(V)}{R^2} \left[-d\tau^2 + R^2 \left( \frac{dV^2}{L^2 f(V)^2} + \frac{V^2}{L^2 f(V)} dU^2 + \frac{V^2 \, \sinh^2 U}{L^2 f(V)} g_{\Sp^{d-n-2}} + \frac{1}{f(V)} g_{\mathbb{S}^{n-1}} \right)\right]\,.
\eeq
If we now change coordinates,
\beq
\sinh u \, \sin \a = \frac{1}{\sqrt{f(V)}}, \qquad \sinh u \, \cos \a = \frac{V \sinh U}{L\sqrt{f(V)}},
\eeq
with $u\in (\infty,\infty)$ and $\a \in [0,\pi/2]$, then the metric in eq.~\eqref{E:boundarymethyp2} becomes
\bea
g_{AdS_{d+1-n}} + g_{\Sp^{n-1}} & =  & \frac{f(V)}{R^2} \left[ - d\tau^2  + R^2 \left( du^2 +\sinh^2 u\left [ d\a^2 + \cos^2\alpha \, g_{\Sp^{d-n-2} }  + \sin^2\alpha \, g_{\Sp^{n-1}}\right] \right)  \right] \nn \\ & = & \frac{f(V)}{R^2} \left [ -d\tau^2 + R^2 g_{\Hy^{d-1}} \right].
\eea
We have thus shown that the metric of $AdS_{d+1-n} \times \Sp^{n-1}$, with $AdS_{d+1-n}$ in hyperbolic slicing, is conformal to the metric of $\mathbb{R} \times \Hy^{d-1}$, where $\mathbb{R}$ is the time $\tau$ and $\Hy^{d-1}$ has radius $R$, as advertised. To implement the Weyl transformation from $AdS_{d+1-n} \times \Sp^{n-1}$ to $\mathbb{R} \times \Hy^{d-1}$, we simply switch from the natural defining function of $AdS_{d+1}$ in $AdS_{d+1-n}$ slicing to that of hyperbolic slicing.

We have now performed the first two steps of our proof. The third and final step is to show that the minimal area surface $Z^2 + \rp^2 = R^2$ coincides with the horizon in the hyperbolic slicing, $V_H$. From eq.~\eqref{E:defectPoinToHyp} we find that at the horizon $V_H=L$,
\beq
\label{E:defectminareasol}
\left . \left[Z^2+r_{||}^2\,\right] \right|_{V_H} = R^2,
\eeq
which is precisely the minimal area surface. Using eq.~\eqref{E:defectPoinToHyp} we can straightforwardly match the cutoffs on the $x$ and $\rp$ integrations. We thus have $\mathcal{A}_H = \mathcal{A}_{\rm min}$, which completes our proof of RT's conjecture: for the BCFT or DCFT in Minkowski space, when $\Sigma$ is a (hemi-)sphere centered on the boundary or defect, $\see$ is identical to the thermal entropy of the same BCFT or DCFT on $\mathbb{R} \times \Hy^{d-1}$ at temperature $T_0 = 1/(2\pi R)$, which via holography is $\A_H/(4G) = \Am/(4G)$.

As discussed in subsection~\ref{S:CHMcft}, in the BCFT or DCFT we can write the R\'enyi entropies $S_q$ as differences of the free energy of the theory on $\mathbb{R} \times \Hy^{d-1}$ at temperatures $T_0$ and $T_0/q$, eq.~\eqref{E:SrenyiF}. Can we compute those free energies, and hence $S_q$, holographically? To do so, we need black hole solutions that describe the BCFT or DCFT on $\mathbb{R}\times\mathbb{H}^{d-1}$ at the temperatures $T_0$ and $T_0/q$. For the metric in eq.~\eqref{genmet}, we have shown that a hyperbolic slicing of the $AdS_{d+1-n}$ slice, eq.~\eqref{E:defectGhyper}, describes the dual BCFT or DCFT on $\mathbb{R}\times\mathbb{H}^{d-1}$ at temperature $T_0$, but how do we describe other temperatures? One na\"ive guess is to introduce a non-zero black brane mass $m$ into the blackening factor $f(V)$ of eq.~\eqref{E:defectGhyper}, in analogy with the $AdS_{d+1}$ case, eq.~\eqref{E:adshypbraneotherT}. That is clearly wrong, however: a straightforward exercise shows that the metric at the boundary is conformal to $\mathbb{R}\times\mathbb{H}^{d-1}$ only when $m=0$. Finding a hyperbolic black brane describing the dual BCFT or DCFT on $\mathbb{R} \times \Hy^{d-1}$ at temperature $T_0/q$ will require more work, which we will leave for the future.

Let us now return to the special cases of BCFTs in $d=2$ and DCFTs with $n=d-1$, where in the field theory the boundary or impurity is point-like. Extending our proof to these special cases requires only minor modifications, so we will be brief. For the dual of a DCFT with $n=d-1$, the metric in eq.~\eqref{genmet} is
\beq
\label{E:Gimpurity}
g = L^2 (dx^2 + A(x)^2 g_{AdS_2} + B(x)^2 g_{\mathbb{S}^{d-2}}), \qquad g_{AdS_2} = \frac{1}{Z^2} \left(dZ^2 - dt^2\right).
\eeq
The warp factors $A(x)$ and $B(x)$ have the same asymptotics as in all other cases with a defect of codimension greater than one, as described below eq.~\eqref{genmet}. For the dual of a BCFT with $d=2$, simply set $d=2$ in eq.~\eqref{E:Gimpurity}, and in the following.

The first step in our proof is to find the minimal area surface at $t=0$ that asymptotically approaches a (hemi-)sphere of radius $R$ centered on the boundary or impurity. That minimal area surface will wrap the $\Sp^{d-2}$ and will trace a curve in the plane spanned by $Z$ and $x$. Upon parameterizing that curve as $Z(x)$, we can write the area functional as
\beq
\label{E:impurityA}
\mathcal{A} = \text{vol}(\mathbb{S}^{d-2})L^{d-1} \int dx\, B(x)^{d-2} \sqrt{1+\frac{A(x)^2 Z'(x)^2}{Z(x)^2}}.
\eeq
In eq.~\eqref{E:impurityA}, the factor under the square root factor is a sum of squares, and hence attains a global minimum when $Z'(x)/Z(x)=0$, that is, when $Z(x)$ is a constant. The minimal area surface we want is thus $Z(x)=R$.

The second step in our proof is to find a slicing of the metric in eq.~\eqref{E:Gimpurity} that describes the dual BCFT or DCFT on $\mathbb{R} \times \Hy^{d-1}$ at temperature $T_0 = 1/(2 \pi R)$. A change of coordinates
\begin{subequations}
\label{E:impuritypointohyp}
\bea
Z & = & \frac{RL}{V + \cosh(\tau/R) \sqrt{V^2-L^2}}, \\ t & = & \frac{R \sinh(\tau/R) \sqrt{V^2 - L^2}}{V + \cosh(\tau/R) \sqrt{V^2-L^2}}.
\eea
\end{subequations}
and coordinate restriction puts the $AdS_2$ metric in ``hyperbolic'' slicing, where we use quotes because $V$ is the only spatial direction, so no hyperbolic plane is present:
\beq
\label{E:gAdS2H}
L^2g_{AdS_2} = \frac{dV^2}{f(V)} -f(V) \frac{L^2}{R^2}d\tau^2\,, \qquad f(V) = \frac{V^2}{L^2}-1.
\eeq
The metric in eq.~\eqref{E:gAdS2H} has a horizon at $V_H=L$, with Hawking temperature $T_0=1/(2\pi R)$. Following the steps explained below eq.~\eqref{E:boundarymethyp2}, with obvious modifications, we find that the dual BCFT or DCFT lives on a spacetime conformal to $\mathbb{R} \times \Hy^{d-1}$, where $\Hy^{d-1}$ has radius $R$.

The third and final step in our proof is to show that the minimal area surface and the horizon are identical. Plugging $V_H=L$ into eq.~\eqref{E:impuritypointohyp}, we indeed find $t=0$ and $\left . Z \right |_{V_H} = R$, which completes our proof for these special cases.

\section{Application: Probe Branes} \label{S:probes}

We will now put the results of the previous section to use: we will use our generalization of CHM to compute EE holographically from probe branes.

To be precise, in this section we consider a bottom-up model of codimension-$n$ branes in Einstein gravity with negative cosmological constant. These systems are defined by the following bulk action:
\begin{subequations}
\label{E:bottomup}
\beq
S_{d,n} = \seh + \sbr,
\eeq
\beq
\seh = \frac{1}{16\pi G}\int d^{d+1}x \sqrt{-\text{det }g} \left( R_g +\frac{d(d-1)}{L^2}\right),
\eeq
\beq
\label{E:sbr}
\sbr = - \frac{\mu_n}{16\pi G}\int d^{d-n+1}\xi \, \sqrt{-\text{det P}[g]},
\eeq
\end{subequations}
where $g$ is the bulk metric, $R_g$ is the Ricci scalar built from $g$, $\mu_n/(16 \pi G)$ is the tension of the brane, $\xi$ denotes the brane's worldvolume coordinates, and P$[g]$ is the pullback of the bulk metric to the brane worldvolume. To guarantee a well-posed variational principle and finite on-shell action, we must also add boundary terms to the action in eq.~\eqref{E:bottomup}, such as for example the Gibbons-Hawking term. We will not write the boundary terms explicitly, except in subsection~\ref{S:probeCHM}.

In the Einstein's equation arising from eq.~\eqref{E:bottomup}, the right-hand-side includes only two contributions: one from the negative cosmological constant and one from the brane's stress-energy tensor. The latter is proportional to the dimensionless parameter $\mu_n L^{2-n}$. When $\mu_nL^{2-n}=0$, the solution of Einstein's equation is $AdS_{d+1}$ with radius of curvature $L$, and the gravity theory in that background is dual to a CFT. If $\mu_n L^{2-n}$ is non-zero but small, $\mu_n L^{2-n} \ll 1$, then we can take the probe limit: expand all quantities in $\mu_n L^{2-n}$, and retain terms up to linear order in $\mu_n L^{2-n}$. In that limit the brane probes an undeformed $AdS_{d+1}$ spacetime. As we increase $\mu_n L^{2-n}$, eventually we leave the probe limit and the brane back-reacts on the metric, which is then of the form in eq.~\eqref{genmet}, \textit{i.e.} the back-reaction preserves the $SO(2,d-n) \times SO(n)$ subgroup of the $AdS_{d+1}$ isometry. The bulk gravity theory in that background is dual to a DCFT.

Eq.~\eqref{S:bottomup} arises as an effective action for many top-down systems, including various brane intersections in string and M-theory, usually after Kaluza-Klein (KK) reduction. Of course, string theory includes more fields than just the metric and a brane. Indeed, all known supersymmetric (SUSY) branes act as sources not only for the metric but also for other fields, such as the dilaton and RR fields. Eq.~\eqref{S:bottomup} is a reliable effective action whenever the only contributions to the Einstein-frame stress-energy tensor come from a negative cosmological constant and the brane itself, as emphasized in ref.~\cite{Chang:2013mca} and as we discuss in detail below. That may occur only in some limit, such as the probe limit.

In many top-down systems a precise dictionary exists between the parameters in eq.~\eqref{E:bottomup}, $G$, $L$, and $\mu_n$, and field theory parameters, allowing us to translate from the gravity theory to the field theory. For example, when eq.~\eqref{S:bottomup} describes intersecting branes in string or M-theory, typically the dual field theory is a non-Abelian gauge theory coupled to fields in the fundamental representation of the gauge group, \textit{i.e.}\ flavor fields~\cite{Karch:2000gx,DeWolfe:2001pq,Karch:2002sh,Constable:2002xt}. Roughly speaking, $\seh$ is dual to the adjoint degrees of freedom while $\sbr$ is dual to the flavor degrees of freedom, which are restricted to propagate along a defect of codimension $n$. If the rank of the gauge group is $N$ and the number of flavors is $N_f$, then typically $1/G$ counts the number of adjoint degrees of freedom, for example when $d=4$ typically $1/G \propto N^2$, so that $\mu_n / G \propto N_f N$ counts the number of flavor degrees of freedom, and $\mu_n L^{2-n}\propto N_f/N$ measures the ratio of flavor to adjoint degrees of freedom.

Knowing that $\mu_n L^{n-2}$ measures the ratio of flavor to adjoint degrees of freedom, we can translate the various limits of $\mu_n L^{2-n}$ described above from the gravity theory to the field theory. The case $\mu_n L^{2-n} = 0$ translates to $N_f=0$: the theory is a CFT with no flavor fields. The probe limit translates to $N_f \neq 0$ but $N_f \ll N$, where the field theory is a CFT ``probed'' by flavor fields. More precisely, the adjoint degrees of freedom vastly outnumber the flavor degrees of freedom, so we ignore the flavor contribution to any beta functions of single-trace couplings, such as an 't Hooft coupling. Increasing $\mu_n L^{2-n}$ and leaving the probe limit means increasing $N_f$ relative to $N$ such that we cannot neglect the flavor contribution to the beta functions of single-trace couplings. Typically those contributions will be positive, but in some special cases the flavor fields may preserve defect conformal symmetry to leading order, or even to all orders, in $N_f$. In such cases the field theory is a DCFT, with the flavor degrees of freedom restricted to the defect.

In the gravity theory of eq.~\eqref{E:bottomup}, suppose we calculate the metric, including the brane's back-reaction, and then compute $\see$ holographically using RT's prescription, with the minimal area surface in eq.~\eqref{E:defectminareasol}. Suppose also that $\mu_n L^{2-n} \ll 1$ and we expand $\see$ in $\mu_n L^{2-n}$,
\beq
\label{E:eegenform}
\see = \seez + \seeo + \mathcal{O}(\mu_n^2 L^{2(2-n)})\,.
\eeq
Here $\seez$ is the $\mu_n L^{2-n}=0$ result and $\seeo$ is the  contribution linear in $\mu_n L^{2-n}$, hence the superscripts. Our main question in this section is: can we compute $\seeo$ without ever computing back-reaction? In technical terms, can we calclate $\seeo$ from the action of the probe brane $\sbr$ in the undeformed $AdS_{d+1}$ background?

As shown in ref.~\cite{Chang:2013mca}, the answer is ``yes.'' The strategy in ref.~\cite{Chang:2013mca} was to do the ``honest'' calculation: compute the back-reaction of the brane on the metric to order $\mu_n L^{2-n}$, from that determine the order $\mu_n L^{2-n}$ shift in $\Am$, and hence obtain $\seeo$. In fact, that approach is very general, requiring neither (defect) conformal symmetry nor a spherical $\S$, but rather requiring only that eq.~\eqref{E:bottomup} is a reliable effective action in the probe limit. Although the intermediate steps in ref.~\cite{Chang:2013mca}  involved computing back-reaction, the final result for $\seeo$ only involved information from the probe limit. Specifically, $\seeo$ can be written as a double integral over the undeformed ($\mu_n L^{2-n}=0$) minimal area surface and the probe brane worldvolume, with an integrand including three factors: the ``stress-energy tensor'' of the area functional, $\delta\A/\delta g$, and the graviton propagator, both at $\mu_n L^{2-n}=0$, and the stress-energy tensor of the probe brane, evaluated in the undeformed background. In general, that double integral may be difficult to perform, in part because the graviton propagator is known in very few backgrounds, including $AdS_{d+1}$, and even then is not simple. Nevertheless, by performing the double integral, in ref.~\cite{Chang:2013mca} $\seeo$ was obtained for the theory in eq.~\eqref{E:bottomup}, with any $d$ and $n=0$ or $n=1$, and for two top-down systems in type IIB string theory describing CFTs in $d=4$ with $n=0$ and $n=1$ SUSY defect flavor fields.

In subsection~\ref{S:bottomup} we present a method simpler than that of ref.~\cite{Chang:2013mca}, based on our results from section~\ref{S:dcfts}, and hence relying crucially on defect conformal symmetry and a spherical $\S$. We treat eq.~\eqref{E:bottomup} as a self-contained bottom-up system, ignoring the question of whether it arises as an effective action for some top-down system, and proceed through two steps.

The first step is an ``honest'' calculation similar to that of ref.~\cite{Chang:2013mca}: we allow the brane to back-react, compute $\see$ using RT's prescription with the minimal-area surface in eq.~\eqref{E:defectminareasol}, expand the result in $\mu_n L^{2-n} \ll 1$, and extract $\seeo$. We present explicit calculations only for the two cases in ref.~\cite{Chang:2013mca}, branes with $n=0$, in subsection~\ref{S:codimzero}, and $n=1$, in subsection~\ref{S:codimone}.

In subsection~\ref{S:probeCHM} we take the second step, which exploits our results from section~\ref{S:dcfts}. We work in the probe limit from the beginning, so the bulk spacetime is $AdS_{d+1}$, and following CHM we switch from Poincar\'e to hyperbolic slicing, so that $\seeo$ maps to the order $\mu_n L^{2-n}$ contribution to the thermal entropy. Fortunately, in the probe limit the only two contributions to the thermal entropy at order $\mu_n L^{2-n}$ are from $\sbr$, evaluated on the undeformed background, and a boundary term proportional to the order $\mu_n L^{2-n}$ correction to the metric, whose precise form we discuss below. We can show that the latter vanishes, in which case the \textit{only} contribution to the thermal entropy at order $\mu_n L^{2-n}$ is from $\sbr$. We thereby obtain $\seeo$ for any $d$ and $n$. When $n=0$ or $n=1$ we find perfect agreement with the results of the ``honest'' calculations in ref.~\cite{Chang:2013mca}, which provides two non-trivial checks of our method. In subsection~\ref{S:renyi}, we use essentially the same technique to compute the order $\mu_n L^{2-n}$ contributions to the R\'enyi entropies $S_q$, which we denote $\sqo$. Our results for $\seeo$ and $\sqo$ are in eqs.~\eqref{E:een} and eq.~\eqref{E:proberenyi}, respectively.

The main message of this section is: given a DCFT described holographically by the action in eq.~\eqref{E:bottomup}, if $\Sigma$ is a sphere centered on the defect and if $\mu_n L^{2-n} \ll 1$, then using our results from section~\ref{S:dcfts} we can holographically calculate $\seeo$ and $\sqo$ entirely within the probe limit, just using $\sbr$ evaluated on the undeformed ($\mu_n L^{2-n}=0$) background metric, without ever calculating back-reaction or a minimal area surface. That is obviously a very useful simplification.

In subsection~\ref{S:topdown} we apply the results of subsection~\ref{S:bottomup} to top-down systems, namely SUSY brane intersections in string and M-theory. We begin by enumerating the criteria that such branes must satisfy for eq.~\eqref{E:bottomup} to be a reliable effective action. One of these criteria is that the only contribution to the (Einstein-frame) stress-energy tensor at order $\mu_n L^{2-n}$ comes from the brane alone~\cite{Chang:2013mca}. We then compute $\seeo$ in four examples where these criteria are satisfied. These include the two examples of ref.~\cite{Chang:2013mca}, plus two in M-theory. Our fifth and final example is an $n=0$ brane in type IIA supergravity whose back-reaction preserves $AdS_4$. In this example we show that eq.~\eqref{E:bottomup} is not a reliable effective action because the RR two-form sourced by the brane contributes to the Einstein-frame stress-energy tensor at order $\mu_n L^{2-n}$. We then show explicitly, using the fully back-reacted solution, that $\seeo$ receives \textit{two} non-zero contributions, one from the brane action, and one from a boundary term proportional to the order $\mu_0 L^2$ correction to the RR two-form. The lesson of this example is: if eq.~\eqref{E:bottomup} is not a reliable effective action, then we must check whether all boundary terms involving fields sourced by the brane vanish before using our method.

\subsection{Bottom-Up Systems}
\label{S:bottomup}

\subsubsection{Codimension Zero}
\label{S:codimzero}

Follwing ref.~\cite{Chang:2013mca}, let us consider a codimension-zero brane, meaning a space-filling brane, for which $n=0$ in eq.~\eqref{E:bottomup}. The brane's back-reaction is trivial: the brane's tension, $\mu_0/(16 \pi G)$, merely shifts the cosmological constant, which in turn shifts the radius of curvature of $AdS_{d+1}$. We thus define a new $AdS_{d+1}$ radius of curvature, $\ell$, in terms of $\mu_0$ and the old radius of curvature, $L$,
\beq
\label{E:elldef}
\frac{d(d-1)}{\ell^2} \equiv \frac{d(d-1)}{L^2}-\mu_0.
\eeq
The minimal area surface that describes a sphere of radius $R$ at the $AdS_{d+1}$ boundary is eq.~\eqref{zsol}, $z^2 + r^2 = R^2$. The result for $\see$ is eq.~\eqref{E:adsEE} with the substitution $L\to \ell$,
\beq
\label{E:codimzerobackreactEE}
\see = \frac{\ell^{d-1}}{4 \, G}\text{vol}(\mathbb{S}^{d-2})\int dr \frac{R \, r^{d-2}}{(R^2-r^2)^{d/2}},
\eeq
where the $r$ integration is from $r=0$ up to the cutoff in eq.~\eqref{E:reps}, $\reps = \sqrt{R^2 - \varepsilon^2}$.

We now take the probe limit: assuming $\mu_0 L^2 \ll 1$, we expand $\ell$ in $\mu_0 L^2$,
\beq
\label{E:codim0L}
\ell = L\left(1-\frac{\mu_0L^2}{d(d-1)}\right)^{-1/2} = L + \frac{\mu_0L^3}{2d(d-1)} + \Ocal\left(\mu_0^2L^5\right),
\eeq
from which we trivially find the convenient formula
\beq
\label{E:codim0Luseful}
\ell^{d-1} = L^{d-1}\left( 1 + \frac{\mu_0L^2}{2d}+\mathcal{O}(\mu_0^2L^4)\right).
\eeq
Inserting eq.~\eqref{E:codim0Luseful} into eq.~\eqref{E:codimzerobackreactEE}, we find an expansion for $\see$ of the form in eq.~\eqref{E:eegenform}, with
\begin{subequations}
\bea
\label{E:codimzero}
\seez & = & \frac{L^{d-1}}{4 \, G}\vol(\mathbb{S}^{d-2})\int dr \frac{R \, r^{d-2}}{(R^2-r^2)^{d/2}},\\
\label{E:codimzeroprobe}
\seeo & = & \frac{\mu_0 L^2}{2d}\seez,
\eea
\end{subequations}
as found in ref.~\cite{Chang:2013mca}. In subsection~\ref{S:probeCHM} we will reproduce eq.~\eqref{E:codimzeroprobe} from $\sbr$ directly in the probe limit, without computing the back-reaction.

\subsubsection{Codimension One}
\label{S:codimone}

Again following ref.~\cite{Chang:2013mca}, let us now consider a codimension-one brane, $n=1$. When $d=2$, in the field theory a codimension-one defect is point-like. As we saw in section~\ref{S:dcfts}, a point-like defect requires special treatment. For simplicity, in this subsection we will restrict to $d > 2$. For the holographic calculation of $\see$ when $n=1$ and $d=2$, see ref.~\cite{Azeyanagi:2007qj}.  

After we account for the brane's back-reaction, if we integrate Einstein's equation then we obtain the Israel junction conditions at the brane: the extrinsic curvature jumps discontinuously at the brane by an amount $\propto \mu_1 L$~\cite{Karch:2000gx,Azeyanagi:2007qj,Chang:2013mca}. The metric then takes the form in eq.~\eqref{genmet}, with $n=1$,
\beq
\label{E:codim1g}
g = L^2 \left( dx^2 + A(x)^2g_{AdS_d}\right),
\eeq
where the warp factor $A(x)$ is
\beq
\label{E:codim1warp}
A(x) = \cosh(|x|-x_*)\,, \qquad x_* \equiv \textrm{arctanh}\left( \frac{\mu_1 L}{4(d-1)}\right).
\eeq
Clearly $A(x)$ has the asymptotics described below eq.~\eqref{genmet}: $A(x) \to \exp(\pm x-x_*)/2$ as $x \to \pm \infty$, and the spacetime has two asymptotically $AdS_{d+1}$ regions glued together at the brane.

The minimal area surface that asymptotically approaches a sphere of radius $R$ centered on the defect is eq.~\eqref{E:defectminareasol}, $Z(\rp)^2 + \rp^2 = R^2$. The value of the minimal area is then the $\Am$ in eq.~\eqref{adssliceamin}, and hence $\see$ is
\beq
\label{E:codimoneEE}
\see = \frac{L^{d-1}}{4G} \, \text{vol}(\mathbb{S}^{d-3}) \int d\rp \, dx \,A(x)^{d-2} \frac{R \, \rp^{d-3}}{(R^2-\rp^2)^{(d-1)/2}},
\eeq
where the $x$ integration is over the interval $[-\xeps,\xeps]$, with $\xeps$ the cutoff defined in eq.~\eqref{E:xrCutoffs}. Specifically, plugging the $A(x)$ in eq.~\eqref{E:codim1warp} into eq.~\eqref{E:xrCutoffs}, we find
\beq
\label{E:codimonecutoff}
\xeps(\rp,R) = \textrm{arccosh}\left( \frac{1}{\varepsilon}\sqrt{R^2-\rp^2}\right) + \textrm{arctanh}\left( \frac{\mu_1 L}{4(d-1)}\right).
\eeq
After performing the $x$ integral in eq.~\eqref{E:codimoneEE} with the $\rp$-dependent cutoff in eq.~\eqref{E:codimonecutoff}, we integrate $\rp$ from $\rp=0$ up to $\rp=\sqrt{R^2 - \varepsilon^2}$.

We now take the probe limit: we assume $\mu_1 L \ll 1$ and expand the $\see$ in eq.~\eqref{E:codimoneEE} in $\mu_1 L$. In eq.~\eqref{E:codimoneEE}, $\mu_1 L$ appears only in the $x$ integration, both in the integrand, via the $A(x)^{d-2}$ factor, and in the endpoints of the $x$ integration, $\pm\xeps$. We thus focus on the integration over $x$ in eq.~\eqref{E:codimoneEE}, and first expand $A(x)^{d-2}$ in $\mu_1 L$,
\beq
\label{E:codimoneAexp}
\int_{-\xeps}^{\xeps} dx \, A(x)^{d-2} = \int_{-\xeps}^{\xeps} dx \, \left [ \cosh^{d-2}x - \mu_1L \, \frac{(d-2)}{4(d-1)}\cosh^{d-3}x\sinh|x| + \mathcal{O}(\mu_1^2L^2) \right].
\eeq
The first term on the right-hand side of eq.~\eqref{E:codimoneAexp} has the following expansion in $\mu_1 L$:
\beq
\label{E:codimoneAexp1}
\int_{-\xeps}^{\xeps} dx \, \cosh^{d-2}x  = \int_{-x^{(0)}}^{x^{(0)}} dx \, \cosh^{d-2}x  + \frac{\mu_1 L}{2(d-1)} \frac{(R^2-\rp^2)^{(d-2)/2}}{\varepsilon^{d-2}} + \mathcal{O}(\mu_1^2L^2),
\eeq
where $x^{(0)} \equiv \textrm{arccosh}\left( \frac{1}{\varepsilon}\sqrt{R^2-\rp^2}\right)$, the cutoff when the brane is absent. For the second term on the right-hand side of eq.~\eqref{E:codimoneAexp}, we perform the $x$ integration and then expand in $\mu_1 L$,
\bea
\label{E:codimoneAexp2}
-  \mu_1L \, \frac{(d-2)}{4(d-1)} \, \int_{-\xeps}^{\xeps} dx \, \cosh^{d-3}x\sinh|x| & = & - \frac{\mu_1L}{2(d-1)} \left [ \cosh^{d-2} x^{(0)} - 1 \right] + \mathcal{O}(\mu_1^2L^2) \\ & = & -\frac{\mu_1 L}{2(d-1)} \frac{(R^2-\rp^2)^{(d-2)/2}}{\varepsilon^{d-2}} + \frac{\mu_1L}{2(d-1)} + \mathcal{O}(\mu_1^2L^2). \nn
\eea
When we plug eqs.~\eqref{E:codimoneAexp1} and~\eqref{E:codimoneAexp2} into eq.~\eqref{E:codimoneAexp}, the two terms in eqs.~\eqref{E:codimoneAexp1} and~\eqref{E:codimoneAexp2} that are $\propto 1/\varepsilon^{d-2}$ cancel each other, so that
\beq
\int_{-\xeps}^{\xeps} dx \, A(x)^{d-2} = \int_{-x^{(0)}}^{x^{(0)}} dx \, \cosh^{d-2}x + \frac{\mu_1L}{2(d-1)} + \mathcal{O}(\mu_1^2L^2).
\eeq
The expansion in $\mu_1 L$ of the $\see$ in eq.~\eqref{E:codimoneEE} thus takes the form in eq.~\eqref{E:eegenform}, with
\begin{subequations}
\bea
\label{E:codimoneEE2}
\seez & = & \frac{L^{d-1}}{4G} \, \text{vol}(\mathbb{S}^{d-3}) \int d\rp \, \frac{R \, \rp^{d-3}}{(R^2-\rp^2)^{(d-1)/2}} \int_{-x^{(0)}}^{x^{(0)}} dx \, \cosh^{d-2}x, \\
\label{E:codimoneEEprobe}
\seeo & = &  \frac{\mu_1L}{2(d-1)}\frac{L^{d-1}}{4G}\text{vol}(\mathbb{S}^{d-3})\int d\rp \frac{R\rp^{d-3}}{(R^2-\rp^2)^{(d-1)/2}},
\eea
\end{subequations}
as found in ref.~\cite{Chang:2013mca}. In the next subsection we will reproduce eq.~\eqref{E:codimoneEEprobe} from $\sbr$ directly in the probe limit, without computing the back-reaction.

\subsubsection{CHM and Probe Branes}
\label{S:probeCHM}

In this subsection we show how to obtain $\seeo$ via a calculation entirely within the probe limit, with no calculation of the brane's back-reaction or of the minimal area surface, using our results from section~\ref{S:dcfts}. Our result will be valid for any $d$ and $n$. When $n=0$ or $n=1$ we will recover eq.~\eqref{E:codimzeroprobe} or~\eqref{E:codimoneEEprobe}, respectively, providing two non-trivial checks of our result.

In subsection~\ref{S:CHMcft}, we showed that for a DCFT in Minkowski space, when $\Sigma$ is a sphere of radius $R$ centered on the defect, $\see$ is equivalent to the thermal entropy of the DCFT on $\mathbb{R}\times\Hy^{d-1}$ at temperature $T_0=1/(2\pi R)$. To obtain $\seeo$, we can thus compute the order $\mu_n L^{2-n}$ contribution to that thermal entropy.

In general, holography gives us two ways to compute a field theory thermal entropy $S$ from a dual black brane solution with Hawking temperature $T$. The first way is to compute the horizon area and then identify the Bekenstein-Hawking entropy with $S$. The second way begins from the definition of AdS/CFT: the on-shell gravity action is the field theory generating functional. To compute field theory correlation functions, we take variational derivatives of the gravity action with respect to boundary values of bulk fields, and then evaluate the result on a solution. In particular, for a black brane solution, the Euclidean on-shell gravity action is equivalent to $F/T$, where $F$ is the field theory free energy. We can thus compute the Euclidean on-shell gravity action, extract $F$, and then compute $S=-dF/dT$.

For a DCFT on $\mathbb{R} \times \Hy^{d-1}$ at temperature $T_0$ dual to Einstein gravity in the spacetime of eq.~\eqref{E:defectGhyper}, we could in principle use either of these two ways to compute $S$. Indeed, we have already used the first way in subsection~\ref{S:CHMdcftbcft}, when we identified the Bekenstein-Hawking entropy of the hyperbolic horizon at temperature $T_0$ with $\see$. At the moment we cannot use the second way, however: in the second way we must compute $-dF/dT$ and evaluate the result at $T_0$, but to take $d/dT$ we need hyperbolic black brane solutions with $T$ different from $T_0$. As mentioned near the end of subsection~\ref{S:CHMdcftbcft}, currently we only know the hyperbolic black brane solution at $T_0$, eq.~\eqref{E:defectGhyper}.

We only want the order $\mu_n L^{2-n}$ contribution to $S$, however. To be precise, if $\mu_n L^{2-n} \ll 1$, then we can expand $F$ and $S$ in $\mu_n L^{2-n}$ as
\beq
F = \fz + \fo + \mathcal{O}(\mu_n^2 L^{2(2-n)}), \qquad S = \sz + \so + \mathcal{O}(\mu_n^2 L^{2(2-n)}),
\eeq
where, as with $\see$ in eq.~\eqref{E:eegenform}, the superscripts on $\fz$, $\fo$, $\sz$, and $\so$ indicate powers of $\mu_n L^{2-n}$. We want to compute $\fo$ and then $\so = -d\fo/dT$.

Luckily for us, in the probe limit the only possible contributions to the on-shell gravity action at order $\mu_n L^{2-n}$ are the probe brane's action $\sbr$ evaluated on the solution with $\mu_n L^{2-n}=0$, and boundary terms. Let us briefly sketch the proof of this statement, the details of which appear for example in ref.~\cite{Karch:2008uy}. The action in eq.~\eqref{E:bottomup} is a functional of the metric, $g$:
\beq
\label{E:bottomup2}
S_{d,n}[g] = S_{EH}[g] + \sbr[g].
\eeq
When $\mu_n L^{2-n} \ll 1$, we can expand $g$ in $\mu_n L^{2-n}$ as,
\beq
g = \gz +  \go + \mathcal{O}(\mu_n^2 L^{2(2-n)}),
\eeq
where $\gz$ is a solution to Einstein's equation with $\mu_n L^{2-n}=0$, and $\go \propto \mu_n L^{2-n}$ is the leading correction to the metric due to the brane's back-reaction. We can then expand $S_{d,n}[g]$ in $\mu_n L^{2-n}$,
\beq
\label{E:actionexpan}
S_{d,n}[g] = S_{EH}[\gz] + \go \, \frac{\delta S_{EH}}{\delta g}[\gz] + \sbr[\gz] + \mathcal{O}(\mu_n^2 L^{2(2-n)}),
\eeq
where for clarity we suppressed boundary terms. The key observation is that the bulk contribution to $\frac{\delta S_{EH}}{\delta g}[\gz]$ vanishes because $\gz$ solves Einstein's equation with $\mu_n L^{2-n}=0$. As a result, the only contributions to $S_{d,n}[g]$ at order $\mu_n L^{2-n}$ are from $\sbr[\gz]$, the probe brane action evaluated on the undeformed background metric $\gz$, and also possibly from the boundary term in $\go \, \frac{\delta S_{EH}}{\delta g}[\gz]$. That boundary term is of the form $\go$, evaluated at the boundary, times the expectation value of the field theory's stress-energy tensor, evaluated at $\mu_n L^{2-n}=0$. If we demand that the back-reaction does not change the metric of the space on which the field theory lives, then $\go \propto \gz$ at the $AdS_{d+1}$ boundary~\cite{Karch:2008uy}. The boundary term will thus be proportional to the trace of the field theory stress-energy tensor evaluated at $\mu_n L^{2-n}=0$, which vanishes (up to Weyl anomalies). The \textit{only} contribution to the on-shell action at order $\mu_n L^{2-n}$ is thus $\sbr[\gz]$.

Similar arguments apply also in gravity theories with more fields than just the metric and a brane: in general, the only contributions to the on-shell action at order $\mu_n L^{2-n}$ are from the probe brane action evaluated on the undeformed background, and from boundary terms. The latter generally take the form of the order $\mu_n L^{2-n}$ correction to a bulk field, evaluated at the boundary, times the expectation value of that field's dual operator, evaluated at $\mu_n L^{2-n}=0$. Such boundary terms will vanish if we impose a Dirichlet condition on the bulk field, or if the expectation value of the operator vanishes when $\mu_n L^{2-n}=0$. If the boundary terms do not vanish, then we must compute back-reaction, to obtain the order $\mu_n L^{2-n}$ corrections to bulk fields. In other words, in these cases we cannot compute $\so$ or $\seeo$ from the probe brane action alone. In subsection~\ref{S:abjmcodimzero} we will present an example from IIA string theory where indeed a boundary term involving the RR two-form field does not vanish.

The upshot is that for any DCFT described holographically by eq.~\eqref{E:bottomup}, we can compute $\so$ entirely within the probe limit, as follows. First we must find $\gz$, a solution of the Einstein equation in Euclidean signature and with $\mu_n L^{2-n}=0$ describing a black brane, with $T$ a free parameter. Next we compute the Euclidean probe brane action $\seuc[\gz]$, which we identify with $\fo/T$. Finally we compute $\so = -d\fo/dT$.

The result for $\so$ so obtained must agree with an ``honest'' calculation: including the brane's full back-reaction, computing $S$ in either of the two ways described above, expanding the result in $\mu_n L^{2-n}\ll 1$, and extracting $\so$. Indeed, the agreement of the two approaches, via $\sbr[\gz]$ or via back-reaction, has been checked in several cases where $\gz$ was a Poincar\'e-sliced $AdS_{d+1}$ black brane: see for example refs.~\cite{Mateos:2006yd,Bigazzi:2009bk}.

In our case $\gz$ is the metric in eqs.~\eqref{adshyp} and~\eqref{E:adshypbraneotherT}, Wick-rotated to Euclidean signature,
\beq
\label{E:hyperBBeuc}
\gz = \frac{dv^2}{f(v)} + f(v) \, \frac{L^2}{R^2} \, d\tau^2 + v^2 \, g_{\mathbb{H}^{d-1}}, \qquad f(v) = \frac{v^2}{L^2} - 1 - \frac{m}{v^{d-2}},
\eeq
where we also compactify $\tau$ into a circle of length $1/T$, so the dual field theory lives on $S^1 \times \Hy^{d-1}$. The metric in eq.~\eqref{E:hyperBBeuc} has a horizon at $v_H$ with the Hawking temperature in eq.~\eqref{E:adshypbraneotherTvalue},
\beq
\label{E:adshypbraneotherTvalue2}
T = \frac{2v_H^d+(d-2)mL^2}{4\pi L R v_H^{d-1}}.
\eeq
Solving $f(v_H)=0$ for $mL^2$ in terms of $v_H$ and plugging the result into eq.~\eqref{E:adshypbraneotherTvalue2}, we find $T$ in terms of $v_H$, or equivalently $v_H$ in terms of $T$,
\beq
\label{E:adshypbraneotherTvalue3}
v_H = \frac{L}{d} \left( \frac{T}{T_0} + \sqrt{\frac{T^2}{T_0^2} + d(d-2)}\right),
\eeq
which will be useful to us later. For the Euclidean probe brane action evaluated on $\gz$, $\seuc[\gz]$, we then find
\begin{subequations}
\bea
\seuc[\gz] & = & \label{E:probeactioneuc} \frac{\mu_n}{16\pi G} \int d^{d+1-n}\xi \, \sqrt{\text{det P}[\gz]}, 
\\ 
& = & \frac{\mu_n}{16\pi G} \, \frac{L}{R} \, \frac{1}{T} \, \vol(\mathbb{H}^{d-n-1}) \int_{v_H}^{\L} dv \, v^{d-n-1} \\ & = & \label{E:probeactiondiv} \frac{\mu_n}{16\pi G} \, \frac{L}{R} \, \frac{1}{T} \, \vol(\mathbb{H}^{d-n-1}) \frac{1}{d-n} \left( \L^{d-n} - v_H^{d-n} \right)\,,
\eea
\end{subequations}
where $\L$ is a cutoff at large but finite $v$.

Clearly $\seuc[\gz]$ will diverge if we remove the cutoff, $\L \to \infty$. As discussed at the beginning of section~\ref{S:dcfts}, we expect divergences in EE due to correlations among short-distance modes near $\S$, however the large-$\L$ divergence in eq.~\eqref{E:probeactiondiv} is dual to a short-distance divergence in correlators at any point in spacetime, not just at points near $\S$. Indeed, the large-$\L$ divergence in eq.~\eqref{E:probeactiondiv} appears in any calculation of a probe brane on-shell action, and is completely independent of $\S$. To eliminate the large-$\L$ divergence, we must perform holographic renormalization~\cite{Henningson:1998ey,Balasubramanian:1999re,Skenderis:2002wp}, adding diffeomorphism-covariant counterterms to the action of the bulk gravity theory at $v = \L$ to render the bulk variational problem well-posed, eliminating large-$\L$ divergences in the process.

The details of holographic renormalization for probe branes appear in ref.~\cite{Karch:2005ms}, so here we will be brief. We add to $\seuc$ a counterterm action $\sct$ consisting of terms localized at the cutoff surface $v = \L$ and invariant under diffeomorphisms on that surface. For a probe brane with the Euclidean action in eq.~\eqref{E:probeactioneuc}, $\sct$ takes the form~\cite{Karch:2005ms}
\beq
\label{E:sct}
\sct = -\frac{\mu_n}{16\pi G} \int_{\L} d^{d-n}\xi \left [ \frac{L}{d-n}\sqrt{\gamma} + \ldots \right]
\eeq
where $\gamma$ is the determinant of the induced metric on the cutoff surface and the $\ldots$ represents terms sub-leading in $\L$. The sub-leading terms are straightforward to determine, and the first few appear already in ref.~\cite{Karch:2005ms}. Generically they are of the form $\sqrt{\gamma}$ times curvature invariants built from the induced metric at $v = \L$. After inserting $\gz$ from eq.~\eqref{E:hyperBBeuc} into eq.~\eqref{E:sct}, we find non-trivial cancellations among the leading and sub-leading terms, with the result
\beq
\sct = -\frac{\mu_n}{16\pi G} \, \frac{L}{R} \, \frac{1}{T} \, \vol(\Hy^{d-n-1}) \, \frac{1}{d-n} \, \left( \L^{d-n} - \frac{1}{2} \, mL^2 \, \delta_{n0} + \ldots \right),
\eeq
where the $\L$-independent term $\propto m$ appears only when $n=0$, hence the Kronecker $\delta_{n0}$, and the $\ldots$ represents terms that vanish as $\L \to \infty$. We obtain the holographically renormalized probe brane action $\sbrren$ by summing $\seuc$ and $\sct$ and removing the cutoff,
\beq
\label{E:Sren}
\sbrren = \lim_{\L\to\infty}\left( \seuc + \sct \right) = -\frac{\mu_n}{16\pi G} \, \frac{L}{R} \, \frac{1}{T} \, \vol(\mathbb{H}^{d-n-1}) \frac{1}{d-n} \left( v_H^{d-n} - \frac{1}{2} mL^2 \, \delta_{n0}\right).
\eeq

We can now compute $\seeo$. Identifying $\fo = T \sbrren$ and defining
\beq
x_H \equiv \frac{v_H}{L} = \frac{1}{d} \left(\frac{T}{T_0}+\sqrt{\frac{T^2}{T_0^2}+d(d-2)}\right),
\eeq
we find
\beq
\label{E:F1}
\fo = -\frac{\mu_n L^{2-n}}{2(d-n)}\,\frac{L^{d-1}}{8 G}\, T_0 \, \text{vol}(\mathbb{H}^{d-n-1}) \, x_H^{d-n-2}\left( x_H^{2(1-\delta_{n0})} + x_H^2\right).
\eeq
We next compute $\so = - d\fo/dT$ and expand the result in $T$ near $T_0$,
\beq
\so = \frac{\mu_nL^{2-n}}{2(d-n)} \, \frac{L^{d-1}}{8 G} \, T_0 \, \text{vol}(\mathbb{H}^{d-n-1})\left[ \frac{1}{T_0}\frac{2(d-n)}{d-1+\delta_{n0}}+\mathcal{O}(T/T_0-1)\right].
\eeq
Finally, identifying $\so$ evaluated at $T=T_0$ as $\seeo$, we find
\beq
\label{E:een}
\seeo = \frac{\mu_nL^{2-n}}{2(d-1+\delta_{n0})}\frac{L^{d-1}}{4G}\text{vol}(\Hy^{d-n-1}).
\eeq

Eq.~\eqref{E:een} is the main result of this section. It is valid for any bulk gravity theory described by the action in eq.~\eqref{E:bottomup}. To obtain eq.~\eqref{E:een}, we needed only the background metric in eq.~\eqref{E:hyperBBeuc} and $\sbr$. We did not need to calculate the brane's back-reaction or a minimal area surface.

We can use eq.~\eqref{E:hypvol} to express the $\vol(\Hy^{d-1})$ factor in eq.~\eqref{E:een} as an integral over $\rp$. If we then set $n=0$, we find precisely eq.~\eqref{E:codimzeroprobe}, while if we set $n=1$ we find precisely eq.~\eqref{E:codimoneEEprobe}. These two cases thus provide two non-trivial checks of our method, as advertised.

\subsubsection{R\'enyi Entropies and Probe Branes}
\label{S:renyi}

Let $\sqz$ and $\sqo$ denote the $\mu_n L^{2-n}=0$ and the order $\mu_n L^{2-n}$ contributions to the R\'enyi entropies $S_q$, respectively. Expressing $S_q$ in terms of $F(T)$ as in eq.~\eqref{E:SrenyiF} and expanding in $\mu_n L^{2-n} \ll 1$, we find $\sqo$ in terms of $\fo(T)$,
\beq
\sqo = \frac{q}{1-q}\frac{1}{T_0}\left( \fo(T_0)-\fo\left( \frac{T_0}{q}\right)\right)\,.
\eeq
Using our result for $\fo(T)$ in eq.~\eqref{E:F1} to compute $\sqo$, we find
\begin{subequations}
\beq
\label{E:proberenyi}
\sqo = \frac{\mu_n L^{2-n}}{2(d-n)} \, \frac{L^{d-1}}{8 G} \, \vol(\Hy^{d-n-1}) \, \frac{q}{q-1}  \left(2-x_q^{d-n-2}(x_q^{2(1-\delta_{n0})}+x_q^2) \right),
\eeq
\beq
x_q \equiv x_H|_{T = \frac{T_0}{q}} = \frac{1+\sqrt{1+d(d-2)q^2}}{d q}.
\eeq
\end{subequations}
In the rest of this subsection, we will elaborate on some of the physics contained in the R\'enyi entropies in eq.~\eqref{E:proberenyi}.

First, we consider various limits of $\sqo$. As briefly mentioned in section~\ref{S:intro}, for a R\'enyi entropy $S_q$, $\lim_{q \to 1} S_q$ is the entanglement entropy, $\lim_{q \to 0} S_q$ is the logarithm of the total number of non-zero eigenvalues of $\rho$, and $\lim_{q \to \infty} S_q$ is minus the logarithm of the largest eigenvalue of $\rho$. For $\sqo$ in eq.~\eqref{E:proberenyi}, $\lim_{q \to 1} \sqo$ is indeed the $\seeo$ in eq.~\eqref{E:een}, which provides a simple consistency check of eq.~\eqref{E:proberenyi}. For the $q \to 0$ and $q \to \infty$ limits of $\sqo$, we find, respectively,
\begin{subequations}
\begin{align}
\label{E:sqosmallq}
\sqo &= \frac{\mu_n L^{2-n}}{2(d-n)} \frac{L^{d-1}}{4G} \vol(\Hy^{d-n-1}) \, \frac{ 2^{d-n-\delta_{n0}}}{d^{d-n} \, q^{d-n-1}} \, \left[ 1 + \mathcal{O}(q)\right],
\\
\label{E:sqolargeq}
\lim_{q \to \infty} \sqo &= \frac{\mu_n L^{2-n}}{2(d-n)} \frac{L^{d-1}}{4G} \vol(\Hy^{d-n-1}) \left[ 1 - \frac{\left[d(d-2)\right]^{(d-n)/2}}{d^{d-n}} \left(1+ \frac{\delta_{n0}}{d-2}\right) \right].
\end{align}
\end{subequations}
When the defect is extended in at least one spatial direction, $n < d-1$, from eq.~\eqref{E:sqosmallq} we see that the $q \to 0$ limit of $\sqo$ diverges as $q^{-d+n+1}$. This divergence is easy to understand. As $q \to 0$, the temperature $T = T_0/q \to \infty$. To leading approximation, we can then ignore the curvature of $\Hy^{d-n-1}$, and treat $\sqo \approx -\frac{q}{T_0} F^{(1)}(T_0/q)$ as $- q/T_0=-1/T$ times the order $\mu_n L^{2-n}$ contribution to the free energy of the DCFT in Minkowski space~\cite{Galante:2013wta}. By dimensional analysis, that contribution must be $\propto T^{d-n}$. The $q^{-d+n+1}$ divergence of $\sqo$ as $q \to 0$ is thus essentially the same as the divergence of the DCFT's free energy at $T$ high compared to the $\Hy^{d-n-1}$ curvature scale.

When $n=0$, the brane's back-reaction preserves $AdS_{d+1}$, so the dual field theory is a CFT. In that case, we can check our result for $\sqo$ against that of ref.~\cite{Hung:2011nu}: for a CFT dual to Einstein gravity on $AdS_{d+1}$ with radius $\ell$, the result of ref.~\cite{Hung:2011nu} for $S_q$ is
\beq
\label{E:SqMyers}
S_q = \frac{\ell^{d-1}}{8G} \, \text{vol}(\mathbb{H}^{d-1}) \, \frac{q}{q-1} \, (2-x_q^{d-2}(1+x_q^2)).
\eeq
The effect of the codimension-zero brane is simply to shift the $AdS_{d+1}$ radius of curvature as in eq.~\eqref{E:elldef}, so plugging eq.~\eqref{E:codim0Luseful} into eq.~\eqref{E:SqMyers} we find
\begin{subequations}
\label{E:SqCodim0}
\begin{align}
S_q^{(0)} &= \frac{q}{q-1}\frac{L^{d-1}}{8 G}\text{vol}(\mathbb{H}^{d-1})(2-x_q^{d-2}(1+x_q^2)),
\\
\label{E:SqAlt}
\sqo &= \frac{\mu_0L^2}{2d}S_q^{(0)}. \qquad (n=0)
\end{align}
\end{subequations}
Eq.~\eqref{E:SqAlt} precisely matches our result for $\sqo$ with $n=0$ in eq.~\eqref{E:proberenyi}, computed directly from the probe brane action.

For a CFT in Euclidean $\mathbb{R}^d$, let us define $C_T$ as the overall coefficient of the stress-energy tensor's two-point function,
\begin{subequations}
\beq
\langle T_{\mu\nu}(x) T_{\alpha \beta}(0) \rangle = C_T \frac{I_{\mu\nu,\alpha\beta}(x)}{x^{2d}},
\eeq
\beq
I_{\mu\nu,\alpha\beta}(x) \equiv \frac{1}{2} \left(I_{\mu\alpha}(x) I_{\nu\beta}(x) + I(x)_{\mu \beta} I(x)_{\nu\alpha} \right) - \frac{1}{d} \delta_{\mu\nu} \delta_{\alpha\beta}, \qquad I_{\mu\nu}(x) \equiv \delta_{\mu\nu} - 2 \frac{x_{\mu} x_{\nu}}{x^2}.
\eeq
\end{subequations}
In the CFT, when $\S$ is a sphere, $\partial S_q/\partial q$ at $q=1$ takes a universal form, meaning it is determined completely by $d$ and $C_T$~\cite{Hung:2011nu,Galante:2013wta,Perlmutter:2013gua,HMSunpub}
\beq
\label{E:ct}
\left . \frac{\partial S_q}{\partial q}\right |_{q=1} = - \vol(\Hy^{d-1}) \frac{\pi^{\frac{d}{2}+1} \Gamma\left(\frac{d}{2}\right) (d-1)}{(d+1)!} \, C_T\,.
\eeq
Using our result for $\sqo$ in eq.~\eqref{E:proberenyi} with $n=0$ we can thus determine the order $\mu_0 L^2$ contribution to $C_T$, which we denote $\cto$. From eq.~\eqref{E:proberenyi} with $n=0$ we find
\beq
\left . \frac{\partial \sqo}{\partial q} \right |_{q=1} = - \vol(\Hy^{d-1}) \frac{\mu_0 L^2}{2d} \frac{L^{d-1}}{8G},
\eeq
and hence, using eq.~\eqref{E:ct}, we identify
\beq
\label{E:cto}
\cto = \frac{(d+1)!}{\pi^{\frac{d}{2}+1}\Gamma\left(\frac{d}{2}\right)(d-1)} \,\frac{\mu_0 L^2}{2d} \frac{L^{d-1}}{8G}.
\eeq
Whether $\partial S_q / \partial q$ at $q=1$ takes a universal form similar to eq.~\eqref{E:ct} also in BCFTs and DCFTs is an important question that we will leave for future research.

For a CFT in even $d$ dual to Einstein gravity in $AdS_{d+1}$ of radius $\ell$, any central charge\footnote{Our conventions for the Weyl anomalies in $d=2$ and $d=4$ are given by
\begin{align}
\begin{split}
\label{E:2dcc}
d=2: & \hspace{.2cm} \langle T^{~\mu}_{\mu}\rangle = - \frac{c_2}{24\pi} \, R,
\\
d=4: & \hspace{.2cm} \langle T^{~\mu}_{\mu} \rangle = c_4 \, W_{\mu\nu\rho\sigma}W^{\mu\nu\rho\sigma} - a \, E_4\,,
\end{split}
\end{align}
where here $R$ is the Ricci scalar (not the radius of a spherical $\S$), $W_{\mu\nu\rho\sigma}$ is the Weyl tensor, and $E_4 = R_{\mu\nu\rho\sigma}R^{\mu\nu\rho\sigma}-4R_{\mu\nu}R^{\mu\nu}+R^2$ is the Euler density in $d=4$. In writing the $d=4$ Weyl anomaly, we have omitted a term $ \propto \nabla_{\mu}\nabla^{\mu}R$ that is allowed by Wess-Zumino consistency, but may be removed by the addition of a suitable local counterterm.} $c_d$ is proportional to the only dimensionless parameter we can build from $\ell$ and $G$, namely $\ell^{d-1}/G$. For instance, in $d=2$ and $d=4$ (with $c_4 = a$ for Einstein gravity in $AdS_5$~\cite{Henningson:1998ey,Balasubramanian:1999re}),
\beq
c_2  = \frac{3\ell}{2G}, \qquad c_4 = a = \frac{\ell^3}{128\pi G}.
\eeq
For our probe branes, when $d$ is even and $n=0$, let $\cdz$ denote the central charge at $\mu_0 L^{2-n}=0$ and $\cdo$ denote the order $\mu_0 L^2$ contribution to the central charge. Plugging $\ell^{d-1}$'s expansion in $\mu_0 L^2$ from eq.~\eqref{E:codim0Luseful} into $\ell^{d-1}/G$, we find that in general $\cdo = \frac{\mu_o L^2}{2d} \cdz$. For example,
\beq
\label{E:2d4dcc}
c_2 = \frac{3\ell}{2G} = \frac{3 L}{2 G} \left[ 1 + \frac{\mu_0 L^2}{4} + \Ocal\left(\mu_0^2 L^4\right)\right] = c_2^{(0)} + c_2^{(1)} + \Ocal\left(\mu_0^2 L^4\right).
\eeq
From eq.~\eqref{E:proberenyi}, when $n=0$ and $d$ is even we find $\sqo \propto c_d^{(1)}$. As a result, anything we obtain from a linear operation on $\sqo$ will also be $\propto c_d^{(1)}$. One example is $\lim_{q \to 1} \sqo = \seeo \propto c_d^{(1)}$, as we see in eq.~\eqref{E:een}. As a result, $\seeo$ will include a contribution $ \propto \cdo \ln \left(2 R/\varepsilon\right)$. As we mentioned below eq.~\eqref{E:SeeSphere}, the EE for a spherical $\S$ in a CFT in even $d$ includes a contribution $s_L \ln \left(2 R/\varepsilon\right)$, where $s_L$ is universal, and is proportional to the $a$-type central charge. The $\cdo \ln \left(2 R/\varepsilon\right)$ contribution to $\seeo$ is precisely the order $\mu_0 L^2$ contribution to that universal term. Another example is $\lim_{q \to 1} \partial S_q / \partial q \propto \cdo$, which implies $\cto \propto \cdo$. Indeed, for a CFT in even $d$, $C_T \propto c_d$, and specifically if $d=2$ then $C_T = c_2/(2\pi^2)$ while if $d=4$ then $C_T = 640 c_4/\pi^2$~\cite{Perlmutter:2013gua}. Our result for $\cto$ in eq.~\eqref{E:cto} gives $\cto = c_2^{(1)}/(2 \pi^2)$ when $d=2$ and $\cto = 640 c_4^{(1)}/\pi^2$ when $d=4$, as expected. In short, when $n=0$ and $d$ is even, from eq.~\eqref{E:proberenyi} we find $\sqo \propto \cdo$, which produces the expected dependence on the $a$-type central charge in various quantities obtained via linear operations on $\sqo$.

\subsection{Top-Down Systems}
\label{S:topdown}

We now want to compute $\seeo$ in top-down systems, namely SUSY brane intersections in string and M-theory, using our results from subsection~\ref{S:bottomup}.

To do so, we must perform two steps. The first step is to show that eq.~\eqref{E:bottomup} is a reliable effective action for the top-down system, at least within the probe limit. Let us enumerate the criteria for when that will be the case. First, the top-down system must admit a (consistent) KK truncation to $d+1$ dimensions. Furthermore, in string and M-theory all known SUSY branes source not only the metric, but also other supergravity fields, such as the dilaton and RR fields, and also support fields on their worldvolume, such as scalar fields describing motion of the brane in transverse directions. For eq.~\eqref{E:bottomup} to be a reliable effective action in the probe limit, after the KK reduction the supergravity theory's Einstein-frame stress-energy tensor must include only two contributions up to order $\mu_n L^{2-n}$: a negative cosmological constant and the probe brane's stress-energy tensor. Moreover, the latter must take a specific form, namely that obtained from variation of $\int d\xi^{d-n+1}\sqrt{-\textrm{P}[g]}$ with respect to $g$. If that is not the case, then the order $\mu_n L^{2-n}$ backreaction may not preserve $SO(2,d-n) \times SO(n)$ isometry, so the dual field theory may not have defect conformal symmetry, our proof that $\see$ maps to thermal entropy on $\mathbb{R} \times \Hy^{d-1}$ may not be valid, and we cannot guarantee that our method will work. In general if the brane's worldvolume fields are non-trivial then the brane's stress-energy tensor will take a different form, except for special solutions, such as certain worldvolume instantons~\cite{Ammon:2012mu,Gibbons:2000mx}. When these criteria are met, the top-down system's equations of motion will be identical in form to those obtained from eq.~\eqref{E:bottomup}, up to order $\mu_n L^{2-n}$, and so eq.~\eqref{E:bottomup} will be a reliable effective action.

If we can show that eq.~\eqref{E:bottomup} is a reliable effective action, then the second step is to match the parameters of the top-down system, namely Newton's constant, the $AdS_{d+1}$ radius, and the brane's tension, to the parameters in eq.~\eqref{E:bottomup}, $G$, $L$, and $\mu_n$. Having performed these two steps, we can simply plug the specific values of $G$, $L$, and $\mu_n$ of the top-down system into eq.~\eqref{E:een} to obtain $\seeo$. We could also obtain $\sqo$ from eq.~\eqref{E:proberenyi}, but we will only discuss $\seeo$ in the following.

An alternative to our method is the ``honest'' approach of ref.~\cite{Chang:2013mca}. Indeed, in ref.~\cite{Chang:2013mca} $\seeo$ was computed in two top-down systems with $d=4$ and either $n=0$ or $n=1$, each described in the probe limit by eq.~\eqref{E:bottomup}. As mentioned above, the approach of ref.~\cite{Chang:2013mca} is very general, and in particular does not require spherical $\S$ or defect conformal symmetry, and so does not require an $AdS_{d+1}$ background or a special form for the probe brane's stress-energy tensor.

In what follows we perform the two steps above in four examples. For simplicity we only consider branes with trivial worldvolume fields, and we only consider cases where a precise dictionary from gravity to field theory exists, so we can translate our results into field theory quantities. Our first two examples are precisely those of ref.~\cite{Chang:2013mca}, with which we find perfect agreement, followed by two examples from M-theory. We also present a fifth example, from type IIA supergravity, which illustrates how our method can go wrong, due to boundary terms involving fields sourced by the brane, as we explain in detail.

\subsubsection{$\N=4$ SYM with Codimension-Zero SUSY Flavor}

In type IIB string theory, consider $N$ D3-branes intersecting $N_f$ D7-branes in $(3+1)$ dimensions. At low energy the D3-brane worldvolume theory is $(3+1)$-dimensional $\N=4$ SYM with gauge group $SU(N)$ and 't Hooft coupling $\lambda$, coupled to $N_f$ fundamental hypermultiplets in such a way as to preserve $\mathcal{N}=2$ SUSY. $\N=4$ SYM is a CFT, but the flavor fields make a positive contribution $\propto \lambda^2 N_f/N$ to $\lambda$'s beta function, suggesting the existence of a Landau pole. In other words, the theory has a dynamically-generated scale: the D3/D7 theory is not a CFT. At sufficiently low energy and in the probe limit, however, the D3/D7 theory is approximately a CFT.

In the Maldacena limits, $N \to \infty$ with $\lambda$ fixed, followed by $\lambda \to \infty$, $\N=4$ SYM is dual to type IIB supergravity in $AdS_5 \times \mathbb{S}^5$, where the dilaton $\phi$ and hence the string coupling $g_s = e^{\phi}$ are constant, the RR five form has $N_c$ units of flux on the $\mathbb{S}^5$, and all other supergravity fields are zero. In terms of string theory parameters, the string-frame type II gravitational constant, $\kappa_{10}$, the radius $L$ of both $AdS_5$ and the $\Sp^5$, and the 't Hooft coupling are
\beq
\label{E:ads5L}
\kappa_{10}^2 = \frac{1}{2}(2\pi)^7 (\alpha')^4 g_s^2, \qquad L^4 = 4\pi g_s N (\alpha')^2, \qquad \lambda = 4\pi g_s N,
\eeq
where $\alpha'$ is the string length squared. In the probe limit the $N_f$ codimension-zero hypermultiplets are dual to $N_f$ D7-branes extended along $AdS_5 \times \Sp^3$~\cite{Karch:2002sh}. The tension of a D$p$-brane is
\beq
\label{E:dptension}
T_{Dp} = (2\pi)^{-p} \, (\alpha')^{-\frac{p+1}{2}} \, g_s^{-1}.
\eeq
The dimensionless parameter controlling the back-reaction of the D7-branes is thus
\beq
\kappa_{10}^2 \, N_f \, T_{D7} = \frac{\lambda}{8 \pi} \, \frac{N_f}{N}. 
\eeq

Let us now perform the two steps to justify using eq.~\eqref{E:een} in this case. The D7-branes source the metric and the axio-dilaton. When the D7-branes back-react the leading correction to each of these fields will be of order $\lambda N_f/N$. The type IIB Einstein-frame stress-energy tensor is (at least) quadratic in all fields, and the axio-dilaton is trivial in the $AdS_5 \times \Sp^5$ solution, so its leading contribution will be of order $\lambda^2 (N_f/N)^2$. The contribution from the D7-branes will be of order $\lambda N_f/N$. We thus find only two contributions to the Einstein-frame stress-energy tensor up to order $\lambda N_f/N$, an order-one term from the RR five-form, giving rise to a negative cosmological constant after KK reduction, and an order $\lambda N_f/N$ term, from the D7-branes. Eq.~\eqref{E:bottomup} will therefore be a reliable effective action in the probe limit.

Next we need to match string theory parameters to the parameters in eq.~\eqref{E:bottomup}, $G$, $L$, and $\mu_0$. The $AdS_5$ radius $L$ appears in eq.~\eqref{E:ads5L}. Dimensionally reducing on the $\mathbb{S}^5$ and using eq.~\eqref{E:ads5L}, for $G$ and $\mu_0$ we find
\beq
\label{E:IIBG}
\frac{1}{16\pi G} = \frac{L^5 \text{vol}(\mathbb{S}^5)}{2\kappa_{10}^2} =\frac{1}{8\pi^2}\frac{N^2}{L^3}\,, \qquad \mu_0 = (16\pi G) \, L^3 \text{vol}(\mathbb{S}^3) \, N_f T_{D7} = \frac{\lambda}{2\pi^2} \frac{N_f}{N} \frac{1}{L^2}\,.
\eeq
Notice that $L^3/G \propto N^2$ and $\mu_0 L^2 \propto N_f/N$, as advertised. Plugging $d=4$, $n=0$, and the above values of $G$, $L$, and $\mu_0$ into eqs.~\eqref{E:codimzero} and~\eqref{E:codimzeroprobe}, or equivalently eq.~\eqref{E:een}, and performing the integration for $\vol(\Hy^3)$, we find
\begin{subequations}
\begin{align}
\label{E:N4see}
\see^{(0)} & =N^2\left( \frac{R^2}{\varepsilon^2} - \ln \left(\frac{2R}{\varepsilon}\right) - \frac{1}{2} \right) + \Ocal\left(\varepsilon^2/R^2\right),
\\
\label{E:seeoD7}
\seeo &= \frac{\lambda \, N_f N}{16\pi^2} \left( \frac{R^2}{\varepsilon^2} - \ln \left(\frac{2R}{\varepsilon}\right) - \frac{1}{2} \right) + \Ocal\left(\varepsilon^2/R^2\right),
\end{align}
\end{subequations}
An ``honest'' calculation of $\seeo$ for the D3/D7 theory, computing the linearized back-reaction and using RT's prescription, appears in ref.~\cite{Chang:2013mca}. Our result agrees perfectly with that of ref.~\cite{Chang:2013mca}.

As mentioned below eq.~\eqref{E:SeeSphere}, for a CFT in even $d$, when $\S$ is a sphere the coefficient $s_L$ of the $\ln \left(2 R/\varepsilon\right)$ term in the EE is universal and proportional to the $a$-type central charge. In particular, if $d=4$ then $s_L = - 64 \pi^2 a$~\cite{Casini:2011kv}. From eq.~\eqref{E:N4see} we thus extract the central charge $a$ of $\N=4$ SYM without flavor, $a = N^2/(64 \pi^2)$, which of course agrees with eq.~\eqref{E:2d4dcc}. Thanks to $\N=1$ superconformal symmetry, this result for $a$ is independent of $\lambda$~\cite{Anselmi:1997am,Intriligator:2003jj}, and indeed this result for $a$ agrees with the free field result $(N^2-1)/(64 \pi^2)$ in the large-$N$ limit. In contrast, in the $\seeo$ in eq.~\eqref{E:seeoD7} the coefficient of $\ln \left(2R/\varepsilon\right)$ explicitly depends on $\lambda$. The reason is simple: as mentioned above, in the D3/D7 theory the flavor fields break conformal symmetry, so the coefficient of $\ln \left(2R/\varepsilon\right)$ in eq.~\eqref{E:seeoD7} is not protected by $\N=1$ superconformal symmetry, and indeed is not a central charge at all. Although our method for computing $\seeo$ relies on conformal symmetry, we are able to compute EE in the D3/D7 theory, which is non-conformal, by working to leading order in the deformation away from conformality, $\mu_0 L^2 \propto N_f/N$, in a manner similar to conformal perturbation theory.

\subsubsection{$\N=4$ SYM with Codimension-One SUSY Flavor}

In type IIB string theory, consider $N$ D3-branes intersecting $N_f$ D5-branes in $(2+1)$ dimensions. At low energy the D3-brane worldvolume theory is $\N=4$ SYM coupled to $N_f$ $(2+1)$-dimensional $\N=4$ hypermultiplet flavor fields. As shown in refs.~\cite{DeWolfe:2001pq,Erdmenger:2002ex}, these defect flavor fields preserve $SO(2,3)$ defect conformal symmetry for any $N$ and $N_f$, \textit{i.e.}\ the D3/D5 theory is a DCFT.

In the Maldacena and probe limits, the $N_f$ codimension-one hypermultiplets are dual to $N_f$ probe D5-branes extended along $AdS_4 \times \mathbb{S}^2$ inside $AdS_5 \times \Sp^5$~\cite{Karch:2000gx,DeWolfe:2001pq}. The dimensionless parameter controlling the back-reaction of the D5-branes is, using eqs.~\eqref{E:ads5L} and~\eqref{E:dptension},
\beq
\kappa_{10}^2 \, N_f \, T_{D5} \, L^{-2} = \sqrt{\lambda} \, \frac{\pi}{2} \, \frac{N_f}{N}. 
\eeq

The supergravity solution including the full back-reaction of the D5-branes appears in refs.~\cite{D'Hoker:2007xy,D'Hoker:2007xz}. In that solution, the Einstein-frame metric includes an $AdS_{4}$ factor, consistent with the fact that the D3/D5 theory has $SO(2,3)$ defect conformal symmetry for any $N$ and $N_f$. The metric in that solution also includes non-trivial dependence on the coordinates of the internal space, and only approaches the form in eq.~\eqref{genmet} asymptotically. We thus cannot use eq.~\eqref{adssliceamin} to compute $\see$. Instead we will use eq.~\eqref{E:een} to obtain just $\seeo$.

The D5-branes source the metric, dilaton, and RR three-form. When the D5-branes back-react the leading correction to each of these fields will be of order $\sqrt{\lambda} N_f/N$. The dilaton and the RR three-form and both are trivial in the $AdS_5 \times \Sp^5$ solution, hence their leading contributions to the Einstein-frame stress-energy tensor will be of order $\lambda (N_f/N)^2$. The contribution from the D5-branes will be larger, of order $\sqrt{\lambda} N_f/N$. Eq.~\eqref{E:bottomup} with $d=4$ and $n=1$ will therefore be a reliable effective action in the probe limit.

To use eq.~\eqref{E:een} for $\seeo$ we now just need to match string theory parameters to the parameters in eq.~\eqref{E:bottomup}, $G$, $L$, and $\mu_1$. For $G$ and $L$ the matching is identical to the previous case, eqs.~\eqref{E:ads5L} and~\eqref{E:IIBG}, respectively. Dimensionally reducing on the $\mathbb{S}^5$ and using eq.~\eqref{E:ads5L}, for $\mu_1$ we find
\beq
\mu_1 = (16\pi G) L^2 \, \text{vol}(\mathbb{S}^2) \, N_f T_{D5}  = \frac{4}{\pi}\sqrt{\lambda} \, \frac{N_f}{N}\frac{1}{L}.
\eeq
Plugging $d=4$, $n=1$, and the above values of $G$, $L$, and $\mu_1$ into eq.~\eqref{E:een}, and performing the integration for $\vol(\Hy^2)$, we find
\beq
\label{E:seeoD5}
\seeo = \frac{2}{3\pi}\sqrt{\lambda}N_f N \left( \frac{R}{\varepsilon} - 1\right).
\eeq
An ``honest'' calculation of $\seeo$ for the D3/D5 theory appears in ref.~\cite{Chang:2013mca}. Our result agrees perfectly with that of ref.~\cite{Chang:2013mca}. As mentioned in ref.~\cite{Chang:2013mca}, in a $(3+1)$-dimensional theory the characteristic leading divergence in an EE is $\propto 1/\varepsilon^2$, but here the codimension-one flavor fields produce a $1/\varepsilon$ divergence characteristic of an EE in a $(2+1)$-dimensional theory.

\subsubsection{ABJM Theory with Codimension-One SUSY Flavor}

In M-theory consider $N$ M2-branes sitting at the fixed point of an $\mathbb{R}^8/\Zk$ orbifold and intersecting $N_f$ M5-branes in $(1+1)$ dimensions. At low energy the M2-brane worldvolume theory is the ABJM theory, the $(2+1)$-dimensional $\N=6$ SUSY Chern-Simons-matter theory with gauge group $U(N)_k \times U(N)_{-k}$, where the $\pm k$ subscripts denote Chern-Simons levels~\cite{Aharony:2008ug}. When $k=1,2$ the SUSY is enhanced to $\N=8$. The M5-branes introduce $N_f$ $(1+1)$-dimensional hypermultiplets in the fundamental representation in each gauge group. These preserve $\N=(3,3)$ SUSY when $k>2$, which is enhanced to $\N=(4,4)$ SUSY when $k=1,2$~\cite{Ammon:2009wc}.

Taking the 't Hooft limit, $N \to \infty$ with the 't Hooft coupling $\lambda \equiv N/k$ fixed, and then taking $\lambda \gg 1$, when $N \gg k^5$ the ABJM theory is dual to eleven-dimensional supergravity in $AdS_4 \times \Sp^7/\Zk$, with $N$ units of seven-form flux on $\Sp^7/\Zk$, and where the $AdS_4$ has radius $L$ and the $\Sp^7/\Zk$ has radius $2L$~\cite{Aharony:2008ug}. Writing $\Sp^7$ as a $U(1)$ fibration over $\CP^3$, the $\Zk$ orbifold acts on the $U(1)$ fiber. If $k$ increases such that $\lambda \gg 1$ but $N \ll k^5$, then the $U(1)$ fiber shrinks and the approrpiate description is type IIA supergravity in $AdS_4 \times \CP^3$, with $N$ units of RR six-form flux on $\CP^3$ and $k$ units of RR two-form flux on the $\CP^1 \subset \CP^3$. We will call the 't Hooft limit with large 't Hooft coupling and $N \gg k^5$ or $N \ll k^5$ the M-theory or type IIA limit, respectively. We consider only the M-theory limit until subsection~\ref{S:abjmcodimzero}. In terms of the Planck length $l_P$, the gravitational constant in eleven dimensions, $\kappa_{11}$, and the radius $L$ are
\beq
\label{E:ads4L}
\kappa_{11}^2 = \frac{1}{2} (2\pi)^8 l_P^9, \qquad L^6 = \frac{\pi^2}{2} \, N k \, l_P^6.
\eeq
In the probe limit the $N_f$ codimension-one hypermultiplets are dual to $N_f$ M5-branes extended along $AdS_3 \times \Sp^3/\Zk$~\cite{Ammon:2009wc}. The tension of an M$p$-brane is
\beq
\label{E:mptension}
T_{Mp} = (2\pi)^{-p} \,\, l_P^{-(p+1)}.
\eeq
The dimensionless parameter controlling the back-reaction of the M5-branes is thus
\beq
\kappa_{11}^2 \, N_f \, T_{M5} \, L^{-3} =4 \pi^2\sqrt{2 \lambda}\frac{N_f}{N}.
\eeq

The M5-branes source the metric and are a magnetic source for the three form. When the M5-branes back-react the leading corrections to these fields will be of order $N_f\sqrt{\lambda}/N$. The stress-energy tensor of eleven-dimensional supergravity is quadratic in all fields. The key observation is that these M5-branes will source different components of the three-form than that of the background, so the order $N_f\sqrt{\lambda}/N$ correction to the three-form will produce an order $(N_f\sqrt{\lambda}/N)^2$ term in the stress-energy tensor. The contribution from the M5-branes will be of order $N_f\sqrt{\lambda}/N$. Eq.~\eqref{E:bottomup} with $d=3$ and $n=1$ will therefore be a reliable effective action in the probe limit.

To use eq.~\eqref{E:een} for $\seeo$ we now just need to match parameters of eleven-dimensional supergravity to the parameters in eq.~\eqref{E:bottomup}, $G$, $L$, and $\mu_1$. The $AdS_4$ radius $L$ appears in eq.~\eqref{E:ads4L}. Dimensionally reducing on $\Sp^7/\Zk$ and using eq.~\eqref{E:ads4L}, for $G$ and $\mu_1$ we find
\beq
\label{E:M5probe}
\frac{1}{16\pi G} = \frac{(2L)^7 \vol(\Sp^7/\Zk)}{2 \kappa_{11}^2}= \frac{N^2}{12\pi \sqrt{2\lambda}}\frac{1}{L^2}, \qquad \mu_1 = (16\pi G)N_f T_{M5} (2L)^3 \text{vol}(\mathbb{S}^3/\mathbb{Z}_k) = 3 \sqrt{2\lambda} \, \frac{N_f}{N}\frac{1}{L}.
\eeq
Plugging $d=3$, $n=1$, and the above values of $G$, $L$, and $\mu_1$ into eq.~\eqref{E:een}, and performing the integration for $\vol(\Hy^1)$, we find
\beq
\label{E:seeoM5}
\seeo = \frac{1}{2} N_f N \ln \left(\frac{2R}{\varepsilon}\right) + \co\left(\varepsilon^2/R^2\right).
\eeq
As mentioned below eq.~\eqref{E:SeeSphere}, for a CFT in even $d$, with a spherical $\S$ the EE includes a $\ln \left(2R/\varepsilon\right)$ term whose coefficient $s_L$ is universal and proportional to the $a$-type central charge. In particular, if $d=2$ then $s_L = c_2/3$. Eq.~\eqref{E:seeoM5} takes the form in eq.~\eqref{E:SeeSphere} with $d=2$, allowing us to identify a ``central charge'' $\ceo_2$ associated with the defect hypermultiplets,
\beq
\label{E:M5c}
\ceo_2 = \frac{3}{2} N_f N,
\eeq
which is precisely $1/4$ the central charge of $N_f N$ free hypermultiplets in $d=2$.

We must be careful by what we mean by a ``central charge,'' however. Consider a CFT on a curved manifold in $d=2$. We can define the central charge $c_2$ from the coefficient of the Ricci scalar in the trace anomaly, as in eq.~\eqref{E:2dcc}. Now consider a CFT on a curved manifold in $d=3$, with an $n=1$ defect along some spatial curve. We can define a Ricci scalar from the induced metric on the curve. We can also define an extrinsic curvature tensor $K_{\mu\nu}$, with trace $K$. The trace anomaly will receive no contribution from the CFT in $d=3$ but may receive contributions from the defect. Indeed, in general the trace anomaly will be a delta-function at the defect times a linear combination of three things: the induced metric's Ricci scalar, $K_{\mu\nu} K^{\mu\nu}$, and $K^2$. We can thus define \textit{three} central charges from the coefficients of these three terms, although Wess-Zumino consistency conditions may fix some in terms of the others. The central charge $\ceo_2$ in eq.~\eqref{E:M5c} is presumably a linear combination of these three central charges. Exactly which linear combination is an important question that we will leave for future research.

\subsubsection{ABJM Theory with Codimension-Two SUSY Flavor}

In M-theory consider $N$ M2-branes sitting at the fixed point of an $\mathbb{R}^8/\Zk$ orbifold and intersecting $N_f$ M2-branes in $(0+1)$ dimensions. We will denote the latter as M2$'$-branes to distinguish them from the $N$ M2-branes. At low energy the M2-brane worldvolume theory is the ABJM theory coupled to $N_f$ $(0+1)$-dimensional flavor fields preserving four real supercharges of SUSY when $k>2$, which is enhanced to eight when $k=1,2$~\cite{Ammon:2009wc}.

In the M-theory and probe limits the $N_f$ codimension-two flavor fields are dual to $N_f$ M2$'$-branes extended along $AdS_2 \times \Sp^1$ inside $AdS_4 \times \Sp^7/\Zk$~\cite{Ammon:2009wc}. The dimensionless parameter controlling the back-reaction of the M2$'$-branes is, using eqs.~\eqref{E:ads4L} and~\eqref{E:mptension},
\beq
\kappa_{11}^2 \, N_f \, T_{M2} \, L^{-6} = (2\pi)^4 \, \frac{4}{k} \, \frac{N_f}{N}.
\eeq

The M2$'$-branes source the metric and three-form. When the M2$'$-branes back-react the leading corrections to these fields will be of order $N_f/(Nk)$. The M2$'$-branes source different components of the three-form than the M2-branes source, however. The order $N_f/(Nk)$ correction to the three form will thus produce an order $N_f^2/(Nk)^2$ term in the stress-energy tensor, while the contribution of the M2$'$-branes will be of order $N_f/(Nk)$. Eq.~\eqref{E:bottomup} with $d=2$ and $n=2$ will therefore be a reliable effective action in the probe limit.

To use eq.~\eqref{E:een} for $\seeo$ we now just need to match the parameters of eleven-dimensional supergravity to the parameters in eq.~\eqref{E:bottomup}, $G$, $L$, and $\mu_2$. For $G$ and $L$ the matching is identical to the previous case, eqs.~\eqref{E:M5probe} and~\eqref{E:ads4L}, respectively. Dimensionally reducing on $\Sp^7/\Zk$ and using eqs.~\eqref{E:M5probe} and~\eqref{E:ads4L}, for $\mu_2$ we find
\beq
\mu_2 = (16\pi G) N_f T_2 (2L) \text{vol}(\mathbb{S}^1)= 12 \pi \, \frac{N_f}{N}.
\eeq
Plugging $d=3$, $n=2$, and the above values of $G$, $L$, and $\mu_2$ into eq.~\eqref{E:een}, and using $\vol(\Hy^0)=1$, we find
\beq
\label{E:seeoM2}
\seeo = \frac{\pi}{\sqrt{2\lambda}} \, N_f N.
\eeq

When $d=3$ a codimension-two defect is an impurity. In ref.~\cite{Jensen:2011su} a lattice of these probe M2$'$-brane impurities was used to model a Kondo lattice similar to those in heavy fermion compounds. Remarkably, the M2$'$-branes gave rise to an electrical resistivity linear in $T$, characteristic of ``strange metals'' such as the heavy fermion compounds. We can thus interpret eq.~\eqref{E:seeoM2} as an impurity entropy analogous to that in the Kondo effect. As mentioned in section~\ref{S:intro}, when $d=2$ the impurity entropy is strictly non-increasing along an RG flow. Whether the same is true in the system here is an important question that we will leave for future research.

\subsubsection{ABJM Theory with Codimension-Zero SUSY Flavor}
\label{S:abjmcodimzero}

In this final example we present a case where our method does not work. By studying why, we learn some valuable and quite general lessons about the role of boundary terms in our method.

In M-theory consider $N$ M2-branes sitting at the fixed point of an $\mathbb{R}^8/\Zk$ orbifold. We can introduce KK monopoles transverse to the M2-branes in such a way as to deform the orbifold singularity and change the M2-brane worldvolume theory. We will consider a particular orientation of $N_f$ KK monopoles, discussed in detail in ref.~\cite{Gaiotto:2009tk}, that gives rise to the ABJM theory coupled to $N_f$ $(2+1)$-dimensional hypermultiplet flavor fields preserving $\N=3$ SUSY when $k>2$, which is enhanced to $\N=4$ SUSY when $k=1,2$. We can split the $N_f$ flavor fields between the two gauge groups in $U(N)_k \times U(N)_{-k}$ as $N_f = N_1 + N_2$, with $N_1$ and $N_2$ fields in the fundamental representation of $U(N)_k$ and $U(N)_{-k}$, respectively. As shown in refs.~\cite{Gaiotto:2009tk,Hohenegger:2009as} and references therein, these flavor fields preserve superconformal symmetry for any $N$, $k$, and $N_f$.

In the M-theory limit, $N \gg N_f +k$ and $N \gg \left(N_f + k\right)^5$, the ABJM theory with $N_f$ codimension-zero hypermultiplets is dual to eleven-dimensional supergravity on $AdS_4 \times \cm_7$, where $\cm_7$ is a seven-dimensional manifold whose detailed properties are discussed in refs.~\cite{Gaiotto:2009tk,Hohenegger:2009as}. Here we only need to know two properties of $\cm_7$: the volume of $\cm_7$ is
\beq
\label{E:M7vol}
\vol(\cm_7) = \vol(\Sp^7) \frac{N_f + 2k}{2(N_f + k)^2},
\eeq
and if $N_f=0$ then $\cm_7 = \Sp^7/\Zk$. The $AdS_4$ radius $L$ and $(3+1)$-dimensional Newton constant $G$ are then given by
\beq
\label{E:ABJMmL}
L^6 = \frac{\pi^6}{6} \frac{N}{\vol(\cm_7)} \,  l_P^6\,, \qquad \frac{1}{16\pi G} = \frac{(2L)^7 \text{vol}(\mathcal{M}_7)}{2\kappa_{11}^2} = \frac{\pi}{12}\frac{N^{3/2}}{\sqrt{6 \, \text{vol}(\mathcal{M}_7)}} \, \frac{1}{L^2}.
\eeq
Using eq.~\eqref{E:ABJMmL}, we can compute $\see$ for any $N_f$, not just the probe contribution $\seeo$: plugging eq.~\eqref{E:ABJMmL} into the result for $\see$ in $AdS_4$, eq.~\eqref{E:adsEE} with $d=3$, we find
\beq
\label{E:seeN3ABJM}
\see = \frac{L^2}{4G}\text{vol}(\mathbb{H}^2) = \frac{2\pi^3}{3} \frac{N^{3/2}}{\sqrt{6\text{vol}(\mathcal{M}_7)}}\left( \frac{R}{\varepsilon}-1\right).
\eeq
Eq.~\eqref{E:seeN3ABJM} is valid not only in the M-theory limit but also in the type IIA limit, $(k+N_f)^5\gg N\gg k+N_f$, and in the type IIA limit combined with the probe limit $N_f \ll N$ and $N_f \ll k$. In particular, in the type IIA and probe limits we obtain $\seeo$ from eq.~\eqref{E:seeN3ABJM} by expanding $\text{vol}(\mathcal{M}_7)$ in $N_f$ to linear order,
\beq
\label{E:seeD6exact}
\see^{(1)} = \frac{\pi}{2\sqrt{2}}\, \sqrt{\lambda} \, N_f N \,\left( \frac{R}{\varepsilon}-1\right).
\eeq

Can we obtain the $\seeo$ in eq.~\eqref{E:seeD6exact} directly from the probe brane action, using eq.~\eqref{E:een}? In the type IIA and probe limits, the string parameters $g_s$ and $\alpha'$, and the $AdS_4$ radius $L_{\textrm{IIA}}$, are fixed in terms of the parameters of the $AdS_4 \times \Sp^7/\Zk$ M-theory solution as~\cite{Aharony:2008ug}
\beq
\label{E:ads4IIAL}
g_s^2 = \frac{(2L)^3}{k^3 \, l_P^3}, \qquad \a' = l_P^2, \qquad L_{\textrm{IIA}}^2 =\frac{(2L)^3}{4 \, k \, l_P}.
\eeq
In the type IIA and probe limits, the $N_f$ hypermultiplets are dual to $N_f$ probe D6-branes extended along $AdS_4 \times \RP^3$ inside $AdS_4 \times \CP^3$, where the split $N_f = N_1 + N_2$ is encoded in D6-brane worldvolume Wilson loops valued in $\pi_1(\RP^3) = \mathbb{Z}_2$~\cite{Gaiotto:2009tk,Hohenegger:2009as,Hikida:2009tp}. The dimensionless parameter that controls the back-reaction of the D6-branes is, using $\kappa_{10}^2$ from eqs.~\eqref{E:ads5L}, $T_{D6}$ from eq.~\eqref{E:dptension}, and eq.~\eqref{E:ads4IIAL},
\beq
\kappa_{10}^2 \, N_f \, T_{D6} \, L_{\textrm{IIA}}^{-1} = 2\pi \lambda \, \frac{N_f}{N}.
\eeq

If we attempt the two steps to justify using eq.~\eqref{E:een} in this case, then we fail in the first step: in the type IIA and probe limits, we find that eq.~\eqref{E:bottomup} with $d=3$ and $n=0$ is not a reliable effective action. The D6-branes source the metric and dilaton and are a magnetic source for the RR two-form. When the D6-branes back-react the leading correction to each of these fields will be of order $\lambda N_f/N$. The key observation is that the background already has $k$ units of RR two-form flux on the $\CP^1 \subset \CP^3$, and the D6-branes source components of the RR two-form parallel to those of the background. The Einstein-frame stress-energy tensor will thus receive \textit{two} contributions at order $\lambda N_f/N$: one from the D6-branes, and another from the square of the RR two-form, namely from a cross-term involving the product of the background value and the order $\lambda N_f/N$ correction. Eq.~\eqref{E:bottomup} is thus \textit{not} a reliable effective action in the probe limit, and as a result so our method for computing $\seeo$ is not guaranteed to work.

Indeed, if we na\"ively apply our result eq.~\eqref{E:een} in this case, then we find the wrong result for $\seeo$, as we will now explicitly demonstrate. If eq.~\eqref{E:bottomup} were the correct effective action for this system in the probe limit, then we would need to match the parameters of type IIA supergravity to those in eq.~\eqref{E:bottomup}. Dimensionally reducing on $\CP^3$ and using $\kappa_{10}^2$ from eq.~\eqref{E:ads5L}, $T_{D6}$ from eq.~\eqref{E:dptension}, and eq.~\eqref{E:ads4IIAL}, along with $\vol(\CP^3) =\pi^3/3!$ and $\vol(\RP^3)=\pi^2$, we find
\begin{subequations}
\beq
\frac{1}{16 \pi G} = \frac{(2L_{\textrm{IIA}})^6 \vol(\CP^3)}{2 \kappa_{10}^2} = \frac{1}{12 \pi \sqrt{2}} \, \frac{N^2}{\sqrt{\lambda}} \, \frac{1}{L_{\textrm{IIA}}^2},
\eeq
\beq
\mu_0 = (16 \pi G) N_f T_{D6} (2L_{\textrm{IIA}})^3 \vol(\RP^3) = 3 \, \lambda \, \frac{N_f}{N} \, \frac{1}{L_{\textrm{IIA}}^2}\,.
\eeq
\end{subequations}
Plugging $d=3$ and $n=0$ and the above values of $G$, $L_{\textrm{IIA}}$, and $\mu_0$ into eq.~\eqref{E:een}, and performing the integration for $\vol(\Hy^2)$, we find
\beq
\label{E:seeD6incorrect}
\seeo = \frac{\pi}{3 \sqrt{2}} \, \sqrt{\lambda} \, N_f N \left( \frac{R}{\varepsilon} - 1\right),
\eeq
which is clearly incorrect, being $2/3$ of the correct result in eq.~\eqref{E:seeD6exact}.

Suppose that instead of using any effective action, we attempted to compute $\seeo$ directly in supergravity, in either the M-theory or type IIA limit. The ABJM theory with codimension-zero hypermultiplets is a CFT for any $N$, $k$, and $N_f$~\cite{Gaiotto:2009tk,Hohenegger:2009as}, and correspondingly the back-reaction of the D6-branes or KK monopoles preserves $AdS_4$. We can therefore use our mapping of EE to thermal entropy in hyperbolic space: to compute $\seeo$ we could compute the order $\mu_0 L^2$ contribution to the Bekenstein-Hawking entropy of a hyperbolic black brane, $\so$, in a fully back-reacted $AdS_4$ supergravity solution. As discussed below eq.~\eqref{E:actionexpan}, the only contributions to the on-shell gravity action at order $\mu_0 L^2$, and hence to $\so$, come from the brane action evaluated on the undeformed background and from boundary terms. The fact that $\seeo$ computed directly from the brane action, eq.~\eqref{E:seeD6incorrect}, is only $2/3$ of the correct result, eq.~\eqref{E:seeD6exact}, indicates that the contributions from the boundary terms must in fact be non-vanishing, and indeed must contribute the remaining $1/3$ of the correct result. In the type IIA limit, using arguments similar to those below eq.~\eqref{E:actionexpan}, we find that the boundary term involving the metric vanishes. The only non-vanishing boundary term therefore involves the RR two-form, which must produce the missing $1/3$ of the correct result.

The RR two-form's contribution to $\seeo$ must be the same whether we use our mapping of EE to a thermal entropy on hyperbolic space or we use RT's prescription. In the latter case, however, the RR two-form's contribution is not a boundary term, but a bulk term: the RR two-form makes an order $\mu_0 L_{\textrm{IIA}}^2$ contribution to the Einstein-frame stress-energy tensor, as mentioned above, which must eventually produce an order $\mu_0 L_{\textrm{IIA}}^2$ correction to $\Am$. From the previous paragraph we can infer that in RT's prescription the RR two-form's bulk contribution ultimately produces exactly $1/3$ of the final result in eq.~\eqref{E:seeD6exact}.

The general lesson of this subsection is: to guarantee that our method will work in a top-down system, not only do we need the order $\mu_n L^{2-n}$ back-reaction of the brane to preserve $SO(2,d-n) \times SO(n)$ isometry, but also we need all boundary terms involving order $\mu_n L^{2-n}$ corrections to bulk fields to vanish, so that the only contribution to the on-shell bulk action at order $\mu_n L^{2-n}$ is from $\sbr$. When the bulk gravity theory is described to order $\mu_n L^{2-n}$ by eq.~\eqref{E:bottomup}, that is indeed the case, as we explained below eq.~\eqref{E:actionexpan}. If the bulk gravity theory is not described by eq.~\eqref{E:bottomup}, then we must show that the boundary terms involving order $\mu_n L^{2-n}$ corrections to bulk fields vanish before applying our method.

\section*{Acknowledgements}

We thank H.~Liu and M.~van Raamsdonk for useful conversations and correspondence. We especially thank H.-C.~Chang and A.~Karch for sharing a copy of ref.~\cite{Chang:2013mca} with us prior to publication, R.~Myers and E.~Perlmutter for reading and commenting on the manuscript, and J.~Estes, E.~Tsatis, and T.~Wrase for collaboration on closely related subjects. K.~J. would also like to thank the organizers of ``Holography and Applied String Theory'' workshop at the Banff International Research Station and the organizers of ``Relativistic Hydrodynamics and the Gauge Gravity Duality'' workshop at the Technion for their hospitality while a part of this work was completed. The research leading to these results has received funding from NSERC, Canada, the National Science Foundation under grant PHY-0969739, and the European Research Council under the European Community's Seventh Framework Programme (FP7/2007-2013) / ERC grant agreement no. 247252.

\begin{appendix}
\section{Global Minimization of the Area Functional}
\label{A:globmin}

In this appendix, we provide short proofs that each of the surfaces in eqs.~\eqref{zsol} and~\eqref{E:defectminareasol} produces the global minimum of the area functional $\A$ in their respective background geometries.

We begin with pure $AdS_{d+1}$, with the metric in eq.~\eqref{adspoin}. Our goal is to prove that among surfaces that extremize the $\A$ in eq.~\eqref{afunc} and that approach a sphere of radius $R$ at the $AdS_{d+1}$ boundary, the extremal surface in eq.~\eqref{zsol}, $z^2+r^2=R^2$, produces the global minimum of $\A$.

\begin{figure}[t]
\begin{center}
\includegraphics[width=3.5in]{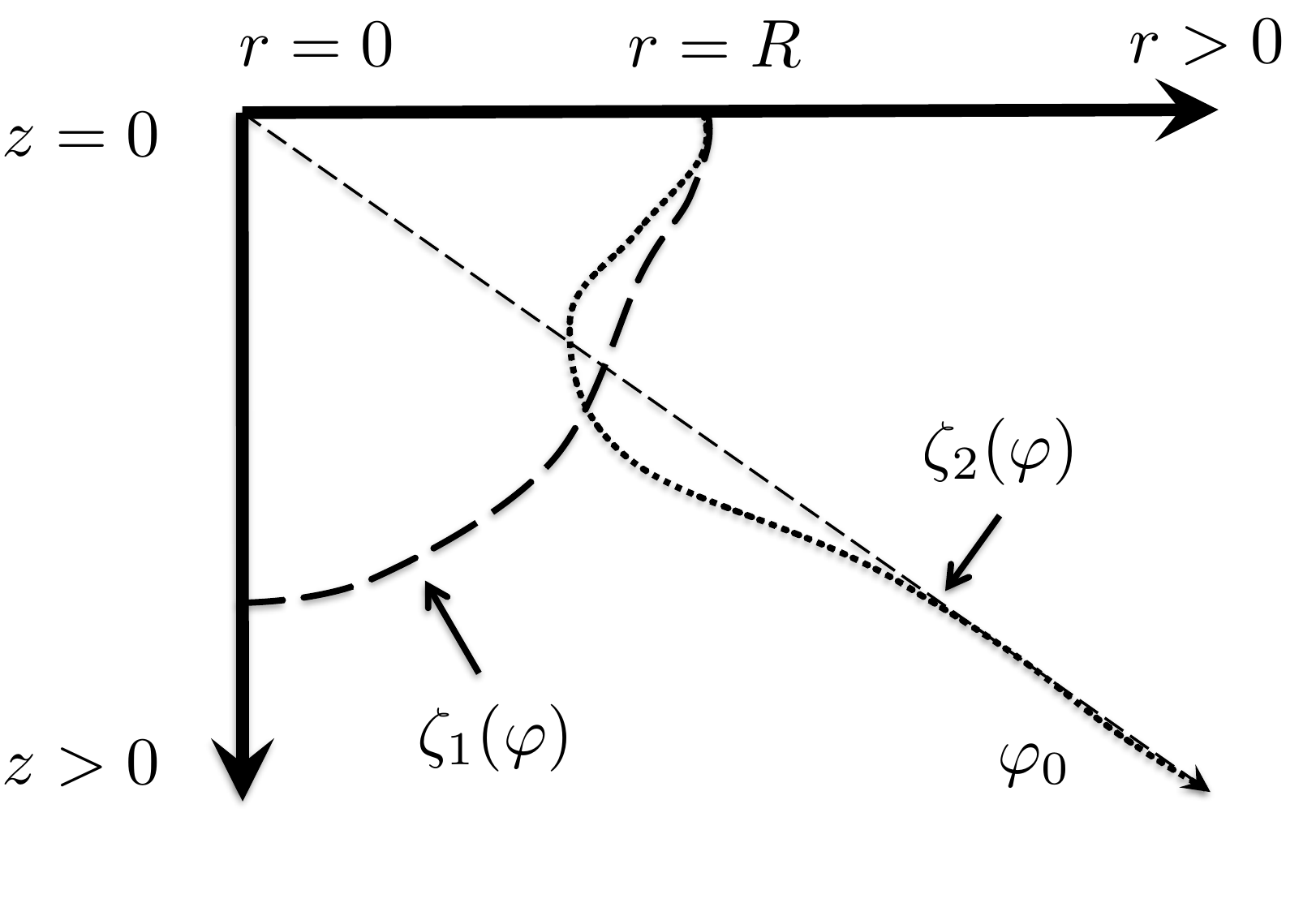}
\caption{\label{F:globalMin} A schematic representation of the two classes of solutions appearing in our proof that for surfaces in $AdS_{d+1}$ approaching a sphere of radius $R$ at the boundary, eq.~\eqref{zsol} produces the global minimum of the area functional $\A$. The solutions are curves parameterized in terms of their radius $\zeta$ as a function of the angular coordinate $\varphi \in [0,\pi/2]$ in the quadrant spanned by $r$ and $z$. The first class of solutions, of which the heavy-dashed curve $\zeta_1(\varphi)$ is a representative, are valued on the domain $[0,\pi/2]$. In the second class, of which the dotted curve $\zeta_2(\varphi)$ is a representative, $\zeta(\varphi) \to \infty$ at some angle $\varphi_0 \in (0,\pi/2)$, denoted by the thin-dashed line. The solution producing the global minimum of $\A$ is in the first class, and is simply the constant function $\zeta(\varphi)=R$.}
\end{center}
\end{figure}

In the quadrant spanned by $z$ and $r$, let us switch to polar coordinates $\zeta$ and $\varphi$,
\beq
z = \zeta \sin\varphi\,, \qquad r = \zeta \cos\varphi, \nn
\eeq
where $\zeta \in [0,\infty)$ and $\varphi \in [0,\pi/2]$. We will assume that the minimal area surface is simply-connected, tracing a curve that we parameterize as $\zeta(\varphi)$. The area functional then becomes
\beq
\label{E:generalArea}
\mathcal{A} =L^{d-1} \text{vol}(\mathbb{S}^{d-2}) \int d\varphi \,\frac{\cot^{d-2}\varphi}{\sin\varphi}\sqrt{1+\frac{\zeta'(\varphi)^2}{\zeta(\varphi)^2}},
\eeq
where the range of $\varphi$ integration depends on the endpoints of the curve $\zeta(\varphi)$. For the curve to describe a sphere of radius $R$ at the $AdS_{d+1}$ boundary, we demand that one and only one endpoint sits at $z=0$, that is, on the $r$ axis, at $r=R$. In other words we only consider curves that reach $\varphi=0$, such that as $\varphi \to 0$ the curve is single-valued and approaches $R$.\footnote{For all such solutions $\mathcal{A}$ diverges due to the $(\text{cot}\,\varphi)^{d-2}/(\sin\varphi)$ factor in eq.~\eqref{E:generalArea}. As above, we regulate the divergence by introducing a cutoff $z=\varepsilon$, or equivalently $\varphi \approx \varepsilon/R + \Ocal\left(\varepsilon^2/R^2\right)$.}  We can place the other endpoint of the curve in only two places, as shown in fig.~\ref{F:globalMin}. One option is to place the other endpoint on the $z$ axis, at finite, non-zero $z$, meaning $\zeta(\varphi)$ reaches a finite, non-zero value at $\varphi=\pi/2$. In that case, the solution with minimal area is easy to find. In the integrand of eq.~\eqref{E:generalArea}, the factor under the square root is a sum of squares, and attains a global minimum only when $\zeta'(\varphi)/\zeta(\varphi) = 0$ for all $\varphi$. The only such solution obeying $\lim_{\varphi \to 0} \zeta(\varphi)=R$ is the constant solution $\zeta(\varphi)=R$. The second option is to place the other endpoint at infinity: $\zeta(\varphi) \to \infty$ at some angle $\varphi_0 \in (0,\pi/2]$. In that case, $\zeta'(\varphi)/\zeta(\varphi)$ must diverge at $\varphi_0$ also: $\zeta(\varphi) = \textrm{exp} \int d\varphi \, \zeta'(\varphi)/\zeta(\varphi)$ diverges at $\varphi_0$ if and only if $\zeta'(\varphi)/\zeta(\varphi)$ diverges there. As a result, near the $\varphi_0$ endpoint, the integral in eq.~\eqref{E:generalArea} approaches $\int d\varphi |\zeta'(\varphi)/\zeta(\varphi)|=\ln \zeta(\varphi)$, which diverges at $\varphi_0$ by assumption. The unique solution with minimal area is thus $\zeta(\varphi)=R$, or equivalently $z^2 + r^2 = R^2$, which completes our proof.

We now turn to the metrics of the form in eq.~\eqref{genmet}, which are the backgrounds for gravity theories dual to BCFTs or DCFTs. Here we consider BCFTs and DCFTs in which the defect or boundary is extended in at least one spatial directions. The cases of point-like boundaries or defects are discussed at the end of section~\ref{S:CHMdcftbcft}. Our goal is to prove that among surfaces that extremize the area functional $\A$ in eq.~\eqref{adssliceA} and approach a (hemi-)sphere of radius $R$ centered on the defect or boundary, the extremal surface $Z^2+r_{||}^2=R^2$, with $Z$ and $\rp$ defined in eq.~\eqref{E:defectminareasol}, produces the global minimum of $\A$.

In the quadrant spanned by $Z$ and $\rp$, let us switch to polar coordinates $\zeta$ and $\varphi$,
\beq
Z = \zeta \sin\varphi, \qquad r_{||} = \zeta \cos\varphi, \nn
\eeq
where $\zeta \in [0,\infty)$ and $\varphi \in [0,\pi/2]$. We then choose to parameterize the solution as $\zeta(\varphi,x)$, which we assume is simply-connected. The area functional is then
\beq
\label{E:adssliceApolar}
\mathcal{A} = \text{vol}(\mathbb{S}^{n-1})\text{vol}(\mathbb{S}^{d-n-2})L^{d-1}\int dxd\varphi A(x)^{d-n-1}B(x)^{n-1}\frac{\cot^{d-n-2}\varphi}{\sin\varphi}\sqrt{1+ \frac{(\partial_{\varphi} \zeta)^2}{\zeta^2} + \frac{A(x)^2}{\sin^2\varphi}\frac{(\partial_x \zeta)^2}{\zeta^2}},
\eeq
where the ranges of the $x$ and $\varphi$ integrations depend on $\zeta(\varphi,x)$'s boundary conditions. For the solution to describe a sphere of radius $R$ in the field theory, we demand that $\zeta(\varphi,x)$ is single-valued and approaches $R$ as either $\varphi \to 0$ at any finite $x$ or as $|x| \to \infty$ at any fixed $\varphi$. The surface then has only two options. First, the surface can extend all the way to $\varphi = \pi/2$ and the other boundary of $x$. In that case, the minimal area surface is easy to find. In the integrand of eq.~\eqref{E:adssliceApolar}, the factor under the square root is a sum of squares, and attains a global minimum if and only if $(\partial_{\varphi} \zeta)^2/\zeta^2=0$ and simulateously $(\partial_x \zeta)^2/\zeta^2=0$. The only such solution obeying the boundary conditions is the constant solution $\zeta(\varphi,x)=R$. The second option is for $\zeta(\varphi,x)\to \infty$ along some curve in the space of $\varphi$ and $x$. That curve then determines the endpoints of integration in eq.~\eqref{E:adssliceApolar}. For such solutions, $(\partial_x \zeta)^2/\zeta^2$ must diverge at any fixed $\varphi$ on the curve, and $(\partial_{\varphi} \zeta)^2/\zeta^2$ must diverge at any fixed $x$ on the curve. As a result, $\A$ will diverge at the corresponding endpoints of integration. The unique solution with minimal area is therefore $\zeta(\varphi,x)=R$ or equivalently $Z^2 + \rp^2 = R^2$, which completes our proof.

\section{Cutoffs in the Holographic Duals of BCFTs and DCFTs}
\label{B:cutoffs}

In this appendix we explain how to implement the Poincar\'e-patch cutoff $z = \varepsilon$ for the minimal area integral $\Am$ in eq.~\eqref{adssliceamin} in the holographic duals of BCFTs and DCFTs, which have metrics of the form in eq.~\eqref{genmet}.

We begin by changing coordinates: in any asymptotically locally $AdS_{d+1}$ region we can write the metric in eq.~\eqref{genmet} in Fefferman-Graham (FG) form,
\beq
\label{E:gFG}
g = L^2 \left[ \frac{dz^2}{z^2}+ \frac{1}{z^2}\left(g^{(0)}_{\mu\nu}(x)dx^{\mu}dx^{\nu} + z\,g^{(1)}_{\mu\nu}(x)dx^{\mu}dx^{\nu} + \hdots\right)\right],
\eeq
where the dots indicate an expansion in powers of $z$ at small $z$.\footnote{In many cases, for example when the metric is the only non-trivial bulk field, only even powers of $z$ appear in the FG expansion. As we mentioned in section~\ref{S:CHMdcftbcft}, the dual of a BCFT or DCFT typically involves non-trivial matter fields, which can introduce odd powers of $z$, so we have allowed such odd powers in eq.~\eqref{E:gFG}.} The symmetries of the BCFT or DCFT restrict the FG expansion of the metric in eq.~\eqref{genmet} to be~\cite{Papadimitriou:2004rz,Estes:2012nx}
\beq
\label{E:genmetFG}
g = L^2 \frac{dz^2}{z^2} + \frac{L^2}{z^2}\left[ f_1\left( \frac{z}{\rn} \right) (-dt^2+dr_{||}^2+r_{||}^2g_{\mathbb{S}^{d-n-2}})+ f_2\left( \frac{z}{\rn}\right)dr_{\perp}^2 +f_3\left( \frac{z}{\rn}\right) r_{\perp}^2 g_{\mathbb{S}^{n-1}}\right],
\eeq
where in particular $f_1$, $f_2$ and $f_3$ depend only on $z/\rn$ because of dilatation invariance. Moreover, because $g$ is asymptotically locally $AdS_{d+1}$, $\lim_{z\to 0} f_1\left( \frac{z}{\rn}\right) = 1$, and similarly for $f_2$ and $f_3$. For the dual of a DCFT with $n=1$, the explicit change of coordinates that puts the metric of eq.~\eqref{genmet} into the FG form of eq.~\eqref{E:genmetFG} in any asymptotically locally $AdS_{d+1}$ region is~\cite{Papadimitriou:2004rz,Estes:2012nx},
\beq
\label{E:FGcoordchange}
z = Z \exp\left( \mp \int_{\pm \infty}^x dx' \sqrt{1-\frac{1}{A(x')^2}}\right)\,, \qquad \rn = Z \exp\left(\mp \int_{\pm\infty}^x \frac{dx'}{A(x')}\frac{1}{\sqrt{A(x')^2-1}} \right)\,,
\eeq
where $x'$ is a dummy variable. The $\pm$ encodes the fact that for a DCFT with $n=1$, the dual geometry has \emph{two} locally asymptotically $AdS_{d+1}$ regions far away from the defect. For the dual of a BCFT or a DCFT with $n>1$, which has only one asymptotically locally $AdS_{d+1}$ region, simply drop the $-\infty$ endpoints of the $x'$ integrals in eq.~\eqref{E:FGcoordchange}. Determining $f_1$, $f_2$, and $f_3$ from the metric in eq.~\eqref{genmet}, which is a function of $Z$ and $x$, requires inverting eq.~\eqref{E:FGcoordchange}~\cite{Papadimitriou:2004rz,Estes:2012nx}.

As mentioned below eq.~\eqref{genmet}, in any asymptotically locally $AdS_{d+1}$ region, $A(x)$ diverges exponentially in $x$, which guarantees that the integrals in eq.~\eqref{E:FGcoordchange} converge and thus are well-defined at large $|x|$. As we move to smaller $|x|$, however, generically $A(x)^2$ decreases until eventually $A(x)^2<1$ and the square roots in eq.~\eqref{E:FGcoordchange} become imaginary. At that point the FG expansion breaks down~\cite{Papadimitriou:2004rz,Estes:2012nx}. (The only exception we know is pure $AdS_{d+1}$, where $A(x) =\cosh x \geq 1$.) The breakdown of the FG expansion is easy to understand from eq.~\eqref{pointoads}: as $|x|$ decreases, $\rn$ decreases, so the expansion parameter $z/\rn$ grows, and we expect the expansion to break down. In other words, the expansion breaks down if in the field theory we move too close to the boundary or defect. We can thus picture the geometries of eq.~\eqref{genmet} as FG patches in each asymptotically locally $AdS_{d+1}$ region, describing physics away from the field theory's boundary or defect, smoothly connected to other regions that describe physics near the boundary or defect.

For the dual of a DCFT with $n=1$, whenever $A(x)\geq 1$ and a FG patch exists, we can implement the cutoff $z = \varepsilon$ by plugging the solution $Z(\rp)^2 + \rp^2 = R^2$ into eq.~\eqref{E:FGcoordchange},
\beq
\label{E:xrCutoffs}
\varepsilon = \sqrt{R^2-\rp^2} \, \exp\left( \mp \int_{\pm\infty}^{x_{\epsilon}(\rp,R)} dx' \sqrt{1-\frac{1}{A(x')^2}}\right).
\eeq
For the dual of a BCFT or a DCFT with $n>1$, which has only one FG patch, simply drop the $-\infty$ endpoint of the $x'$ integration in eq.~\eqref{E:xrCutoffs}. Although a solution-dependent cutoff sounds suspicious, we have introduced an $x$ cutoff that depends on the solution, $\xeps(\rp,R)$, which we adjust to maintain the solution-independent cutoff $z = \varepsilon$. An explicit example of such a cutoff surface, for a particular warp factor $A(x)$, appears in subsection~\ref{S:codimone}.

The cutoff surface $\xeps(\rp,R)$ couples the $x$ and $\rp$ integrations in eqs.~\eqref{adssliceA2} and~\eqref{adssliceamin}: we first integrate in $x$ up to $\xeps(\rp,R)$, and then integrate in $\rp$ from $\rp=0$ up to $\rp = \sqrt{R^2 - \varepsilon^2}$. In other words, those integrals do not actually factorize, as we claimed below eq.~\eqref{adssliceA2}. In the variation of $\A$ in eq.~\eqref{adssliceA}, any cutoff will only affect boundary terms, however, and not the Euler-Lagrange equation, so our argument that $Z(\rp)^2 + \rp^2 = R^2$ is a solution remains unchanged.

\end{appendix}

\bibliography{chmdefect}
\bibliographystyle{JHEP}

\end{document}